\documentclass[pdflatex,sn-mathphys-num]{sn-jnl}

\usepackage{graphicx}%
\usepackage{multirow}%
\usepackage{amsmath,amssymb,amsfonts}%
\usepackage{amsthm}%
\usepackage{mathrsfs}%
\usepackage[title]{appendix}%
\usepackage{xcolor}%
\usepackage{textcomp}%
\usepackage{manyfoot}%
\usepackage{booktabs}%
\usepackage{algorithm}%
\usepackage{algorithmicx}%
\usepackage{algpseudocode}%
\usepackage{listings}%
\usepackage{hyperref}

\unnumbered

\begin{document}


\title{Imaging how fluctuations destroy superconductivity in two dimensions
}


\author*[1,2]{\fnm{Logan} \sur{Bishop-Van Horn}}\email{logan.bvh@gmail.com}

\author[3,4]{\fnm{Teng} \sur{Zhang}}

\author[4,5]{\fnm{Sara} \sur{Metti}}

\author[3,7]{\fnm{Tyler} \sur{Lindemann}}

\author[3,4,5,6,7]{\fnm{Michael J.} \sur{Manfra}}

\author*[1,2]{\fnm{Kathryn A.} \sur{Moler}}\email{kmoler@stanford.edu}

\affil[1]{\orgdiv{Department of Physics}, \orgname{Stanford University}, \orgaddress{\city{Stanford}, \state{California} \postcode{94305}, \country{USA}}}

\affil[2]{\orgdiv{Stanford Institute for Materials and Energy Sciences}, \orgname{SLAC National Accelerator Laboratory}, \orgaddress{\city{Menlo Park}, \state{California} \postcode{94025}, \country{USA}}}

\affil[3]{\orgdiv{Department of Physics and Astronomy}, \orgname{Purdue University}, \orgaddress{\city{West Lafayette}, \state{Indiana} \postcode{47907}, \country{USA}}}

\affil[4]{\orgdiv{Birck Nanotechnology Center}, \orgname{Purdue University}, \orgaddress{\city{West Lafayette}, \state{Indiana} \postcode{47907}, \country{USA}}}

\affil[5]{\orgdiv{Elmore Family School of Electrical and Computer Engineering}, \orgname{Purdue University}, \orgaddress{\city{West Lafayette}, \state{Indiana} \postcode{47907}, \country{USA}}}

\affil[6]{\orgdiv{School of Materials Engineering}, \orgname{Purdue University}, \orgaddress{\city{West Lafayette}, \state{Indiana} \postcode{47907}, \country{USA}}}

\affil[7]{\orgname{Microsoft Quantum Lab West Lafayette}, \orgaddress{\city{West Lafayette}, \state{Indiana} \postcode{47907}, \country{USA}}}

\abstract{\unboldmath
Two-dimensional superconductors test our understanding of quantum phases. Low-temperature ``anomalous metal'' states are widely observed where conventional theories predict only superconducting or insulating ground states. In model systems based on proximity-coupled Josephson junction arrays, the critical temperature falls far faster with decreasing normal-state conductance than predicted by existing models. Whether these observations indicate a new quantum phase, a need to refine models of disorder and fluctuations, or both remains unresolved. Using scanning magnetic susceptometry, we image the local superfluid phase stiffness $\rho_s$ in gate-tunable Josephson junction arrays consisting of epitaxial Al islands on an InAs quantum well. Across six arrays and more than two orders of magnitude in the zero-temperature phase stiffness $\rho_{s0}$, the diamagnetic critical temperature $T_{c,\varphi}$ is suppressed more strongly than can be explained solely by thermal phase fluctuations, scaling approximately as $T_{c,\varphi}\propto\rho_{s0}^{0.3}$. While $\rho_s$ is remarkably uniform at high carrier densities, at low carrier densities we find that anomalous metal transport emerges together with large-scale inhomogeneity that transport alone cannot directly detect. These results are inconsistent with a classical XY description within standard junction models, instead pointing to quantum fluctuations that suppress superconductivity throughout the superconducting phase and amplify inhomogeneity near the superconductor-insulator transition, producing anomalous metal behavior.
}

\maketitle


Superconductivity in two dimensions is generally understood to be bracketed by the superconductor-to-insulator quantum phase transition at zero temperature~\cite{Goldman1998-hs,Sondhi1997-ab} and the Berezinskii–Kosterlitz–Thouless (BKT) topological phase transition at finite temperature~\cite{Kosterlitz1973-ou,Berezinskii1971-lz,Beasley1979-jw}.
Theories describing these phase transitions in two-dimensional (2D) superconductors are largely based on a picture of Josephson-coupled superconducting grains forming a 2D Josephson junction (JJ) array. These models yield clear predictions for the fate of superconductivity in zero magnetic field at both low and high temperature. At zero temperature, a variety of theories predict the collapse of superconductivity at a universal critical value of the normal state resistance between superconducting grains close to the pair resistance quantum $R_Q=h/(4e^2)=6.45\,\mathrm{k}\Omega$, where $h$ is the Planck constant and $e$ is the elementary charge. This transition may be driven by a disorder-induced enhancement of Coulomb repulsion that competes with Cooper pairing~\cite{Skvortsov2005-ru, Sacepe2020-eo}, or by zero-point fluctuations arising from the conjugate relationship between the Cooper pair number and phase on each grain~\cite{Orr1986-jz,Sondhi1997-ab}, or by dissipation, which stabilizes phase order by damping quantum fluctuations~\cite{Fisher1986-sq,Chakravarty1986-wv,Chakravarty1987-ad,Chakravarty1988-dl}. At high temperature, JJ array models reduce to the classical XY model which predicts a BKT transition characterized by a discontinuous drop in the superfluid phase stiffness $\rho_s$ at a universal value $\rho_s(T_\text{BKT})=(2/\pi)T_\text{BKT}$~\cite{Beasley1979-jw,Nelson1977-dh,Halperin1979-rq}.

The experimental situation is less clear. The existence of a low-temperature transition tuned by film thickness~\cite{Orr1986-jz,Jaeger1986-ym,Haviland1989-sn,Jaeger1989-kn}, proximity to a dissipative bath~\cite{Rimberg1997-it}, or electrostatic gating~\cite{Bollinger2011-fy,Garcia-Barriocanal2013-mf,Han2014-yv,Chen2018-uj,Bottcher2018-fx} has been robustly established. However, the observed critical properties often differ from their predicted universal values~\cite{Goldman1998-hs,Sacepe2020-eo}. In transport measurements of proximity effect JJ arrays, the critical temperature collapses much more rapidly with decreasing normal state conductivity than predicted by theory~\cite{Han2014-yv,Bottcher2018-fx,Bottcher2024-lu}. Moreover, in many 2D systems superconductivity is terminated not by an insulating state but by an ``anomalous metal'' regime in which the resistance saturates at a finite value with decreasing temperature~\cite{Kapitulnik2019-ao}. It is an open question whether this anomalous metal transport is due to the onset of a Bose metal phase~\cite{Das1999-fa,Phillips2003-va}, macroscopic quantum tunneling of vortices~\cite{Lin2012-ti}, spurious heating~\cite{Leonard2026-qp}, emergent spatial inhomogeneity~\cite{Kapitulnik2019-ao,Feigel-man1998-tb,Spivak2001-ad,Spivak2008-mf}, or some other mechanism.

In this work, we directly image the phase stiffness in a continuously tunable model system consisting of aluminum (Al) islands deposited epitaxially on a high-mobility indium arsenide (InAs) two-dimensional electron gas (2DEG) (Figure~\ref{fig:fig1}). The phase stiffness $\rho_s$ is a thermodynamic measure of the strength of superconducting order. It is defined by the free energy cost of a small twist in the phase $\varphi$ of the superconducting order parameter $\psi=|\psi|e^{i\varphi}$: $k_\text{B}\rho_s=\left.\partial^2F/\partial\varphi^2\right|_{\varphi=0}$, where $k_\text{B}$ is the Boltzmann constant and $F$ is the free energy~\cite{Fisher1973-jp,Rudnick1977-ap}. $\rho_s$ determines the linear response of a superconductor to applied magnetic fields. We measure this response using a scanning superconducting quantum interference device (SQUID) microscope.

We find that for all values of the gate-tuned carrier density in the 2DEG, $\rho_s(T)$ is nearly temperature-independent at low temperature, then decreases quasi-linearly to zero above a crossover temperature $T_*$, with no clear signature of a BKT transition at $\rho_s(T)=(2/\pi)T$ (Figure~\ref{fig:fig2}). Both $\rho_s(T)$ and the onset temperature for measurable diamagnetic response $T_{c,\varphi}$ are suppressed compared to a classical model of phase order in JJ arrays (Figure~\ref{fig:fig3}). This suppression cannot be solely due to thermal phase fluctuations~\cite{Emery1995-xq} because $E_\text{J}/k_\text{B}\geq \rho_{s}\gg T_{c,\varphi}$ over nearly the entire experimental parameter space, where $E_\text{J}$ is the Josephson coupling.
The critical temperature scales approximately as $T_{c,\varphi}\propto\rho_{s0}^{0.3}$, where $\rho_{s0}$ is the zero-temperature phase stiffness (Fig.~\ref{fig:fig3}c).

Using an applied gate voltage $V_g$, the arrays can be tuned at low temperature from superconductor to anomalous metal to insulator (Figure~\ref{fig:fig4}). At sufficiently high carrier density, $\rho_s$ is spatially uniform within our sensitivity (Figure~\ref{fig:fig5}d). In contrast, large scale spatial inhomogeneity in $\rho_s$ emerges at low carrier density (Fig.~\ref{fig:fig5}e). In this regime, the resistance saturates at a finite value at low temperature (Fig.~\ref{fig:fig4}b and Fig.~\ref{fig:fig5}). Eventually, $\rho_s$ vanishes at a gate voltage corresponding to a normal state sheet resistance $R_N$ close to $R_Q=h/(4e^2)$. Thus, there is a clear correlation between anomalous metal transport and spatial inhomogeneity in the phase stiffness near the putative dissipation-driven quantum phase transition~\cite{Fisher1986-sq,Chakravarty1986-wv,Chakravarty1987-ad,Chakravarty1988-dl,Rimberg1997-it}.

\section{InAs/Al Josephson junction arrays}

\begin{figure}[h!]
    \centering
    \includegraphics[width=4.76in]{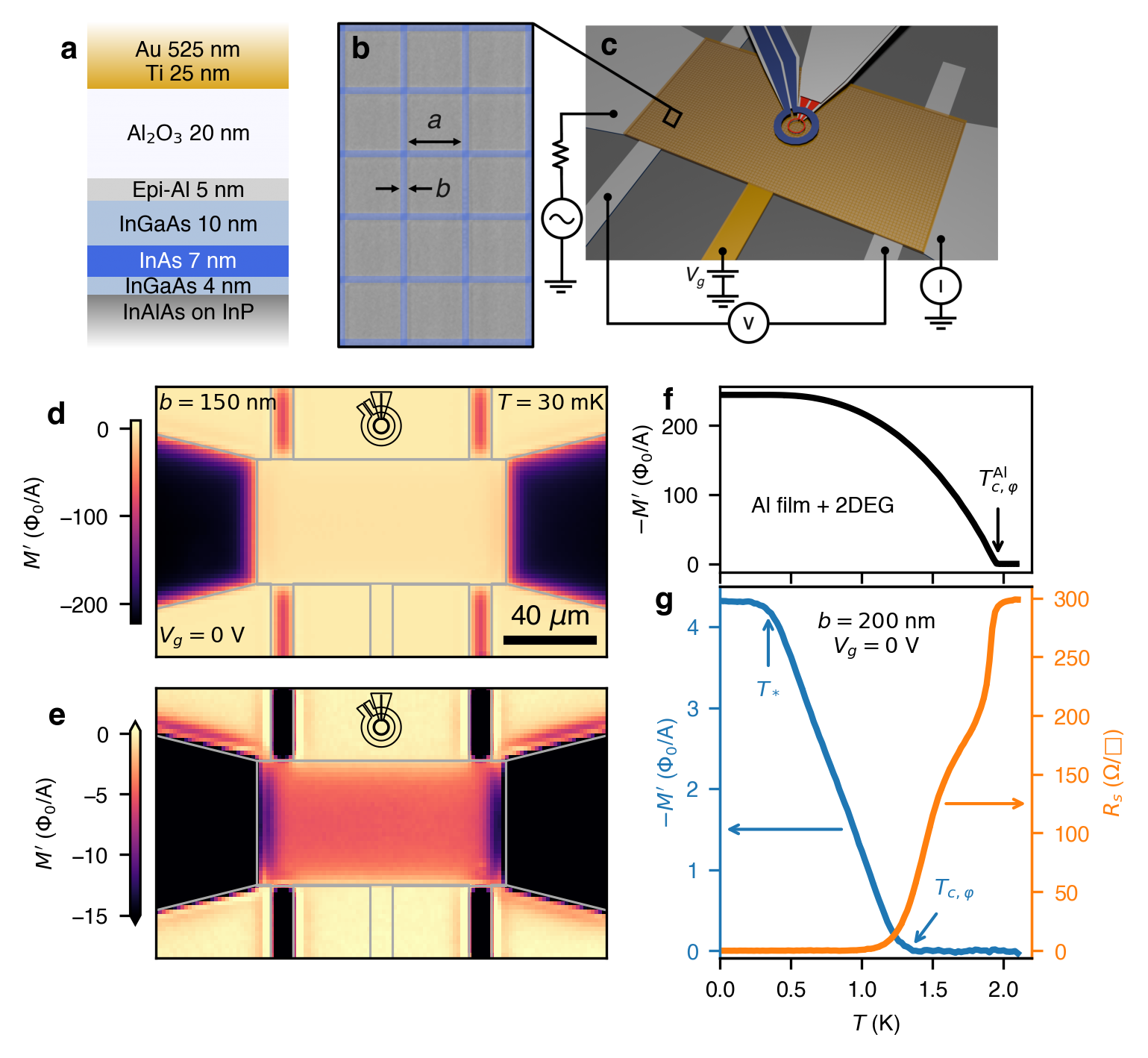}
    \caption{
    {\bf Measuring magnetic response and transport in gate-tuned InAs/Al Josephson junction arrays.}
    ({\bf a}) Schematic of the InAs/epitaxial Al heterostructure showing layer thickness and composition. The two-dimensional electron gas (2DEG) sits in the InAs layer. The epitaxial Al layer (Epi-Al) serves as current and voltage leads and creates superconductivity in the 2DEG through the proximity effect.
    ({\bf b}) False color scanning electron micrograph of a test device after deposition of the $\mathrm{Al}_2\mathrm{O}_3$ gate dielectric. The Al layer (gray) is etched to create a pattern of square Al islands of size $a\times a$ separated by regions of exposed 2DEG (blue) of width $b$. Here, $a=1\,\mu\mathrm{m}$.
    ({\bf c}) Schematic of the measurement setup. The Ti/Au top gate (gold) covers the array. A gate voltage $V_g$ tunes the carrier density in the 2DEG. A scanning SQUID with a pickup loop/field coil pair (red and blue respectively) measures magnetic susceptibility. Transport measurements are made in a standard four-point configuration, with the Al leads shown in light gray.
    ({\bf d}) SQUID susceptometry image of the $b = 150$ nm array at $V_g = 0$ V and $T = 30$ mK. The device design is drawn in light gray lines.
    ({\bf e}) Same data as (d) with the color scale saturated at [-15, 0.25] $\Phi_0/\mathrm{A}$ to show the diamagnetic response of the array. At this value of $T$ and $V_g$, the diamagnetic response of the array is uniform within our resolution.
    ({\bf f}) Temperature dependence of the diamagnetic response measured in a region of continuous Al film.
    ({\bf g}) Temperature dependence of the diamagnetic response (blue, left axis) and sheet resistance $R_s$ (orange, right axis) measured near the center of the $b=200$ nm array at $V_g=0$ V.
    }
    \label{fig:fig1}
\end{figure}

The JJ arrays (Figure~\ref{fig:fig1}) are formed from a 5 nm thick Al film deposited epitaxially on a shallow InAs quantum well (Fig.~\ref{fig:fig1}a). The devices are patterned into Hall bars (Fig.~\ref{fig:fig1}c) and on each device the Al layer is patterned into a $50 \times 100$ array of square islands of size $a=1\,\mu\mathrm{m}$ (Fig.~\ref{fig:fig1}b). The islands are separated by a distance $b$, resulting in a square lattice with lattice constant $a+b$. We measured six arrays fabricated on a single die with $b=150$, 200, 250, 300, 400, and 500 nm. A Ti/Au top gate is deposited over a 20 nm thick Al$_2$O$_3$ dielectric to tune the carrier density in the 2DEG between the Al islands. We estimate the island and junction charging energies to be $E_C/k_\mathrm{B} \approx 0.20$ K and $E_{C_\Sigma}/k_\mathrm{B} \approx 0.39$ K, respectively.

Fig.~\ref{fig:fig1}(d,e) shows a scanning SQUID susceptometry image of the array with island spacing $b=150$ nm. In scanning SQUID susceptometry~\cite{Gardner2001-gr,Huber2008-il,Kirtley2016-zz}, a field coil locally applies a small, low-frequency ($<1$ kHz) AC magnetic field to a sample. A pickup loop detects the sample's magnetic response as a change in the complex mutual inductance $M=M'+iM''$ between the field coil and pickup loop. In a superconductor, $M'$ is related to the superfluid response and $M''$ is related to dissipation, e.g., from vortex motion~\cite{Bishop-Van_Horn2023-ce}.
$M'$ is proportional to the inverse effective penetration depth $\Lambda^{-1}$ in the weak screening limit (i.e., when $\Lambda \gg r_\text{FC}$, where $r_\text{FC}$ is the radius of the field coil)~\cite{Kirtley2012-od}.
$\Lambda^{-1}$ is proportional to the phase stiffness $\rho_s=\hbar^2\Lambda^{-1}/(4\mu_0k_\text{B}e^2)$, where $\hbar=h/(2\pi)$ and $\mu_0$ is the vacuum permeability.
The constant of proportionality relating $M'$ to $\Lambda^{-1}$ depends on the geometry of the SQUID and the sample.
Uncertainty in this quantity results in a systematic relative uncertainty in $\rho_s$ that we estimate to be $-10\%/+20\%$ (Supplementary Information). We set the current in the field coil such that the applied flux per plaquette in the array is $<0.01\Phi_0$, where $\Phi_0=h/(2e)$ is the flux quantum, to probe the quasi-static linear magnetic response of the arrays.

At $V_g=0$ V the arrays exhibit a weak but spatially uniform magnetic response. For example, Figure~\ref{fig:fig1}(d,e) shows a scanning SQUID map of the magnetic response $M'$ in the array with $b=150$ nm at $V_g=0$ V and $T=30$ mK.
Fig.~\ref{fig:fig1}f shows the diamagnetic response $-M'$ measured in a region of continuous Al film, with a sharp onset at the diamagnetic critical temperature of the Al, $T_{c,\varphi}^\text{Al}=1.95$ K. Fig.~\ref{fig:fig1}g shows the temperature dependence of $-M'$ and the sheet resistance $R_s$ for the $b=200$ nm array. With decreasing temperature, $R_s$ drops in a two-step transition characteristic of proximity effect JJ arrays~\cite{Resnick1981-ow,Abraham1982-uh,Lobb1983-jf,Eley2011-id,Han2014-yv,Bottcher2018-fx}. The higher temperature transition occurs at $T_{c,\text{Al}}=1.90$ K (just below $T_{c,\varphi}^\text{Al}$) and is associated with the onset of phase coherence within each superconducting island. The lower temperature transition coincides with the diamagnetic critical temperature of the array, $T_{c,\varphi}$, and is associated with the onset of phase order between the islands.
Below $T_{c,\varphi}$, the diamagnetic response of the array, $-M' \propto \rho_s$, increases roughly linearly before saturating at a temperature $T_*$. At $V_g=0$ V, $R_s$ vanishes approximately 300 mK below $T_{c,\varphi}$.

\section{Gate voltage and temperature dependence of the phase stiffness}

\begin{figure}
    \centering
    \includegraphics[width=4.76in]{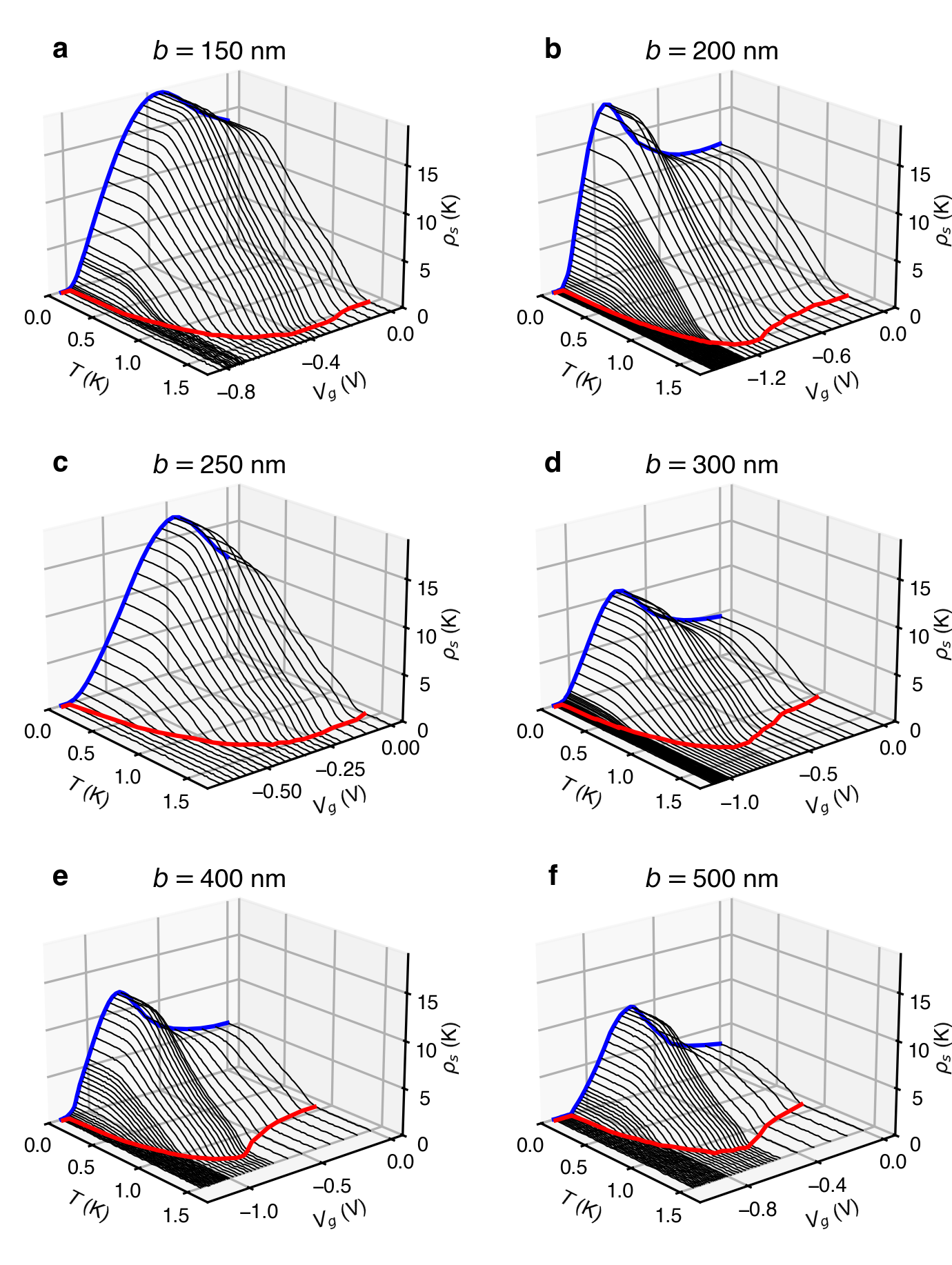}
    \caption{
    {\bf Gate voltage and temperature dependence of the phase stiffness $\rho_s$.} In the limit of weak magnetic screening in two dimensions, $\rho_s$ is proportional to the diamagnetic response $-M'$. ({\bf a}--{\bf f})  $\rho_s$ as a function of gate voltage $V_g$ and temperature $T$ for arrays with island spacing $b = 150$, 200, 250, 300, 400, and 500 nm. $\rho_s(T)$ has a similar shape at each value of gate voltage for each array. At low temperatures, $\rho_s(T)$ is nearly temperature-independent, approaching the value $\rho_{s0}\equiv\rho_s(T\to0)$. With increasing temperature, $\rho_s(T)$ decreases smoothly to zero at the diamagnetic critical temperature, $T_{c,\varphi}$. The gate voltage dependence of $\rho_{s0}$ and $T_{c,\varphi}$ are shown in blue and red, respectively. Both peak at an intermediate value of gate voltage similar to a peak in the normal-state sheet conductance (not shown) due to a peak in the mobility of the two-dimensional electron gas. We define $V_{g,\text{peak}}$ as the gate voltage at which  $\rho_{s0}$ is maximized.}
    \label{fig:fig2}
\end{figure}

Figure~\ref{fig:fig2} shows the phase stiffness $\rho_s(V_g, T)$ for the six arrays with varying island spacing $b$. For all temperature-dependent phase stiffness data, we have subtracted a temperature-dependent background that arises from coupling between portions of the SQUID circuit and Al film on the JJ array die far from the region of interest (Supplementary Information).
All devices exhibit a peak in $\rho_{s0}$ at an intermediate gate voltage, $V_{g,\text{peak}}$, which is due to a peak in the 2DEG mobility related to the presence of a second electronic subband (Supplementary Information).
The peak in $\rho_{s0}(V_g)$ corresponds to a peak in $T_{c,\varphi}(V_g)$ and the normal state sheet conductance $\sigma_N(V_g)=R_N^{-1}(V_g)$, where $R_N$ is measured at $T=2.1\,\text{K}>T_{c,\text{Al}}$.
The value of $V_{g,\text{peak}}$ varies between the six devices with different island spacing $b$, most likely as a result of differences in electrostatic screening by the islands or an offset in the carrier density at $V_g=0\,\mathrm{V}$ due to charge transfer from the Al~\cite{Zhang2023-gw,Chauhan2022-il}. Below, we demonstrate that the measured $\rho_s(V_g,T)$ is inconsistent with a classical XY description for all values of $V_g$.

The phase stiffness of an isotropic 2D system is $k_\text{B}\rho_s=\left.\partial^2F/\partial\varphi^2\right|_{\varphi=0}$, where $F$ is the free energy~\cite{Fisher1973-jp,Rudnick1977-ap}.
The free energy of a JJ with current-phase relation (CPR) $I_s(\varphi)$ is given by $F_\text{J}(\varphi)=\Phi_0/(2\pi)\int_0^\varphi I_s(\varphi')\,\mathrm{d}\varphi'$~\cite{Golubov2004-zq}.
The phase stiffness of a uniform JJ array in the fully phase-ordered state is therefore equal to the single-junction Josephson coupling, $k_\mathrm{B}\rho_s^\mathrm{max} = \left.\partial^2F_\text{J}/\partial\varphi^2\right|_{\varphi=0}=\Phi_0/(2\pi)\left.\partial I_s/\partial\varphi\right|_{\varphi=0}=E_\text{J}$. The actual phase stiffness $k_\text{B}\rho_s$ may be suppressed relative to $k_\mathrm{B}\rho_s^\mathrm{max}$ by phase fluctuations.
The temperature-dependent CPR of a short JJ with $N_\text{ch}$ conduction channels is given by
\begin{equation}
    I_s(\varphi, \{\tau_p\}, T)=\frac{e\Delta^2(T)}{2\hbar}\sin\varphi\sum_{p=1}^{N_\text{ch}}\frac{\tau_p}{\varepsilon_p(\varphi, T)}\tanh\left(\frac{\varepsilon_p(\varphi, T)}{2k_\text{B}T}\right),
    \label{eq:cpr}
\end{equation}
where $\Delta(T)$ is the gap of the junction leads, and $\tau_p\in[0, 1]$ and $\varepsilon_p(\varphi, T)=\Delta(T)\sqrt{1-\tau_p\sin^2(\varphi/2)}$ are respectively the transparency and Andreev bound state energy of conduction channel $p$~\cite{Beenakker1992-wj,Haberkorn1978-ii,Golubov2004-zq}.
Scanning SQUID measurements of the low-temperature CPR of individual InAs/Al nanowire Josephson junctions agree well with Eq.~\ref{eq:cpr}, except at specific values of the gate voltage at which the CPR is modified by charging effects~\cite{Spanton2017-zy,Hart2019-zg}.

Here, the relevant energy gap $\Delta(T)$ is the proximity-induced gap in the 2DEG beneath the Al islands, $\Delta_\text{ind}(T)$, rather than the gap of the islands themselves, $\Delta_\text{Al}(T)$. For a superconductor-2DEG interface, the induced gap is determined implicitly by
\begin{equation}
    \Delta_\text{ind}(T)=\Delta_S(T)\left(1+\frac{\gamma_\text{B}}{\pi k_\text{B}T_c}\sqrt{\Delta^2_S(T)-\Delta^2_\text{ind}(T)}\right)^{-1},
    \label{eq:induced-gap}
\end{equation}
where $\Delta_S$ and $T_c$ are respectively the Bardeen–Cooper–Schrieffer (BCS) gap and critical temperature of the ``parent'' superconductor, and $\gamma_\text{B} \geq 0$ quantifies the barrier strength at the superconductor-2DEG interface~\cite{Aminov1996-qs,Chrestin1997-vi,Schapers2001-jo,Kjaergaard2017-dh}. For a perfectly transparent interface, $\gamma_\text{B}=0$ and $\Delta_\text{ind}(T)=\Delta_S(T)$. For $\gamma_\text{B}>0$, the induced gap is suppressed relative to the parent gap and the two gaps converge as $T \to T_c$. The induced gap inferred from multiple Andreev reflection measurements of epitaxial planar InAs/Al junctions agrees well with Eq.~\ref{eq:induced-gap}~\cite{Kjaergaard2017-dh}.

\begin{figure}[h]
    \centering
    \includegraphics[width=4.5in]{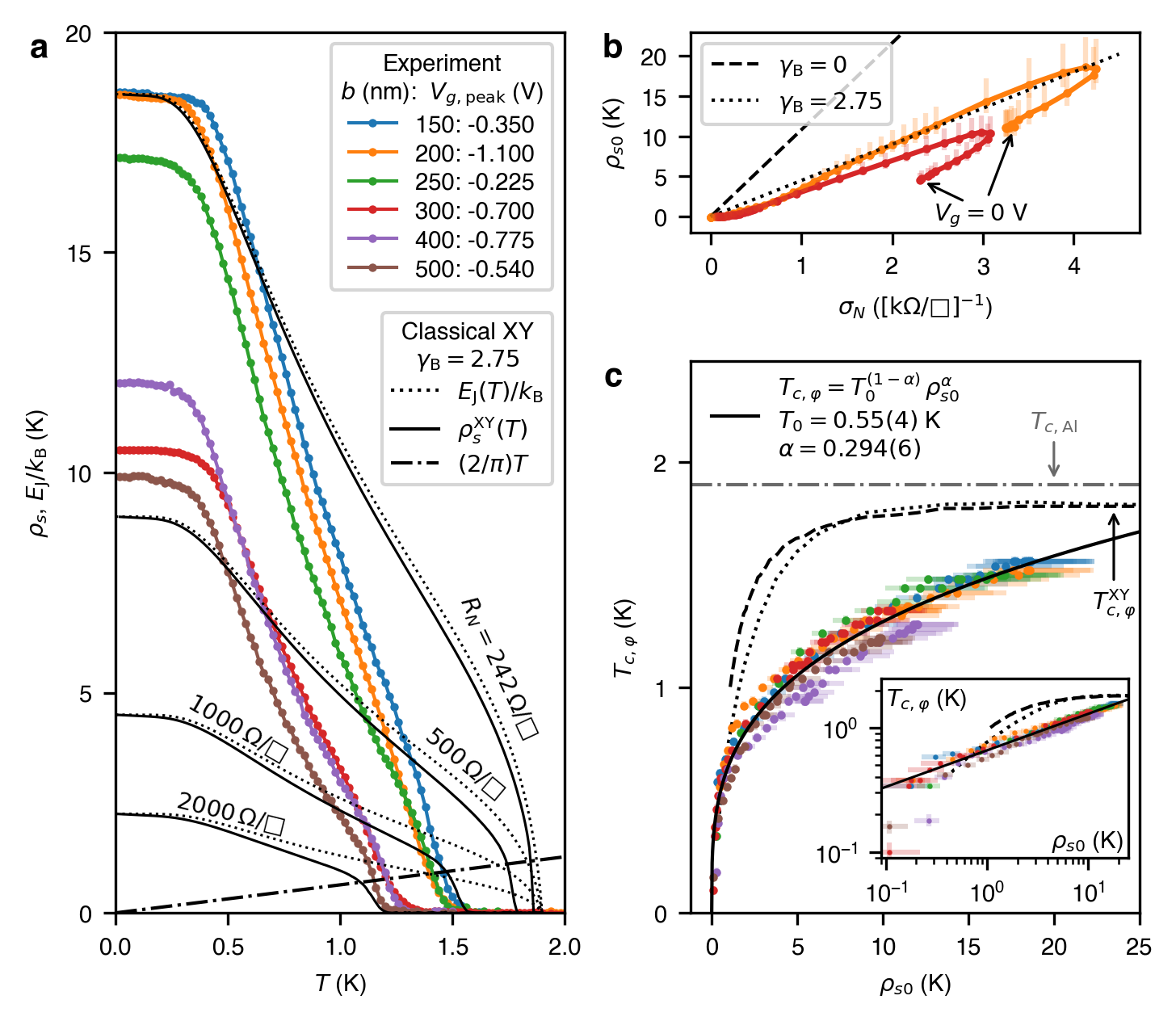}
    \caption{
    {\bf Incompatibility of the measured phase stiffness and critical temperatures with a purely thermal phase fluctuation scenario.}
    ({\bf a}) Measured phase stiffness at $V_{g,\text{peak}}$ for each array (colored circles), calculated Josephson coupling $E_\text{J}/k_\text{B}$ (dotted black lines), and calculated classical XY phase stiffness $\rho_s^\text{XY}$ (solid black lines) vs. temperature, showing that the large suppression of $T_{c,\varphi}$ relative to $T_{c,\text{Al}}$ is inconsistent with a classical XY description. $E_\text{J}/k_\text{B}$ and $\rho_s^\text{XY}$ are calculated for $\gamma_\text{B}=2.75$ and $R_N=\{242, 500, 1000, 2000\}\,\Omega/\square$. 
    ({\bf b}) $\rho_{s0}$ vs. $\sigma_N=R_N^{-1}$ for $b=200$ nm and $b=300$ nm, compared to the $\rho_{s0}^\text{XY}$ predicted from Eq.~\ref{eq:EJ} for $\gamma_\text{B}=0$ (dashed) and $2.75$ (dotted). $\gamma_\text{B}=2.75$ is most consistent with the measured $\rho_{s0}$ and $R_N$ for $b=200$ nm. However, note that the observed non-single-valued relationship is inconsistent with a naive classical XY picture, which predicts $\rho_{s0}=E_{\text{J}0}/k_\text{B} \propto \sigma_N$.
    ({\bf c}) $T_{c,\varphi}$ vs. $\rho_{s0}$ for all $b$ and $V_g$ both above and below $V_{g,\text{peak}}$ (colored dots), compared to the classical XY simulation for $\gamma_\text{B}=0$ (dashed black line) and $\gamma_\text{B}=2.75$ (dotted black line). 
    The solid black line shows a fit of the experimental data to the power law $T_{c,\varphi}=T_0^{(1-\alpha)}\rho_{s0}^\alpha$, with best-fit parameter values and uncertainties indicated in the upper left. 
    The error bars indicate the systematic uncertainty in the linear coefficient relating $M'$ and $\rho_s$. Inset: the same data and fit shown on a log-log scale. Individual fits to the six array datasets are shown in Fig.~\ref{fig:fig3-individual-fits}.
    }
    \label{fig:fig3}
\end{figure}

Combining Equations \ref{eq:cpr} and \ref{eq:induced-gap}, the effective Josephson coupling for a short planar superconductor-2DEG-superconductor junction is
\begin{equation}
\begin{split}
    E_\text{J}(T)&=\frac{\Phi_0}{2\pi}\left.\frac{\partial I_s}{\partial\varphi}\right|_{\varphi=0}\\
    &=\frac{\Phi_0\Delta_\text{ind}(T)}{4eR_N}\tanh\left(\frac{\Delta_\text{ind}(T)}{2k_\text{B}T}\right),
    \label{eq:EJ}
\end{split}
\end{equation}
where $R_N^{-1}=\sigma_N=(2e^2/h)\sum_{p=1}^{N_\text{ch}}\tau_p$ is the normal state conductance of the junction~\cite{Beenakker1992-wj}. Similar to the case for tunnel junctions~\cite{Ambegaokar1963-tr}, $E_\mathrm{J}$ is determined solely by the gap and $R_N$. For a given $R_N$, $E_\mathrm{J}$ is independent of the number and transparency of conduction channels in the junction.
The dashed and dotted black lines in Fig.~\ref{fig:fig3}b show the predicted zero-temperature Josephson coupling $E_{\text{J}0}/k_\text{B}$ for $\gamma_\text{B}=0$ and $\gamma_\text{B}=2.75$, respectively, assuming a BCS gap $\Delta_\text{Al}(0)=1.764\,k_\text{B}T_{c,\text{Al}}$ in the Al islands. $\gamma_\text{B}=2.75$ is chosen to match $\rho_{s0}$ for the $b=200$ nm array at $V_{g,\text{peak}}$ using the measured $R_N(V_{g,\text{peak}})=242\,\Omega/\square$. Because $\rho_{s0}$ is a lower bound on $E_{\text{J}0}/k_\text{B}$, $\gamma_\text{B}=2.75$ is an upper bound on the barrier strength within this model.

The effect of thermal phase fluctuations on a JJ array is described by the 2D classical XY model: $H_\text{XY}=-E_\text{J}(T)\sum_{\langle ij \rangle}\cos(\varphi_i-\varphi_j)$, where $\langle ij \rangle$ indicates nearest neighbor islands~\cite{Beasley1979-jw,Nelson1977-dh,Halperin1979-rq}. Note that while Eq.~\ref{eq:EJ} is valid for any CPR described by Eq.~\ref{eq:cpr}, the classical XY Hamiltonian $H_\text{XY}$ assumes a sinusoidal CPR.
In the classical XY model, the zero-temperature phase stiffness is $\rho_{s0}^\text{XY}=E_{\text{J}0}/k_\text{B}$. For $E_\text{J}(T)=E_{\text{J}0}$, $\rho_s^\text{XY}(T)$ decreases linearly with increasing temperature due to thermal phase fluctuations. The BKT transition occurs when $\rho_s^\text{XY}$ reaches a universal value $\rho_s^\text{XY}(T_\text{BKT})=(2/\pi)T_\text{BKT}\approx 0.56 E_{\text{J}0}/k_\text{B}$, at which point $\rho_s^\text{XY}$ drops abruptly to zero~\cite{Weber1988-ss}.
In our arrays, $E_\mathrm{J}(T)/k_\text{B} \geq \rho_s(T) \gg T_{c,\text{Al}} > T_{c,\varphi}$ over nearly all of the experimental parameter space. Thus, thermal phase fluctuations are expected to be important only for the lowest values of $\rho_s(T)$, where $\rho_s(T)$ is comparable to $T$.
To confirm this, we performed Monte Carlo simulations of the classical XY model (Supplementary Information) with $E_\mathrm{J}(T)$ given by Eq.~\ref{eq:EJ}, the results of which are shown in Fig.~\ref{fig:fig3}a and described below.

For $E_\text{J}(T)/k_\text{B}\gg T_{c,\text{Al}}$, thermal phase fluctuations are small and $\rho_s^\text{XY}(T)$ (solid black lines) closely tracks $E_\text{J}(T)/k_\text{B}$ (dotted black lines). By varying $R_N$ while keeping $T_{c,\text{Al}}$ and $\gamma_\text{B}=2.75$ fixed, we find that for any value of $E_{\text{J}0} / (k_\text{B}T_{c,\text{Al}})$ a BKT transition occurs when $\rho_s^\text{XY}(T)$ reaches $(2/\pi)T$. However, for $E_{\text{J}0} / (k_\text{B}T_{c,\text{Al}}) \gg 1$, $\rho_s^\text{XY}(T)$ is primarily driven toward $(2/\pi)T$ by $E_\mathrm{J}(T)$ rather than by thermal fluctuations. Although the transition at $T_{c,\varphi}$ may ultimately be a BKT transition, the impact of thermal BKT physics is largely incidental when $\rho_{s0}\gg T_{c,\varphi}$.
The difference between the dotted and solid black lines in Fig.~\ref{fig:fig3}a shows the suppression of phase stiffness due to thermal phase fluctuations.

Fig.~\ref{fig:fig3}d shows $T_{c,\varphi}(V_g)$ vs. $\rho_{s0}(V_g)$ for all values of $V_g$ (above and below $V_{g,\text{peak}})$ and island spacing $b$. Across all six devices, we observe a power law-like relationship between $T_{c,\varphi}$ and $\rho_{s0}$, which we fit to $T_{c,\varphi}=T_0^{(1-\alpha)}\rho_{s0}^\alpha$.
Individual fits to the six array datasets (Fig.~\ref{fig:fig3-individual-fits}) give $\alpha = 0.28 \pm 0.02$ (mean $\pm$ standard deviation; range $0.24–0.31$) and $T_0 = (0.59 \pm 0.08)$ K (range $0.47\,\text{K}-0.70\,\text{K}$). The aggregated fit (Fig.~\ref{fig:fig3}c) gives $T_0=0.55(4)$ K and $\alpha=0.294(6)$, but the spread between devices exceeds this uncertainty, so we regard the power law as an approximate empirical scaling. We note that the data systematically deviate from this power law relationship at the lowest values of $\rho_{s0}$, where thermal phase fluctuations become important and large-scale spatial inhomogeneity begins to emerge.
Fig.~\ref{fig:fig3}b shows $\rho_{s0}(V_g)$ plotted against the normal state conductivity $\sigma_N(V_g)$ for the two devices on which we performed a complete transport characterization ($b=200$ nm and $300$ nm). The $\rho_{s0}(\sigma_N)$ curves in Fig.~\ref{fig:fig3}b ``turn around'' at $V_{g,\text{peak}}$, resulting in a non-single-valued relationship that is inconsistent with the classical XY picture, which predicts $\rho_{s0}=E_{\text{J}0}/k_\text{B}\propto\sigma_N$.
The observed relationship between $\rho_{s0}(V_g)$ and $\sigma_N(V_g)$ therefore indicates that $\sigma_N$ not only determines $E_\text{J}$, but also modulates some other parameter that influences $\rho_{s0}$.
The critical temperature $T_{c,\varphi}^\text{XY}$ and zero-temperature phase stiffness $\rho_{s0}^\text{XY}$ extracted from the classical XY simulations for $\gamma_\text{B}=0$ and $\gamma_\text{B}=2.75$ are shown in Fig.~\ref{fig:fig3}c as dashed and dotted curves, respectively. The classical XY model does not capture the observed power law relationship between $T_{c,\varphi}$ and $\rho_{s0}$.

To summarize: $\rho_s$ collapses more rapidly with increasing temperature than expected from thermal phase fluctuations based on the predicted temperature dependence of the induced gap (Fig.~\ref{fig:fig3}a). $T_{c,\varphi}$ is primarily related to $\rho_{s0}$ rather than to $\sigma_N$, and $T_{c,\varphi}$ decreases much more rapidly with decreasing $\rho_{s0}$ than expected from thermal phase fluctuations (Fig.~\ref{fig:fig3}c). If $T_{c,\varphi}$ were determined only by thermal phase fluctuations then for any $\rho_{s0} > 6.5$ K, $T_{c,\varphi}$ would be within 10\% of $T_{c,\text{Al}}$. 
Thus, assuming a CPR resembling Eq.~\ref{eq:cpr} and an induced gap resembling Eq.~\ref{eq:induced-gap}, thermal phase fluctuations alone cannot explain the observed $\rho_s(T)$ or $T_{c,\varphi}$. This conclusion holds for any value of the Al/2DEG interface barrier strength $\gamma_\text{B}$.

We hypothesize that $\rho_s$ and $T_{c,\varphi}$ are suppressed by quantum phase fluctuations. For $\rho_{s0}\gg T_{c,\varphi}$, $\rho_s(T)$ is driven toward zero with increasing temperature by a combination of $E_\mathrm{J}(T)$ and quantum fluctuations. As $\rho_s(T)$ approaches $(2/\pi)T$ there may be a thermally-driven BKT transition, although it is not clearly resolved in the data. Depleting the carrier density below $V_{g,\text{peak}}$ decreases $\sigma_N$, which reduces both $E_{\text{J}0}$ and the strength of dissipation, ultimately destroying phase order. The significant impact of quantum phase fluctuations is unexpected given that over nearly all of the experimental parameter space the arrays are in what is generally considered to be the ``classical regime,'' $E_\text{J} \gg E_{C_\Sigma}$.

\section{Anomalous metal transport and emergent spatial inhomogeneity}

\begin{figure}
    \centering
    \includegraphics[width=4.76in]{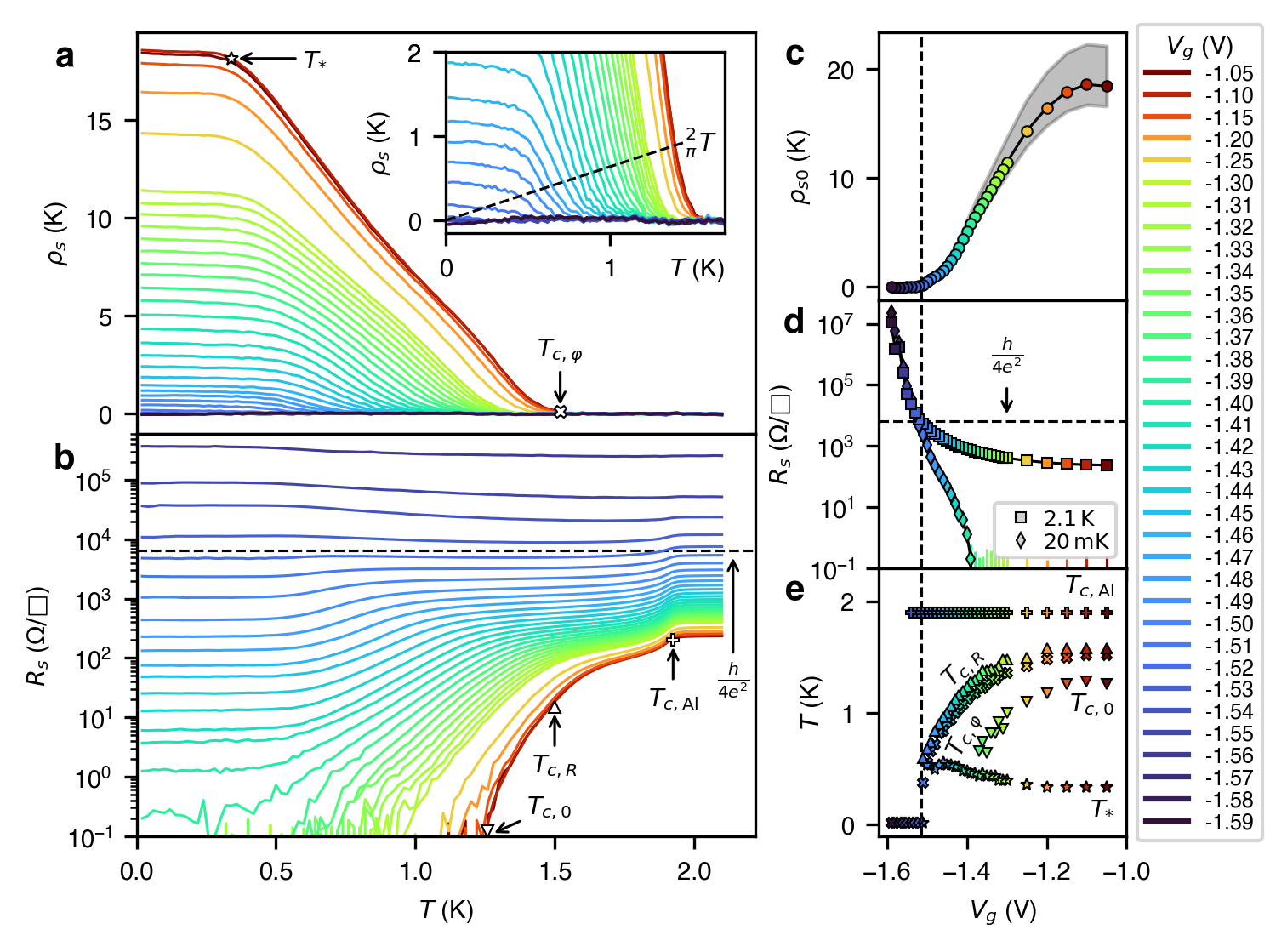}
    \caption{
    {\bf Phase stiffness and transport in the superconducting and anomalous metal regimes.}
    ({\bf a}) Phase stiffness $\rho_s$ and ({\bf b}) sheet resistance $R_s$ vs. temperature $T$ and gate voltage $V_g$ for the array with $b=200$ nm.
    ({\bf c}) $\rho_{s0}$ vs. $V_g$. The shaded region indicates the systematic uncertainty in the linear coefficient relating $M'$ to $\rho_s$ (Supplementary Information).
    ({\bf d}) Normal state sheet resistance $R_N$ ($\square$, $T=2.1\,\text{K}$) and low temperature sheet resistance $R_{s0}$ ($\diamond$, $T=20\,\text{mK}$) vs. $V_g$.
    ({\bf e}) Gate voltage dependence of the temperature scales indicated in (a, b).
    $T_{c,\text{Al}}$ ($+$): critical temperature of the Al film, identified as the high temperature maximum in $\partial R_s/\partial T$. 
    $T_{c,R}$ ($\triangle$): resistive transition of the array, identified as the lower temperature maximum in $\partial R_s/\partial T$.
    $T_{c,0}$ ($\triangledown$): temperature at which $R_s$ falls below the measurement sensitivity $R_s\lesssim 0.1\,\Omega/\square$.
    $T_{c,\varphi}$ ($\times$): onset of diamagnetic response, defined as the highest temperature for which $\rho_s>0.1$ K.
    $T_*$ ($\star$): crossover temperature below which $\rho_s(T)$ saturates.
    The vertical dashed line in (c--e) indicates the value of $V_g$ at which $R_N$ is closest to $R_Q$. For clarity, the most insulating curves are not shown in (b).
    }
    \label{fig:fig4}
\end{figure}

Figure~\ref{fig:fig4} shows the measured $\rho_s$ and $R_s$ as a function of  $T$ and $V_g$ for island spacing $b=200$ nm. The sheet resistance exhibits a two-step transition as a function of $T$, consistent with previous transport studies of similar InAs/Al JJ arrays~\cite{Bottcher2018-fx,Bottcher2024-lu,Sasmal2025-kv}. We empirically identify these two transitions by peaks in $\partial R_s/\partial T$. The higher temperature transition occurs at $T_{c,\text{Al}} = 1.9$ K, just below the temperature $T_{c,\varphi}^\text{Al}$ at which the diamagnetic response of the Al film vanishes (Fig.~\ref{fig:fig1}f).
The lower temperature resistive transition, labeled $T_{c,R}$, shifts down in temperature with decreasing carrier density and closely tracks the diamagnetic critical temperature of the array, $T_{c,\varphi}$.
For $V_g > -1.4$ V, $R_s$ drops to zero at a temperature labeled $T_{c,0}$. The five temperatures extracted from this data ($T_{c,\varphi}$, $T_*$, $T_{c,\text{Al}}$, $T_{c,R}$, and $T_{c,0}$) are shown as a function of $V_g$ in Fig.~\ref{fig:fig4}e. For $-1.52\,\text{V} < V_g \leq -1.4\,\text{V}$, a resistive transition occurs at $T_{c,R}$, but rather than dropping to zero the resistance saturates at a finite value at low temperature. This is the anomalous metal regime. 

\begin{figure}
    \centering
    \includegraphics[width=\linewidth]{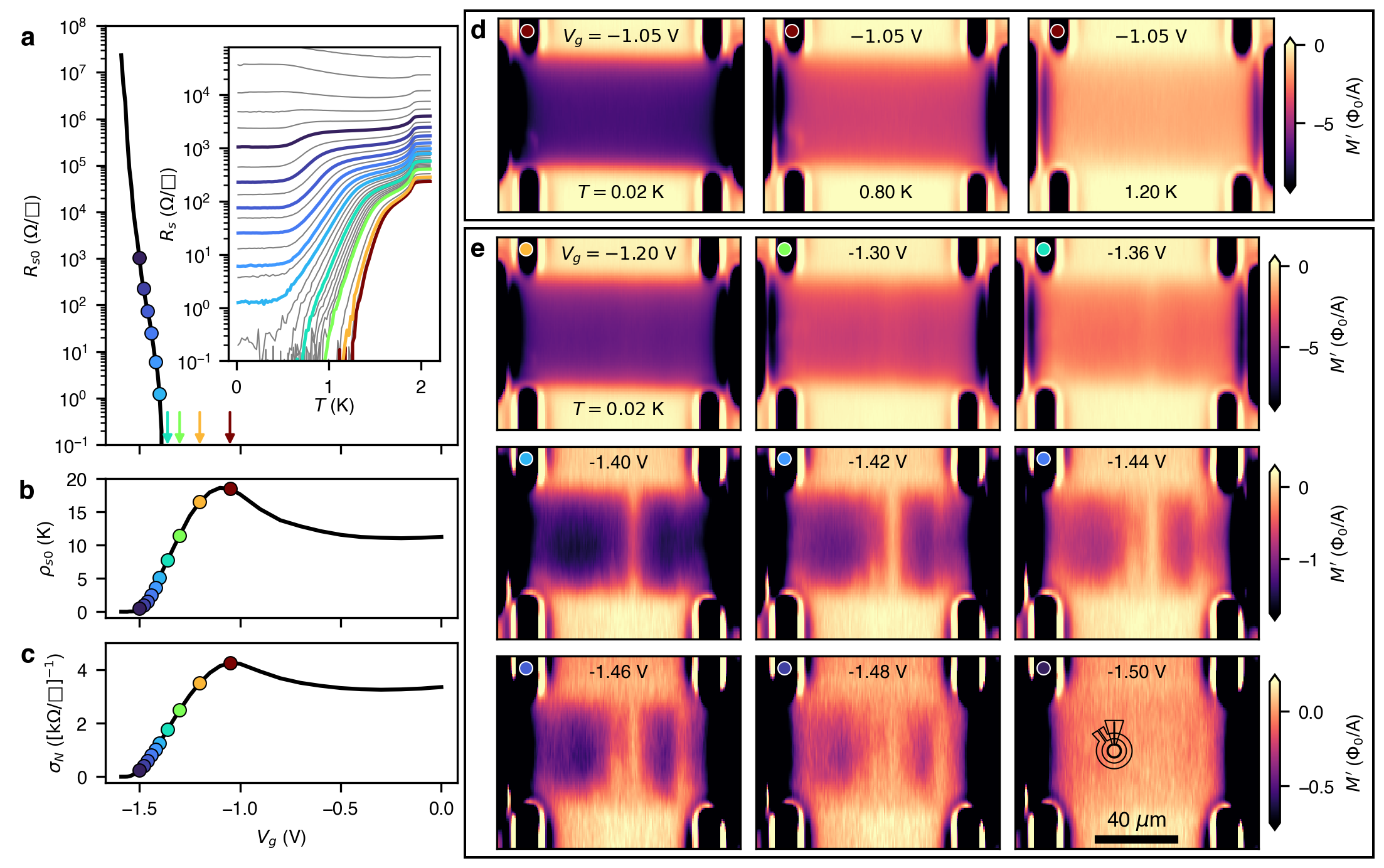}
    \caption{
    {\bf Emergence of anomalous metal transport and spatial inhomogeneity in the phase stiffness in the $b=200$ nm array.}
    Corresponding data for the other arrays can be found in the Supplementary Information.
    ({\bf a}) Low-temperature ($T=20$ mK) sheet resistance $R_{s0}$ vs. $V_g$ in the anomalous metal regime. Inset: temperature-dependent sheet resistance $R_s(T)$. The colored curves correspond to the colored dots and arrows. ({\bf b})  $\rho_{s0}$ vs. $V_g$, showing a peak at $V_{g,\text{peak}}$ and the existence of measurable phase stiffness in the anomalous metal regime.
    ({\bf c}) $\sigma_N$ vs. $V_g$, showing a similar peak near $V_{g,\text{peak}}$. 
    ({\bf d}) Scanning SQUID maps of the superfluid response $M'$ as a function of temperature at $V_g=-1.05\,\text{V}\approx V_{g,\text{peak}}$. At high carrier densities, the magnetic response is uniform within our resolution at all temperatures.
    ({\bf e}) Scanning SQUID maps of $M'$ as a function of gate voltage at $T=20$ mK in the low carrier density regime. Inhomogeneity emerges near the transition to anomalous metal behavior. The black lines in the lower right panel of (e) show the position of the SQUID field coil and pickup loop for the measurements shown in (a,b) and Fig.~\ref{fig:fig4}. Note that each row in (e) has a different color scale. From top to bottom, the color scales are saturated at [-8.5, 0.2], [-1.75, 0.2], and [-0.75, 0.2] $\Phi_0/\mathrm{A}$.
    }
    \label{fig:fig5}
\end{figure}

To explore the coexistence of finite local phase stiffness and finite resistivity, we imaged the magnetic response of the arrays as a function of $V_g$.
Figure~\ref{fig:fig5}d shows scanning SQUID maps of the superfluid response $M'$ of the $b=200$ nm array as a function of temperature outside of the anomalous metal regime, namely at $V_g=-1.05\,\mathrm{V}\approx V_{g,\text{peak}}$. At large carrier densities, $M'$ is spatially homogeneous within our spatial resolution and sensitivity at all temperatures, indicating that the Josephson coupling and critical temperature are uniform throughout the array. In contrast, at lower carrier density, stripe-like spatial fluctuations in the superfluid response begin to emerge. Fig.~\ref{fig:fig5}(a,b,c) shows $R_s(T=20\,\text{mK})$, $\rho_s(T=20\,\text{mK})$, and $\sigma_N$ as a function of $V_g$. Fig.~\ref{fig:fig5}e shows the superfluid response $M'$ as a function of position for $V_g\leq V_{g,\text{peak}}$ at $T=20$ mK. In the gate voltage range where both $\rho_{s0}$ and $R_{s0}$ are finite, the array breaks up into large superconducting regions separated by relatively narrow vertical stripes of reduced or vanishing phase stiffness.

The structure of this inhomogeneity is stable over time, repeatable as a function of both gate voltage and temperature, and is independent of the scan direction and the frequency of the AC current in the field coil. This inhomogeneity is present both when all device leads are grounded and when a current bias is applied to measure transport, i.e., the effect is not caused by the applied current bias. We have observed similar large-scale spatial inhomogeneity at low carrier density for all values of the island spacing (Supplementary Information). Our results demonstrate that the anomalous metal regime in these arrays is correlated with a spatial break-up of the system into phase-ordered and non-phase-ordered regions.

The static and repeatable nature of the observed heterogeneity suggests an origin related to ``baked in'' disorder in the arrays. All 2D carrier systems host a random distribution of charged impurity centers. At large carrier density these charged impurities are effectively screened, resulting in a uniform carrier density distribution. However, as the carrier density is depleted, it is thought that screening of the disorder potential breaks down, leading to inhomogeneity in the form of puddles of carriers~\cite{Ilani2000-uh,Ilani2001-gr,Sarma2005-wq,Manfra2007-qz}. This breakdown of screening eventually drives a metal-insulator transition at a density corresponding to the percolation threshold of the network of puddles~\cite{Efros1988-xk,Efros1989-zp,Das_Sarma2005-vb,Tracy2006-bd,Manfra2007-qz}.

Considering this, the simplest interpretation of Fig.~\ref{fig:fig5} is that the local phase stiffness serves as a sensitive indicator for density inhomogeneity in the 2DEG. As the carrier density is depleted, slight density inhomogeneity may develop even for densities significantly above the normal state percolation threshold.  As a result, in some regions of the array the carrier density and normal state conductance between the islands will eventually fall below a critical value. These regions may then act as nucleation sites for quantum phase slips.
As the carrier density is depleted further, quantum phase slips proliferate, eventually driving the phase stiffness to zero everywhere.
The Al islands themselves likely act as an additional source of spatially correlated disorder at the Al/2DEG interface. 

Our measurements highlight the value of sensitive real space imaging in studies of quantum phase transitions, particularly in dilute and low-dimensional systems where the effects of disorder are strongly enhanced. This work provides two insights into the destruction of superconductivity in two dimensions.
First, both the critical temperature and the temperature-dependence of the phase stiffness are incompatible with a generic model of proximity effect Josephson junction arrays subject to thermal phase fluctuations.
Second, at low carrier density there is an anomalous metal regime that is correlated with the onset of large-scale spatial inhomogeneity in the phase stiffness, which we attribute to a proliferation of quantum phase slips driven by density inhomogeneity in the 2DEG.
These results suggest that quantum phase fluctuations may suppress the phase stiffness and critical temperature throughout the superconducting phase and amplify the role of inhomogeneities near the notional superconductor-insulator transition.


\section{Methods}\label{sec:methods}

\subsection{Data acquisition and processing}

A detailed description of our millikelvin scanning SQUID microscope is provided in the Supplementary Information. We used a SQUID field coil current of $I_\text{FC}=100\,\mu\text{A}$ RMS for SQUID susceptometry measurements of samples $b=150$ nm, 200 nm, 250 nm, and 300 nm. We used a smaller $I_\text{FC}=50\,\mu\text{A}$ RMS for $b=400$ nm and 500 nm. The field coil has an inner (outer) radius of 6 $\mu\text{m}$ (8 $\mu\text{m}$) and the flux sensitive pickup loop has an inner (outer) radius of 3 $\mu\text{m}$ (3.5 $\mu\text{m}$). The maximum field applied to the sample per unit field coil current for a SQUID-sample standoff distance of $z_0=1\,\mu\mathrm{m}$ is approximately $0.085\,\mu\text{T}/\mu\text{A}$ (Figure~\ref{fig:squid-field}). These values of $I_\text{FC}$ correspond to an applied flux per plaquette in the array of $<0.01\,\Phi_0$. We verified that the magnetic response of the arrays was linear with respect to the field coil current in this range (Figures~\ref{fig:sample200-nonlinear} and \ref{fig:sample150-nonlinear}). Transport measurements were performed with an RMS source-drain current ranging from $<1$ nA to 10 nA. The spatial distribution of the magnetic response was not sensitive to the applied transport current even for currents well above 10 nA (Figures~\ref{fig:sample200-5-80nA} and \ref{fig:sample300-1V}).

During high-sensitivity scanning SQUID imaging (e.g., Figure~\ref{fig:fig5}), the gain of the lock-in amplifier used to measure the magnetic response was set such that the amplifier saturated when the SQUID was above the Al leads (where the signal is much larger, see Fig.~\ref{fig:fig1}). The lock-in amplifier time constant and SQUID scan speed and direction were chosen so that the amplifier recovered from the saturated state by the time the SQUID was above the JJ array. The SQUID scan direction for all images shown in this work was top to bottom (fast axis) and left to right (slow axis), however none of the measurement results were sensitive to the scan direction.

We were only able to measure electrical transport in three out of the six devices due to difficulty making reliable wire bond connections to the other devices. The entire sample chip, including the Al bond pads, is covered in a 20 nm thick layer of Al${}_2$O${}_3$ dielectric. To wire bond to the Al bond pads, one must break through this insulating layer without causing too much damage to the 5 nm thick Al layer underneath. Often, one out of the five total wire bonds per device would fail while cooling down (likely due to differential thermal contraction), precluding four-probe transport measurements.

\subsection{Temperature sweeps}

For measurements performed as a function of temperature at a fixed SQUID position (e.g. Figure~\ref{fig:fig2}), the setpoint of the sample stage temperature controller (Lake Shore 372) was swept from base temperature to the target maximum temperature, then back to base temperature, at a rate of 50 mK/minute. While the setpoint was swept, all measurement channels (the DC flux through the SQUID, all demodulated signals from the lock-in amplifiers, and the sample stage temperature) were sampled synchronously at a rate of 20 Hz using a multi-channel digitizer (National Instruments USB-6363). This time-series data was then binned by temperature into 20 mK wide bins, and the mean and standard error on the mean within each temperature bin were calculated. The mean within each temperature bin is the quantity shown in the figures. The statistical uncertainty in $\rho_s$ and $R_s$ is determined by the standard error on the mean within each temperature bin. Vertical error bars indicating this statistical uncertainty are excluded for clarity in the main text figures, but are included in Figures~\ref{fig:sweep-150}, \ref{fig:sweep-200}, \ref{fig:sweep-250}, \ref{fig:sweep-300}, \ref{fig:sweep-400}, and \ref{fig:sweep-500} below.

When cooling from 2.1 K to 20 mK, the cooling rate below approximately 250 mK was limited by the cooling power of the MXC to less than the 50 mK/minute ramp rate of the temperature setpoint, at which point the heater output current was turned off by the temperature control loop. A typical plot of sample stage temperature and mixing chamber temperature vs. time for a single heating and cooling cycle is shown in Figure~\ref{fig:thermal-hysteresis}a. At the heater output range necessary to reach sample temperatures above 1.5 K, the temperature controller PID loop oscillated slightly when heating the sample from base temperature to a few hundred mK, as can be seen in the lower left of Figure~\ref{fig:thermal-hysteresis}a. For this reason, in the main text we present only data acquired during the cooling portion of the temperature sweep. Thermal hysteresis in the resistive transition of the Al film at $T_{c,\text{Al}}$ is measured to be  $<10$ mK (Fig.~\ref{fig:thermal-hysteresis}b). Figure~\ref{fig:sample-200nm-heating-cooling} shows the phase stiffness and resistance for the $b=200$ nm array from both heating (dashed lines) and cooling (solid lines) portions of the temperature sweep.

For each measurement of $\rho_s(T)$, we have subtracted a temperature-dependent background that arises from coupling between portions of the SQUID circuit and Al film on the JJ array die far from the region of interest, as described in the Supplementary Information and illustrated in Figures~\ref{fig:background-subtraction} and \ref{fig:chip-layout}.

\subsection{Model for the temperature-dependent Josephson coupling}

To estimate the temperature-dependent Josephson coupling, we start with Equations~\ref{eq:cpr}-\ref{eq:EJ} from the main text.
We assume a weak coupling isotropic $s$-wave BCS form for the gap of the Al islands, which can be approximated as
\begin{equation}
    \Delta_\text{Al}(T) = \Delta_\text{Al}(0)\tanh\left(\frac{\pi k_\text{B} T_{c,\text{Al}}}{\Delta_\text{Al}(0)}\sqrt{\frac{T_{c,\text{Al}}}{T}-1}\right),
    \label{eq:bcs-gap}
\end{equation}
where $\Delta_\text{Al}(0)=1.764\,k_\text{B}T_c$ with $T_c=T_{c,\text{Al}}=1.9$ K~\cite{Prozorov2006-zu}. Note that $T_{c,\text{Al}}$ is larger than the $T_c$ of bulk Al ($\approx 1.2$ K) due to the small film thickness~\cite{Meservey1971-xw,Lopez-Nunez2025-gp}.
To find the induced gap $\Delta_\text{ind}(T)$ for a given value of the Al/2DEG interface barrier strength $\gamma_\text{B}$, we solve Eq.~\ref{eq:induced-gap} via fixed point iteration (\texttt{scipy.optimize.fixed\_point}) with $\Delta_S(T)=\Delta_\text{Al}(T)$ (see Fig.~\ref{fig:induced-gap}).

We simulated the classical XY model,
\begin{equation}
\begin{split}
    H_\text{XY}=-E_\mathrm{J}(T)\sum_{\langle ij \rangle}\cos(\varphi_i-\varphi_j),
    \label{eq:classical-xy}
\end{split}
\end{equation}
where $\langle ij \rangle$ indicates nearest neighbor islands. We calculate the temperature-dependent coupling $E_\text{J}(T)$ given by Eq.~\ref{eq:EJ} and compute the resulting thermodynamic phase stiffness (helicity modulus) as a function of temperature using standard Monte Carlo methods. In a 2D classical XY system, the helicity modulus is given by
\begin{equation}
\begin{split}
    \Upsilon_{\hat{\mu}} &= \frac{E_\text{J}}{N}\Bigg\langle\sum_{\langle{ij}\rangle}\cos(\varphi_i-\varphi_j)(\hat{e}_{ij}\cdot\hat{\mu})^2\Bigg\rangle - \frac{E_\text{J}^2}{k_\text{B}TN}\Bigg\langle\left(\sum_{\langle{ij}\rangle}\sin(\varphi_i-\varphi_j)(\hat{e}_{ij}\cdot\hat{\mu})\right)^2\Bigg\rangle,
    \label{eq:helicity}
\end{split}
\end{equation}
where $\hat{e}_{ij}$ is a unit vector pointing from site $i$ to site $j$, $\hat{\mu}\in\{\hat{x},\hat{y}\}$ is the direction of the applied phase twist, and $N$ is the number of sites in the system~\cite{Li1989-yg}. The $(\hat{e}_{ij}\cdot\hat{\mathbf{\mu}})$ terms pick out only bonds in the $\hat{\mu}$-direction. Note that Eq.~\ref{eq:helicity} applies only to the classical XY model with periodic boundary conditions. In an isotropic system, $\Upsilon_{\hat{x}}=\Upsilon_{\hat{y}}$. We identify $\Upsilon_{\hat{\mu}}$ with the phase stiffness $k_\text{B}\rho_s$. Further details regarding the Monte Carlo methods used to evaluate Eq.~\ref{eq:helicity} are provided in the Supplementary Information. In the classical XY model, finite-size effects broaden the transition but do not affect $\Upsilon_{\hat{\mu}}$ for $T$ significantly below $T_\text{BKT}$~\cite{Schultka1994-hz}. Spatial disorder in the coupling $E_{\text{J}0}$ slightly reduces $\Upsilon_{\hat{\mu}}(0)$ and $T_\text{BKT}$, but does not qualitatively change $\Upsilon_{\hat{\mu}}(T)$~\cite{Maccari2019-be}. The BKT transition can be significantly broadened by spatially correlated disorder~\cite{Maccari2017-rh,Maccari2026-cx}.

\subsection{Estimated Josephson junction properties}

The 2DEG has a peak mobility of $\mu_\text{2D}=57{,}000\,\text{cm}^2\text{V}^{-1}\text{s}^{-1}$ at carrier density $n_\text{2D}=4\times 10^{11}$~cm$^{-2}$ as measured at $\sim$20 mK in a Hall bar fabricated from the same wafer with the Al layer removed. Therefore, at $V_{g,\text{peak}}$ we estimate the electron mean free path between the Al islands to be $\ell=\hbar\mu_\text{2D} k_\text{F}/e\approx 595$ nm and the number of transverse conduction channels in each planar junction to be $N_\text{ch}\approx ak_\text{F}/\pi\approx 50$, where $k_\text{F}=\sqrt{2\pi n_\text{2D}}$ is the Fermi wavevector and $a=1\,\mu\mathrm{m}$ is the junction width.
Low temperature transport in a quantum point contact fabricated from a similar InAs/Al structure indicated a ``hard'' proximity-induced gap $\Delta_\text{ind}\approx 150\,\mu\text{eV}$~\cite{Zhang2023-gw}, which is roughly half the estimated zero-temperature BCS gap of the Al islands $\Delta_\text{Al}(0)=1.764 k_\text{B}T_{c,\text{Al}}=290\,\mu\text{eV}$.
Assuming an effective mass for InAs of $m^*\approx 0.026m_e$~\cite{Yuan2020-re,Zhang2023-gw} (where $m_e$ is the bare electron mass), this induced gap corresponds to an induced BCS coherence length in the 2DEG of $\xi_\text{ind}=\hbar^2k_\text{F}/(\pi \Delta_\text{ind}m^*)\approx 985$ nm and a disordered effective coherence length $\xi_\text{eff}=\sqrt{\xi_\text{ind}\ell}\approx 765$ nm.
Thus, we expect that at $V_{g,\text{peak}}$ our devices are in the short junction regime, $b<\xi_\text{eff}$, and the ballistic regime, $b<\ell$.

As mentioned in the Main Text, the peak in phase stiffness and normal state conductance as a function of $V_g$ coincides with a peak in the 2DEG mobility.
For carrier densities above the peak ($V_g>V_{g,\text{peak}}$), there are two occupied electronic subbands in the 2DEG, one of which is closer to the surface~\cite{Zhang2023-gw}. In this regime, the mobility is suppressed by inter-subband scattering and interface roughness scattering. Decreasing the carrier density by applying more negative $V_g$ depopulates the second subband, resulting in an increase in mobility.
Decreasing $V_g$ further below $V_{g,\text{peak}}$, decreases both the carrier density and mobility, eventually resulting in an insulating 2DEG.

\subsection{Estimate of the charging energy}

In the context of Josephson junction arrays, the \textit{island} charging energy is typically defined as $E_C=(e^2/2)C_{ii}^{-1}$, where $C_{ii}^{-1} = (\mathbf{C}^{-1})_{ii}$ and $\mathbf{C}$ is the Maxwell capacitance matrix of the circuit~\cite{Fazio2001-tj}. The Maxwell capacitance matrix is defined such that $C_{ii}$ is the total capacitance of island $i$ and $-C_{ij}=-C_{ji}$ is the mutual capacitance between islands $i$ and $j$. In general, the island capacitance $C_{ii}$ depends on all mutual island capacitances and the capacitance of each island to ground. In vector form, the island charges $\mathbf{q}$ are related to the island voltages $\mathbf{v}$ by $\mathbf{q}=\mathbf{C}\mathbf{v}$. In our case the top gate acts as a nearby ground plane, screening the long-range Coulomb interaction between the islands. Note that some authors define the island charging energy to be $E_0=(2e)^2C^{-1}_{ii}=8E_C$~\cite{Chakravarty1986-wv,Chakravarty1987-ad,Fisher1987-wa,Chakravarty1988-dl,Sondhi1997-ab}.

A related quantity is the \textit{junction} charging energy $E_{C_\Sigma}=e^2/(2C_\Sigma)$ where $C_\Sigma$ is the total effective shunt capacitance across a given junction (i.e., between two nearest neighbor islands). $C_\Sigma$ can be obtained by calculating the total electrostatic energy resulting from a unit voltage applied between two nearest neighbor islands with all other islands held neutral~\cite{Schuster2007-kl}:
\begin{align}
\begin{split}
    E=\frac{1}{2}\mathbf{v}_\mathrm{J}^T\mathbf{C}\mathbf{v}_\mathrm{J}=\frac{1}{2}C_\Sigma V_\mathrm{J}^2.
    \label{eq:junction_EC}
\end{split}
\end{align}
Here $\mathbf{v}_\mathrm{J}$ is the vector of island voltages corresponding to the charge configuration with $V=\pm V_\mathrm{J}/2$ on the two islands of interest and $q=0$ on all other islands.

We calculated $\mathbf{C}$ using a commercial 3D finite element electromagnetic parameter extraction tool (Ansys Q3D). The model consists of a $10 \times 10$ array of $a=1\,\,\mu\mathrm{m}$ square metallic islands, each 5 nm thick, separated by a distance $b$. The islands sit on an insulating substrate with dielectric constant $\kappa_\text{InAs}$. Above the islands is a 20 nm thick layer of $\text{Al}_2\text{O}_3$ gate dielectric (dielectric constant $\kappa_{\text{Al}_2\text{O}_3}=9$). On top of the gate dielectric is a 525 nm thick metallic top gate. We calculated $\mathbf{C}$ for island spacings $b=100$ nm to 500 nm assuming the high frequency value of the dielectric constant of InAs, $\kappa_\text{InAs}=12.3$. We used a solver convergence criterion of $<0.1\%$ change in $\mathbf{C}$ between adaptive meshing passes, with a minimum of three converged passes before terminating.

In Table~\ref{table:capacitance}, we report the extracted total island capacitance $C_{ii}$, the island-to-gate capacitance $C_g$, the mutual capacitance between nearest neighbor islands $C_\text{nn}$, and the mutual capacitance between next nearest neighbor islands (i.e., islands with adjacent corners) $C_\text{nnn}$. We also report the diagonals of the inverse capacitance matrix $C_{ii}^{-1} = (\mathbf{C}^{-1})_{ii}$, the island charging energy $E_C/k_\text{B}$, the effective junction shunt capacitance $C_\Sigma$, and the junction charging energy $E_{C_\Sigma}$. As expected for this geometry, the island capacitance is dominated by the gate capacitance: $C_g \gg C_\text{nn} \gg C_\text{nnn}$. $C_g$ increases with island spacing $b$, whereas the mutual capacitances $C_\text{nn}$ and $C_\text{nnn}$ decrease. The two effects cancel out resulting in a charging energy $E_C/k_\text{B}=0.20$ K for all values of $b$ (or $E_0/k_\text{B}=8E_C/k_\text{B}=1.60$ K). Similarly, the junction charging energy is roughly independent of $b$ at $E_{C_\Sigma}/k_\text{B}=0.39$ K.
Therefore, except at very negative $V_g$, the arrays are in the regime $E_{\text{J}0} > \Delta_\text{Al}(0) > E_C$, where $E_{\text{J}0}=E_\text{J}(0)$.

\section{Extended Data}

\begin{figure}[h]
    \centering
    \includegraphics[width=0.5\linewidth]{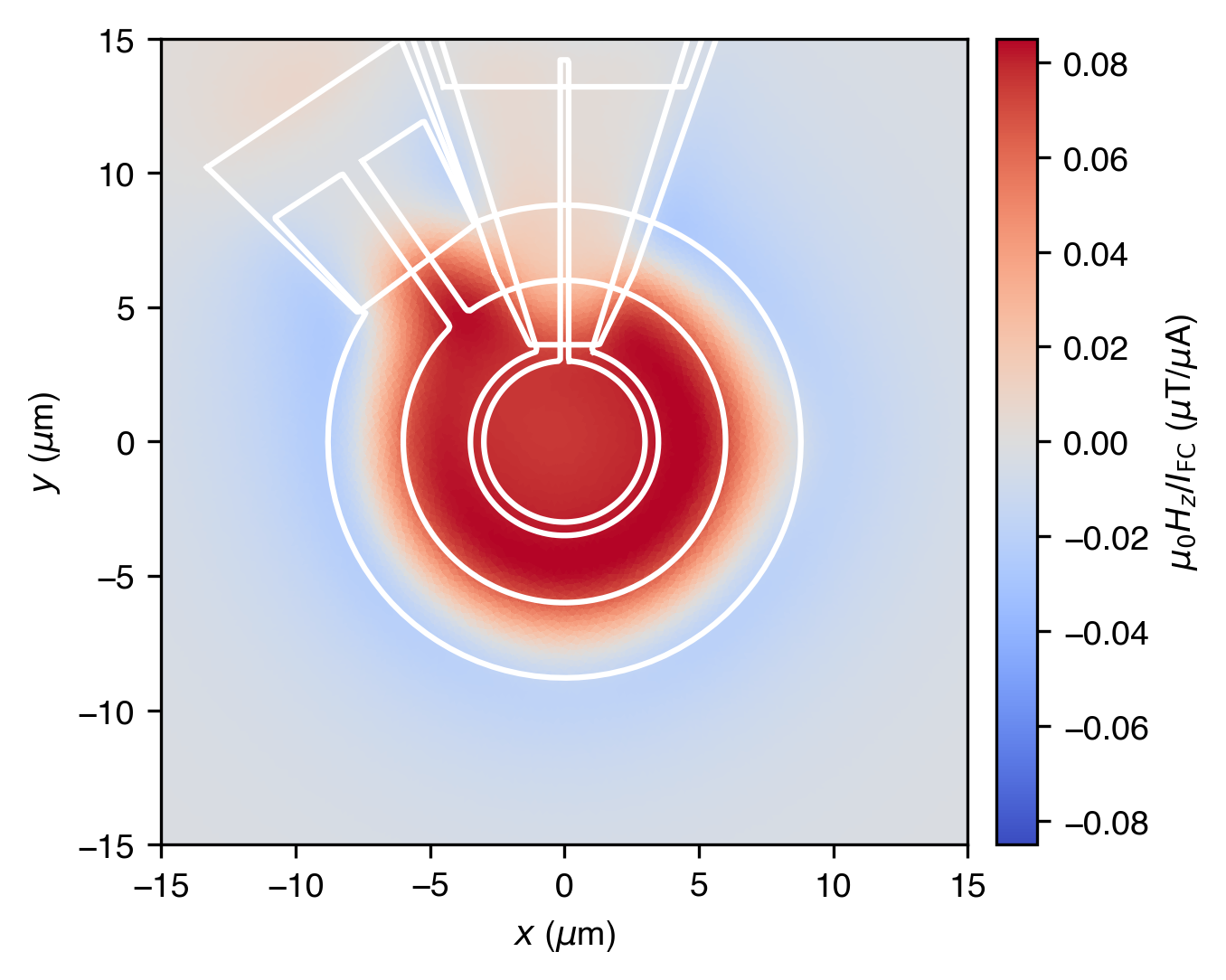}
    \caption{{\bf London-Maxwell simulation of the magnetic field generated by the SQUID field coil.} The magnetic field is evaluated at a standoff distance $z_0=1\,\mu\mathrm{m}$ and alignment angle $\phi=0^\circ$. The simulation method is described in Ref.~\cite{Bishop-Van_Horn2022-sy}.}
    \label{fig:squid-field}
\end{figure}

\begin{figure}[h]
    \centering
    \includegraphics[width=0.8\linewidth]{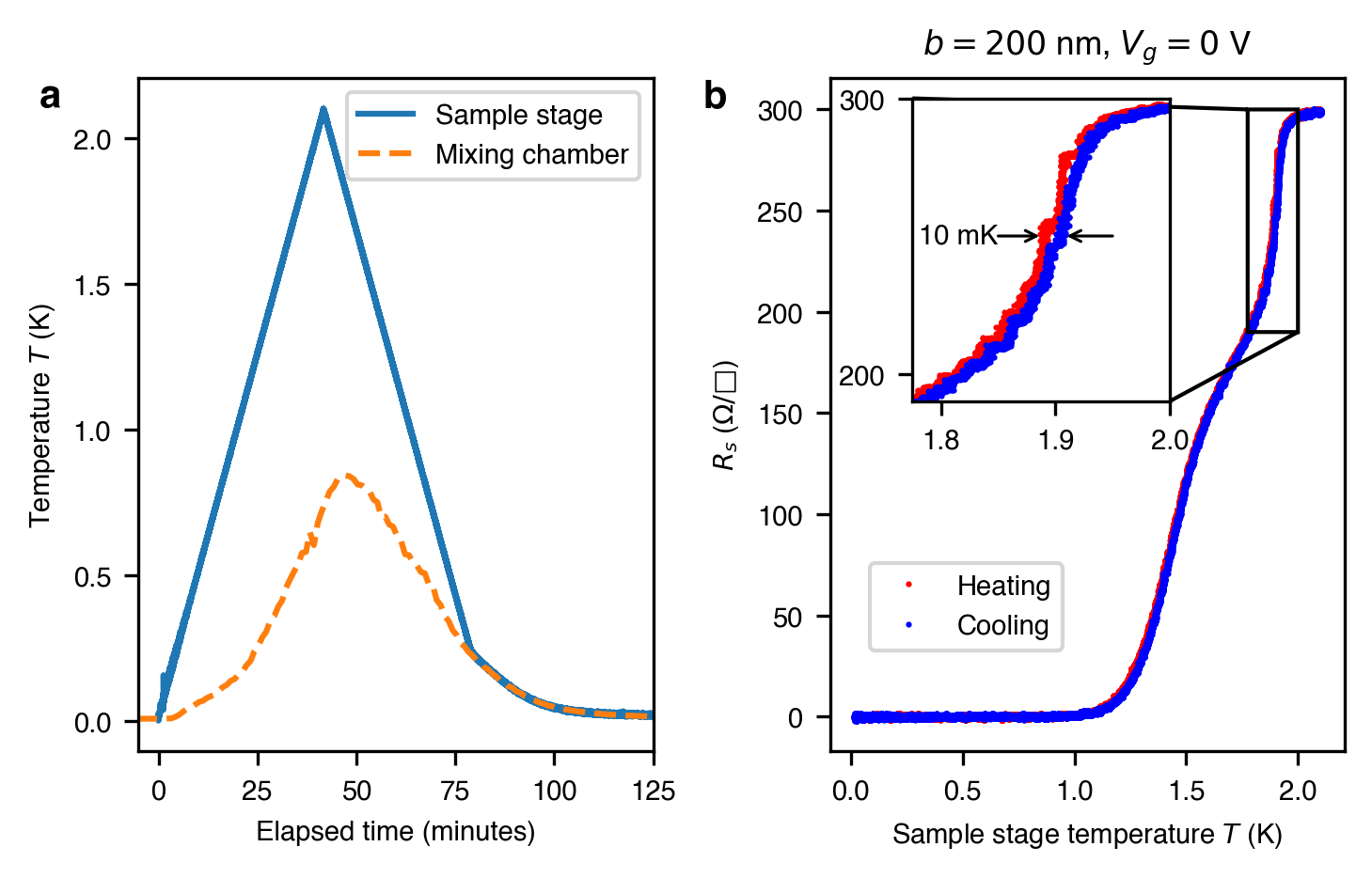}
    \caption{{\bf Details of the temperature sweep measurements.} ({\bf a}) Sample stage temperature and mixing chamber temperature as a function of time for a typical temperature sweep. ({\bf b}) Sheet resistance of the $b=200$ nm array as a function of temperature at $V_g=0$ V for both heating (red) and cooling (blue). The thermal hysteresis in the transition of the Al film is less than 10 mK. The red and blue points show the raw time series $R_s$ plotted as a function of the raw time series sample stage temperature (as opposed to the binned $R_s$ and $T$).
    }
    \label{fig:thermal-hysteresis}
\end{figure}

\begin{figure}[h]
    \centering
    \includegraphics[width=0.6\linewidth]{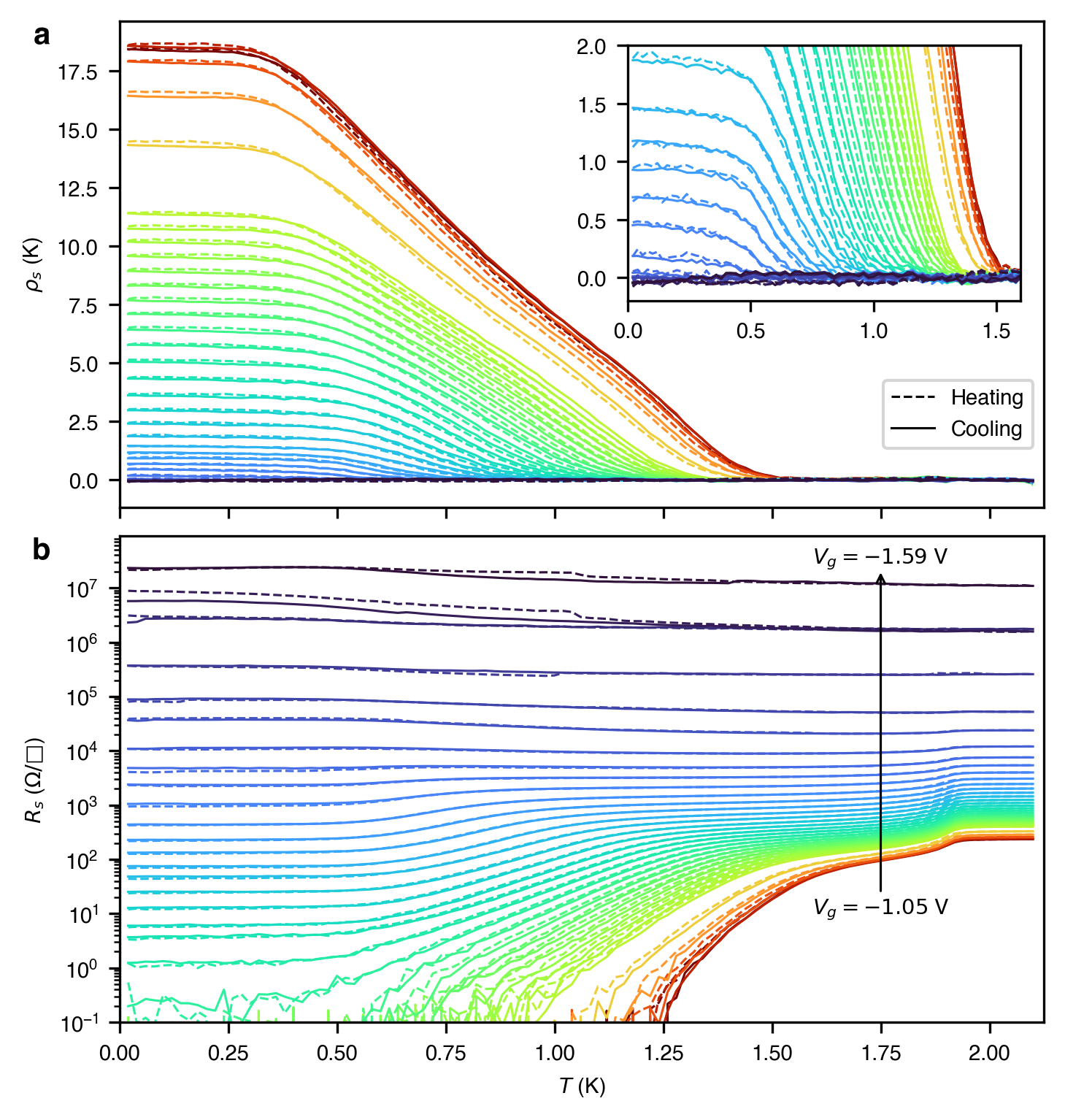}
    \caption{{\bf Phase stiffness and transport for both heating and cooling in the $b=200$ nm device.} ({\bf a}) Phase stiffness $\rho_s$ and ({\bf b}) sheet resistance $R_s$ vs. gate voltage $V_g\leq -1.05$ V and temperature for both heating (dashed lines) and cooling (solid lines). The solid lines (cooling) are the same data shown in Fig.~\ref{fig:fig4} and Fig.~\ref{fig:sweep-200}.}
    \label{fig:sample-200nm-heating-cooling}
\end{figure}

\begin{figure}
    \centering
    \includegraphics[width=\linewidth]{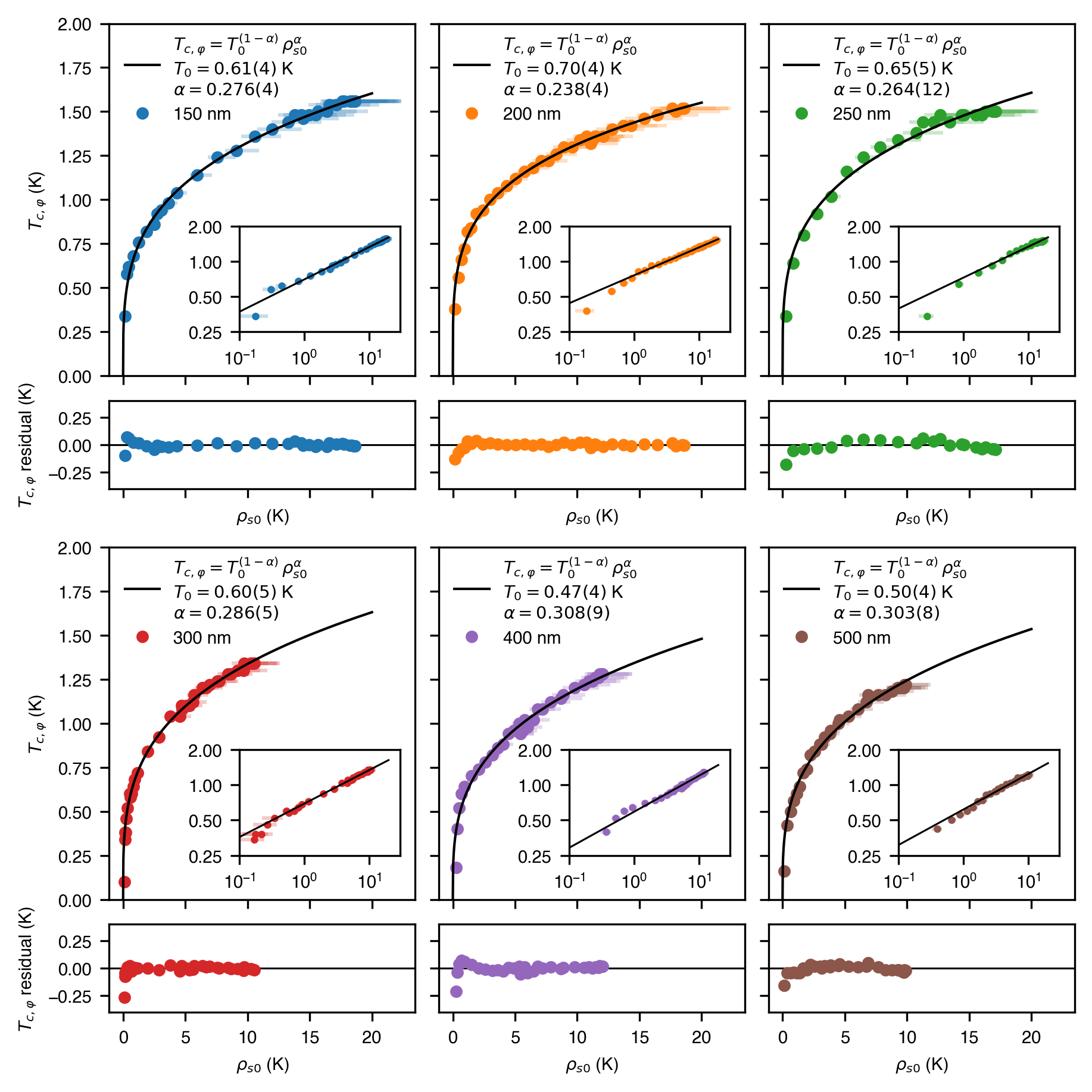}
    \caption{
    {\bf Power law fits to $T_{c,\varphi}$ vs. $\rho_{s0}$ for the individual arrays.} Data for arrays $b=$ 150, 200, 250, 300, 400, and 500 nm are shown in blue, orange, green, red, purple, and brown, respectively. For each array, we have plotted the data and power law fit on a linear scale in the upper panel (log-log scale in the inset), and the fit residual ($\text{data} - \text{fit}$) in the lower panel. Each dataset is fairly well described by a power law $T_{c,\varphi}=T_0^{(1-\alpha)}\rho_{s0}^\alpha$ with $T_0$ in the range $[0.47,0.70]$ K and $\alpha$ in the range $[0.24,0.31]$. The data systematically deviate from this power law relationship at the lowest values of $\rho_{s0}$, where (1) thermal phase fluctuations become important and (2) large-scale spatial inhomogeneity begins to emerge. The horizontal error bars indicate the estimated $-10\%/+20\%$ systematic uncertainty in the coefficient relating $M'$ to $\rho_s$.
    }
    \label{fig:fig3-individual-fits}
\end{figure}

\begin{figure}
    \centering
    \includegraphics[width=\linewidth]{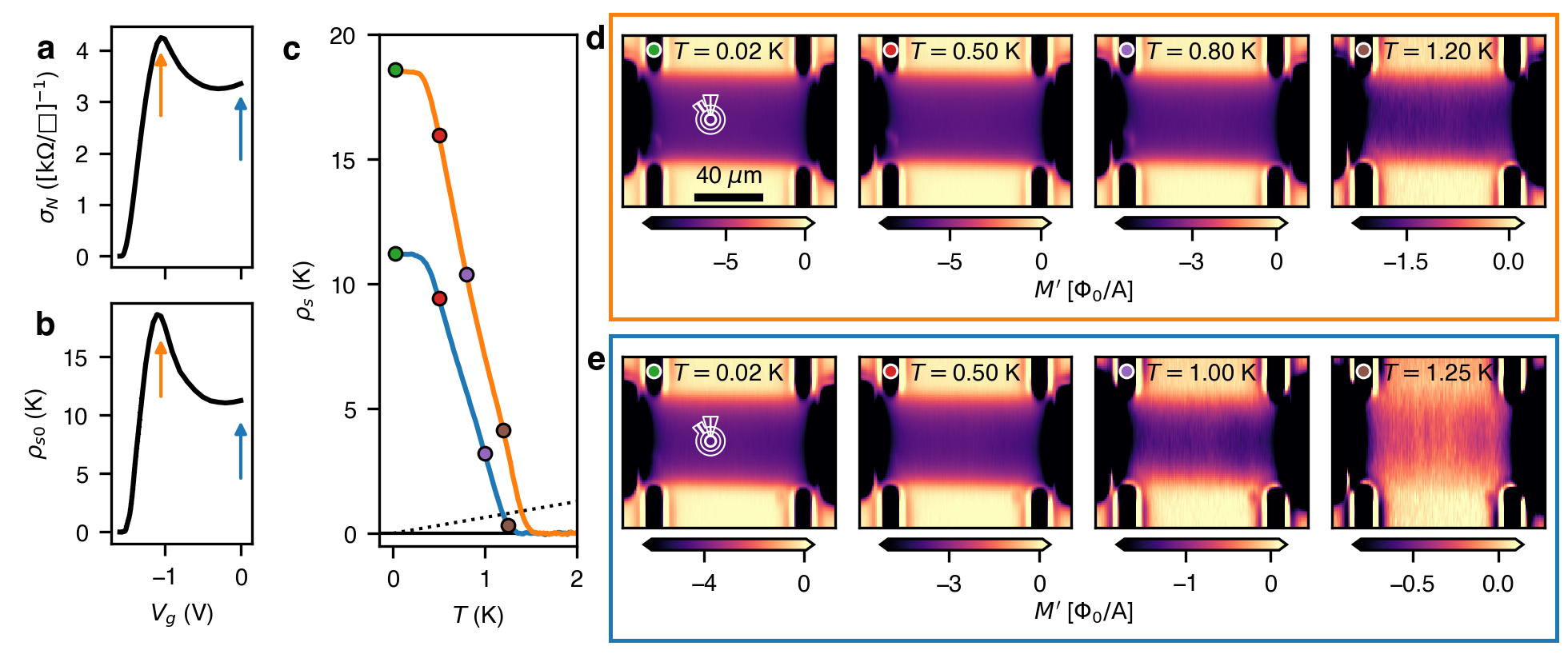}
    \caption{
    {\bf Temperature dependence of the magnetic response at high carrier density in the $b=200$ nm array.} ({\bf a}) Normal state sheet conductance $\sigma_N$ and ({\bf b}) low-temperature phase stiffness $\rho_{s0}$ vs. gate voltage $V_g$. ({\bf c}) Temperature dependence of $\rho_s$ at $V_g=-1.05\,\text{V}\approx V_{g,\text{peak}}$ (orange) and $V_g=0\,\text{V}$ (blue). The black dotted line indicates $\rho_s=(2/\pi)T$. ({\bf d}) Spatial maps of the superfluid response $M'$ at $V_g=-1.05\,\text{V}$ and $T=\{0.02, 0.50, 0.80,1.20\}\,\text{K}$. ({\bf e}) Spatial maps of $M'$ at $V_g=0\,\text{V}$ and $T=\{0.02,0.50,1.00,1.25\}\,\text{K}$. The white lines in (d,e) show the position of the SQUID sensor for the measurements shown in (a--c).
    }
    \label{fig:high-density-scans-vs-T}
\end{figure}

\begin{figure}[h]
    \centering
    \includegraphics[width=0.6\textwidth]{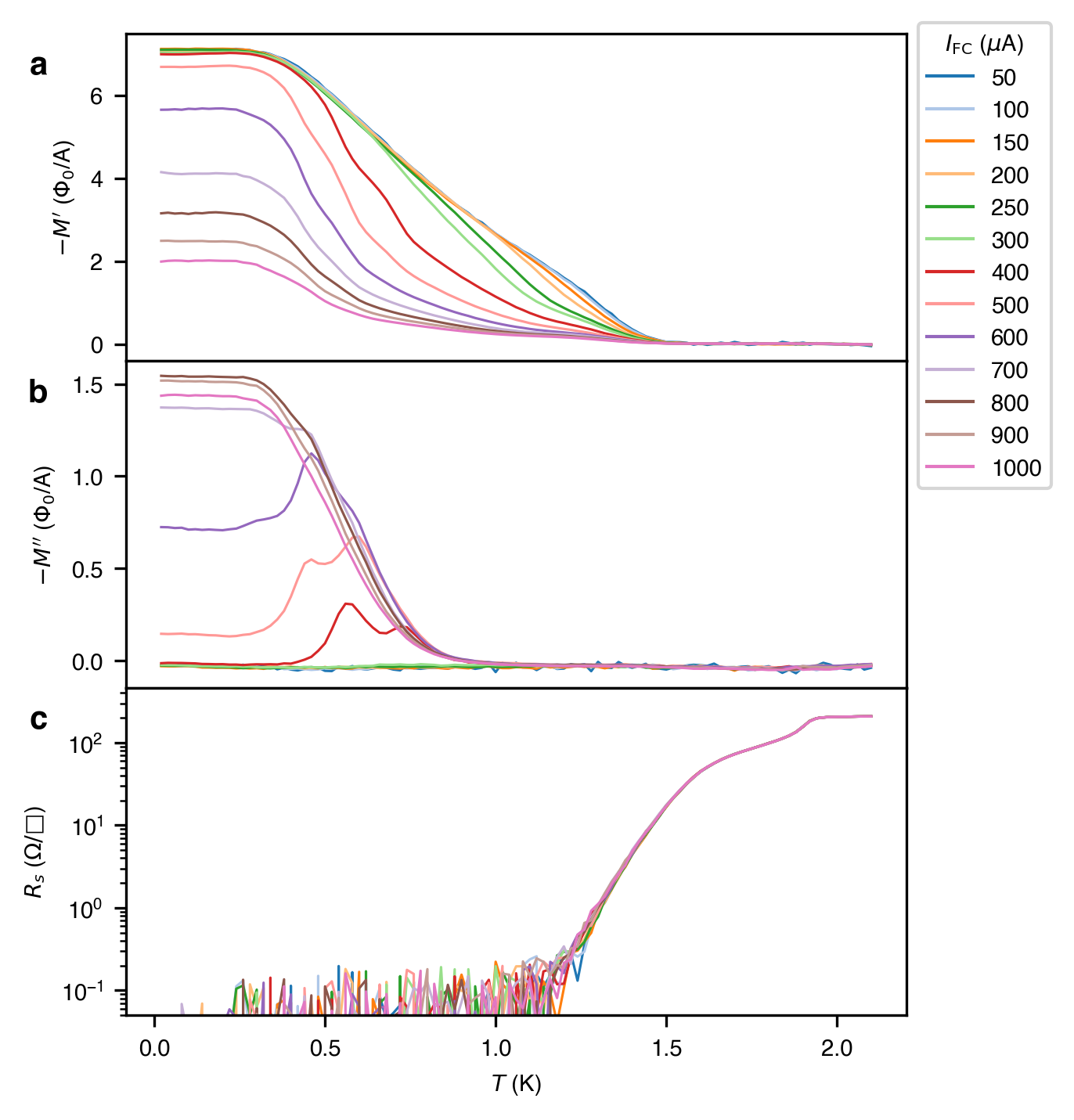}
    \caption{{\bf Temperature-dependent magnetic response of the $b=200$ nm array at $V_g=-1.05$ V as a function of the SQUID field coil current $I_\text{FC}$.} ({\bf a}) In-phase magnetic response $-M'$. ({\bf b}) Out-of-phase magnetic response $-M''$. ({\bf c}) Sheet resistance $R_s$ measured simultaneously with the magnetic response. The magnetic response is linear in $I_\text{FC}$ (i.e., $M=M'+iM''$ is independent of $I_\text{FC}$) at all temperatures for $I_\text{FC}\lesssim 100\,\mu\text{A}$ (dark blue and light blue lines). For larger $I_\text{FC}$, we observe a suppression of the superfluid response $-M'$ and the onset of a measurable out-of-phase response $M''$, which is associated with dissipation due to vortex motion. The magnetic response was measured at 887.7 Hz and the sheet resistance was measured at 589.7 Hz with an RMS source drain current of 5 nA.}
    \label{fig:sample200-nonlinear}
\end{figure}

\begin{figure}
    \centering
    \includegraphics[width=0.6\linewidth]{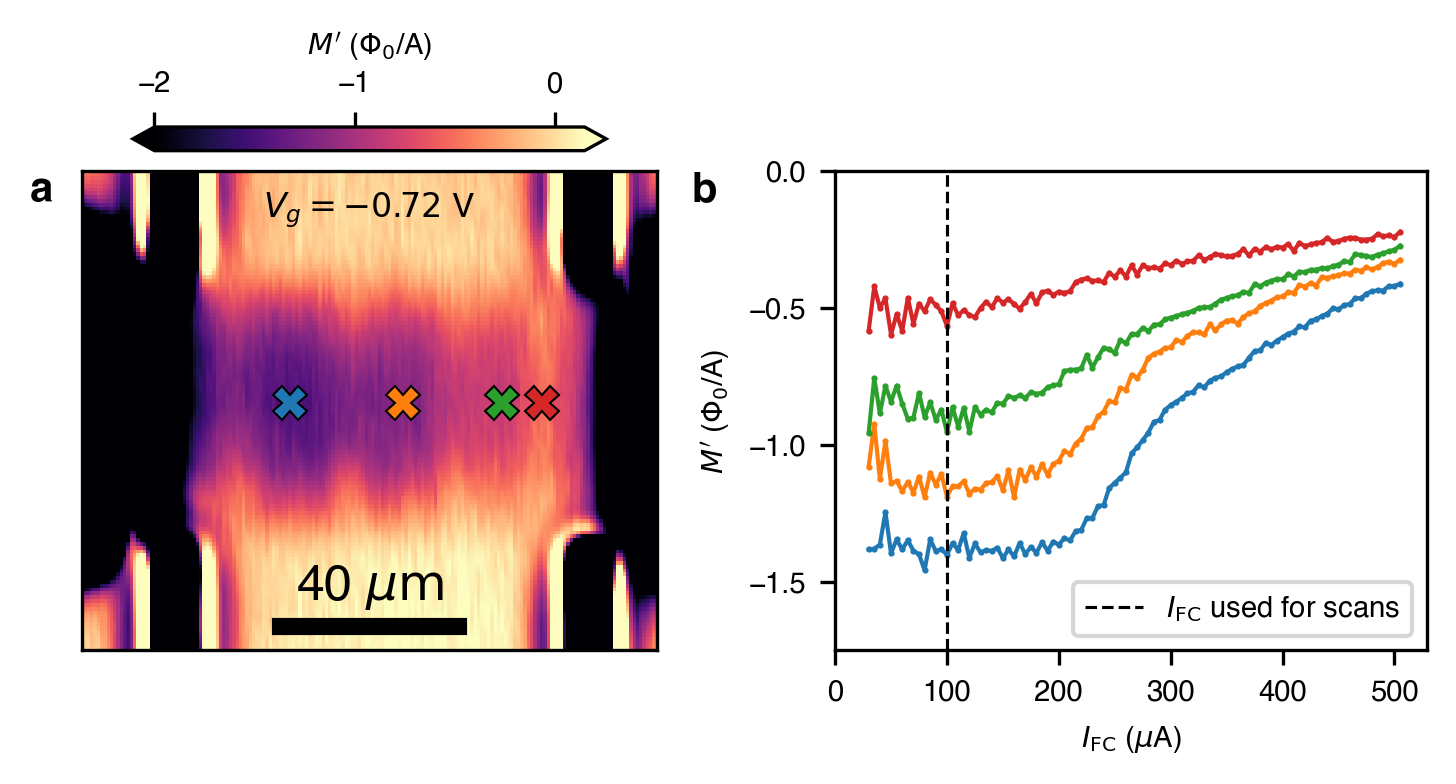}
    \caption{
    {\bf Linearity of the magnetic response in the low density/inhomogeneous regime.}
    ({\bf a}) Scanning SQUID map of the magnetic response $M'$ of the $b=150$ nm array at $T=30$ mK and $V_g=-0.72$ V. ({\bf b}) In-phase magnetic response as a function of RMS field coil current $I_\text{FC}$ at the positions indicated by the colors. The regions with a stronger diamagnetic signal show a linear magnetic response ($M'$ independent of $I_\text{FC}$) up to a larger value of $I_\text{FC}$. The rapid variations in $M'(I_\text{FC})$ are random noise. In the linear regime, the signal-to-noise ratio is proportional to $I_\text{FC}$.
    }
    \label{fig:sample150-nonlinear}
\end{figure}

\begin{figure}
    \centering
    \includegraphics[width=\linewidth]{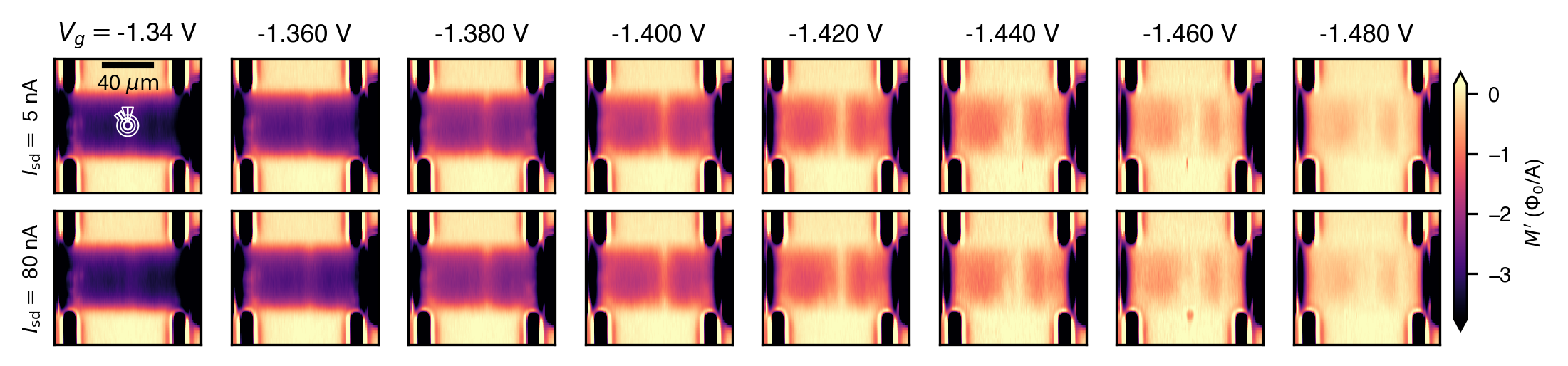}
    \caption{
    {\bf Spatial inhomogeneity at low carrier density is not sensitive to applied transport current.}
    Scanning SQUID maps of the in-phase magnetic response $M'$ of the $b=200$ nm array measured at $T=20$ mK with RMS source-drain current 5 nA (top row) and 80 nA (bottom row). The structure of the spatial inhomogeneity is not sensitive to the applied transport current.
    }
    \label{fig:sample200-5-80nA}
\end{figure}

\begin{figure}
    \centering
    \includegraphics[width=0.6\linewidth]{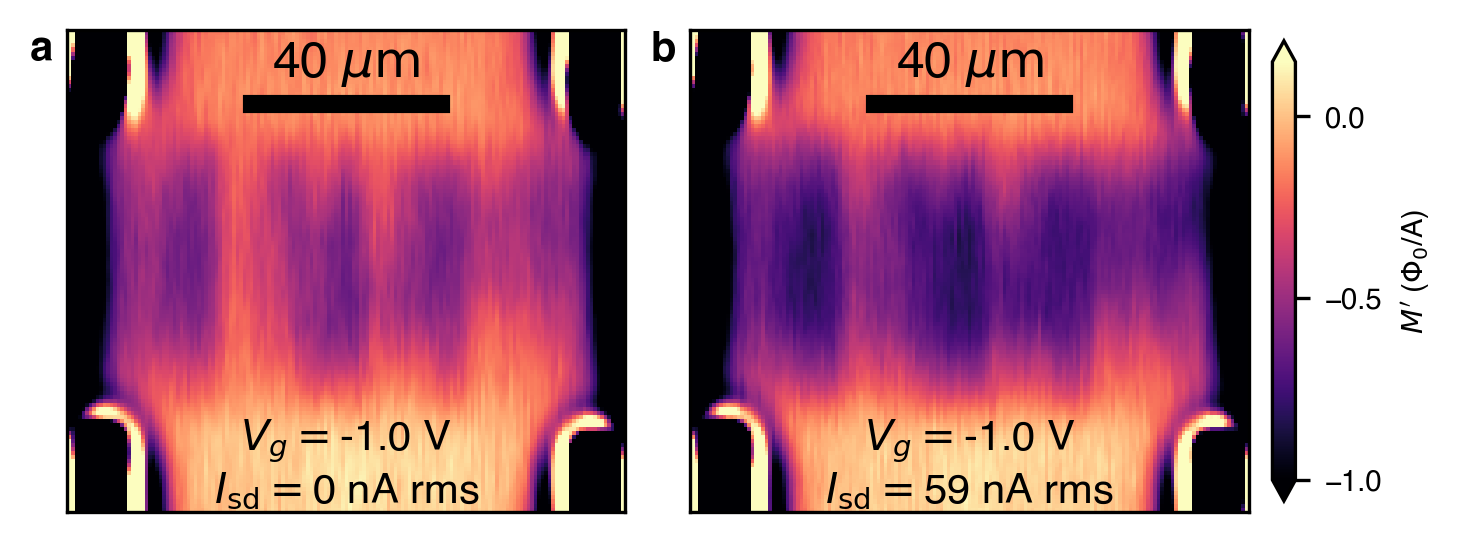}
    \caption{
    {\bf Spatial inhomogeneity at low carrier density is stable over time and is not sensitive to applied transport current.}
    Scanning SQUID maps of the in-phase magnetic response $M'$ of the $b=300$ nm array measured at $T=20$ mK and $V_g=-1.0$ V with RMS source-drain current 0 nA ({\bf a}) and 59 nA ({\bf b}). The structure of the spatial inhomogeneity is not sensitive to the applied transport current. The slight difference in contrast between the two images is likely due to a small offset in the SQUID standoff distance. The two images were acquired 10 days apart, with many gate voltage cycles and thermal cycles above $T_{c,\text{Al}}$ in between.
    }
    \label{fig:sample300-1V}
\end{figure}

\bgroup
\begin{table}[h]
\caption{{\bf Estimated island capacitance and charging energy.} Total island capacitance $C_{ii}$, gate capacitance $C_g$, nearest neighbor mutual capacitance $C_\text{nn}$, next nearest neighbor mutual capacitance $C_\text{nnn}$, diagonal of the inverse capacitance matrix $C_{ii}^{-1}=(\mathbf{C}^{-1})_{ii}$, island charging energy $E_C/k_\text{B}=(e^2/2)C^{-1}_{ii}/k_\text{B}$, effective junction shunt capacitance $C_\Sigma$, and junction charging energy $E_{C_\Sigma}$ as a function of island spacing $b$. These values are estimated from a 3D finite element simulation using Ansys Q3D with $\kappa_{\text{Al}_2\text{O}_3}=9$ and $\kappa_\text{InAs}=12.3$.\\}

\begin{tabular}{|c|c|c|c|c|c|c|c|c|}
\hline
$b$ (nm) & $C_{ii}$ (fF) & \multicolumn{1}{l|}{$C_g$ (fF)} & $C_\text{nn}$ (fF) & $C_\text{nnn}$ (fF) & $C_{ii}^{-1}$ (fF$^{-1}$) & $E_C/k_\text{B}$ (K) & $C_\Sigma$ (fF) & $E_{C_\Sigma}/k_\text{B}$ (K) \\ \hline \hline
100 & 4.6685 & 4.3533 & 0.0537 & 0.0098 & 0.2143 & 0.1992 & 2.3623 & 0.3936 \\ \hline
150 & 4.6597 & 4.4120 & 0.0400 & 0.0082 & 0.2147 & 0.1996 & 2.3503 & 0.3955 \\ \hline
200 & 4.6538 & 4.4558 & 0.0303 & 0.0069 & 0.2149 & 0.1998 & 2.3419 & 0.3969 \\ \hline
250 & 4.6572 & 4.4917 & 0.0241 & 0.0058 & 0.2147 & 0.1996 & 2.3407 & 0.3971 \\ \hline
300 & 4.6464 & 4.5064 & 0.0197 & 0.0050 & 0.2152 & 0.2001 & 2.3332 & 0.3984 \\ \hline
350 & 4.6498 & 4.5278 & 0.0164 & 0.0043 & 0.2151 & 0.1999 & 2.3333 & 0.3984 \\ \hline
400 & 4.6486 & 4.5410 & 0.0142 & 0.0038 & 0.2151 & 0.2000 & 2.3315 & 0.3987 \\ \hline
450 & 4.6466 & 4.5506 & 0.0121 & 0.0034 & 0.2152 & 0.2001 & 2.3294 & 0.3991 \\ \hline
500 & 4.6499 & 4.5643 & 0.0106 & 0.0030 & 0.2151 & 0.1999 & 2.3303 & 0.3989 \\ \hline
\end{tabular}
\label{table:capacitance}
\end{table}
\egroup

\clearpage
\backmatter

\bmhead{Supplementary Information}

Supplementary Information is provided below.

\bmhead{Acknowledgements}

We acknowledge useful conversations with Steve Kivelson, Akshat Pandey, David Goldhaber-Gordon, Elaine Taylor, Charlotte B{\o}ttcher, Yusuke Iguchi, Lila Rodgers, Sergey Tolpygo, and Katja Nowack.

\section{Declarations}

\subsection{Funding}

Work at Stanford (L.B.V.H. and K.A.M.) was supported by the U.S. Department of Energy (DOE), Office of Science, Basic Energy Sciences (BES), under Award DE-SC0021238.
Work at Purdue (T.Z., S.M., T.L., and M.J.M.) was supported by Microsoft Quantum.

\subsection{Author contributions}
L.B.V.H. designed the devices, performed the transport and scanning SQUID measurements, and analyzed the data. T.L. grew the InAs/Al heterostructure. T.Z. and S.M. performed device fabrication and provided feedback on the device design. M.J.M. supervised the sample growth and fabrication. K.A.M. supervised the measurements. L.B.V.H. and K.A.M. wrote the manuscript with input from all authors.

\subsection{Competing interests}
There are no competing interests to declare.

\subsection{Data and code availability}
The source data and analysis code underlying this work are publicly available~\cite{bishop_van_horn_2026_22698999}.

\bibliography{references}

\clearpage
\begin{appendices}


\renewcommand{\thefigure}{S\arabic{figure}}
\renewcommand{\thetable}{S\arabic{table}}
\renewcommand{\theequation}{S\arabic{equation}}
\renewcommand{\thepage}{S\arabic{page}}
\setcounter{figure}{0}
\setcounter{table}{0}
\setcounter{equation}{0}
\setcounter{page}{1} 

\section{Supplementary Information}

\subsection{Millikelvin scanning SQUID microscope}

Our millikelvin scanning SQUID microscope is operated in a BlueFors LD400 cryogen-free dilution refrigerator (DR). The microscope consists of Attocube coarse positioners (ANPx101/LT/HV/BeCu for $x$ and $y$ positioning and ANPz101/LT/HV/BeCu for $z$ positioning) and a home-built S-bender style linear piezoelectric scanner~\cite{Siegel1995-ya}. The scanner is driven by the $\pm$10 V analog outputs of  a multi-channel ADC/DAC (National Instruments USB-6363) through a $20\times$ low noise high voltage amplifier (Attocube ANC250) and has a lateral scan range of roughly 250 $\mu\text{m}$ by 250 $\mu\text{m}$ at low temperature.

The SQUID susceptometer is mounted to a copper cantilever that forms one half of a parallel plate capacitor on a small printed circuit board (PCB) attached to the piezo scanner. We detect mechanical contact between the SQUID and sample by monitoring the capacitance of the cantilever. The SQUID is raster scanned at a fixed standoff distance over the surface of the sample while the sample is held fixed in space. The positioners, scanner, and sample stage are all mounted inside an oxygen-free high conductivity (OFHC) copper “cage” that is suspended from the DR mixing chamber (MXC) plate by a simple spring system with a resonant frequency of approximately 3.5 Hz (similar to the design in \cite{Bishop-Van-Horn2019-qv}). The microscope cage is thermally anchored to the MXC by flexible OFHC copper ribbons. The springs and flexible copper ribbons are used to mechanically isolate the microscope from vibrations of the MXC, especially those caused by the DR pulse tube. The cryostat has a cryogenic mu-metal magnetic shield (Amuneal) attached to the inside of the vacuum can resulting in a residual magnetic field in the sample volume of $<0.5\,\mu\mathrm{T}$ (or $< 5\,\mathrm{mG}$) as measured at room temperature by a 3-axis fluxgate magnetometer. We further null the out-of-plane background magnetic field at the plane of the sample using a small superconducting coil as described below.

\begin{figure}[h]
    \centering
    \includegraphics[width=\textwidth]{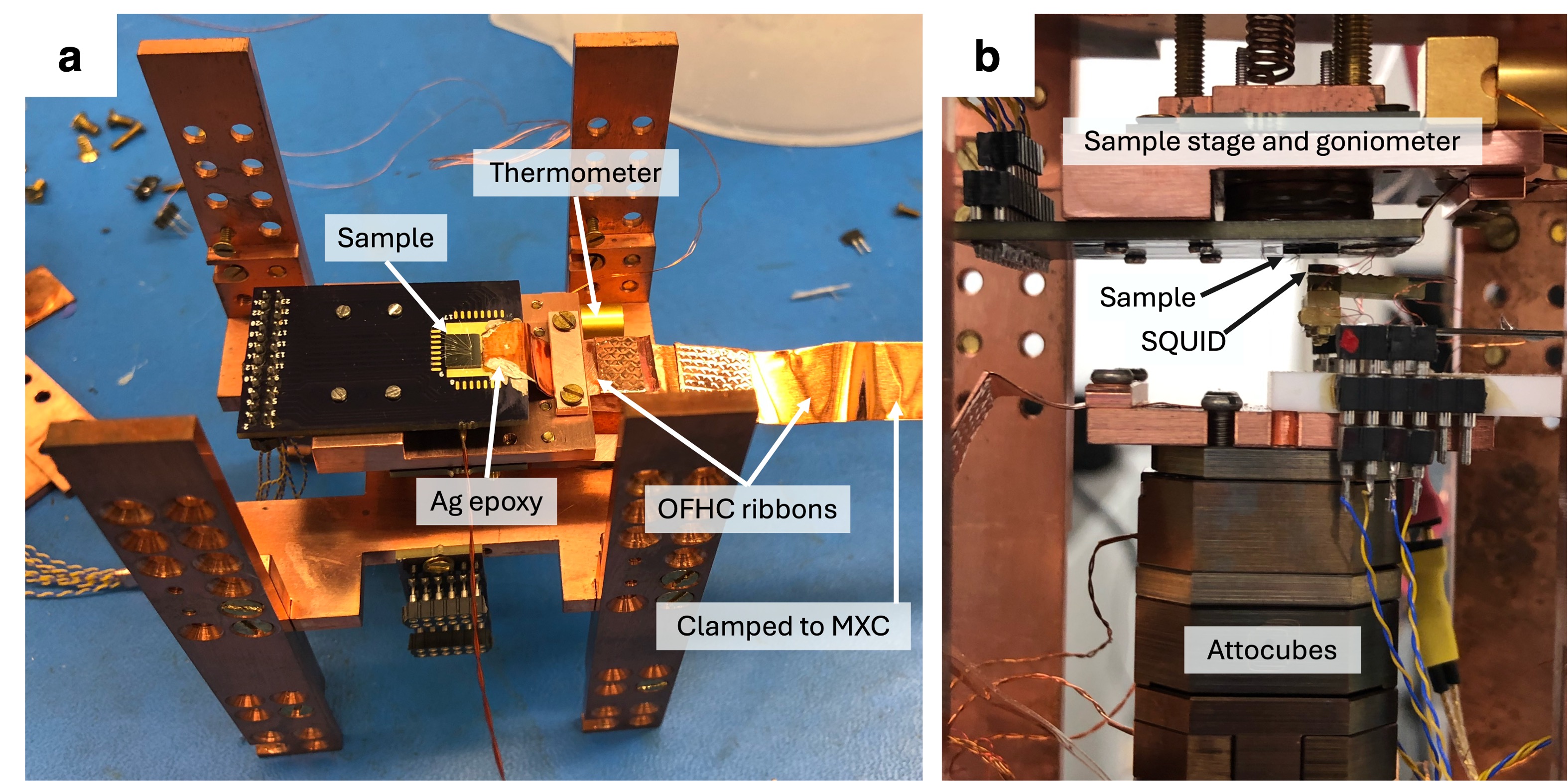}
    \caption{{\bf Scanning SQUID microscope with millikelvin sample stage.} ({\bf a}) Sample stage and PCB. The SQUID is mounted to a gold plated copper PCB pad using a thin layer of GE varnish. The PCB is mounted to an OFHC copper sample stage with four screws, and electrical connections to the sample are made with Al wire bonds. An OFHC copper ribbon is attached to the PCB pad using electrically and thermally conductive epoxy. The other end of the ribbon is clamped firmly to the sample stage. A second stack of several longer OFHC copper ribbons is clamped to both the sample stage and the mixing chamber plate. A calibrated thermometer and resistive heater are also mounted to the sample stage. ({\bf b}) Close-up of the SQUID microscope as mounted on the DR mixing chamber. The SQUID sits atop a piezoelectric scanner (not visible), which is in turn mounted to a stack of Attocube positioners. The sample stage is mounted on a goniometer to set the angle between the SQUID and the sample plane. A small hand-wound superconducting coil sits just between the sample PCB and the sample stage.}
    \label{fig:arrays-microscope}
\end{figure}

The SQUID microscope and millikelvin sample stage are shown in Figure~\ref{fig:arrays-microscope}. The sample is mounted to a gold plated copper pad on a PCB with a thin layer of GE varnish, and electrical contacts between the PCB and sample are made with aluminum wire bonds. The PCB is attached to an OFHC copper sample stage, and the PCB pad on which the sample sits is thermalized to the copper sample stage with an OFHC copper ribbon. One end of the ribbon is attached to the PCB pad with electrically and thermally conductive epoxy (EPO-TEK H20E), and the other end is clamped firmly to the copper sample stage with a copper bar and brass screws. The sample stage itself is thermally anchored directly to the MXC plate with several flexible OFHC copper ribbons. Also attached to the sample stage are a heater (a 5 $\mathrm{k}\Omega$ resistor), thermometer (Lake Shore RX-102B-RS calibrated down to 10 mK), and a small hand-wound superconducting coil used to locally cancel residual out-of-plane magnetic field in the region of interest. We adjust the current through this superconducting coil to minimize the magnetic contrast near the Al leads due to Meissner screening of the background magnetic field.
The heater and thermometer are connected to a Lake Shore Model 372 temperature controller to control the sample temperature. With this configuration, the base temperature of the MXC plate is $<10$ mK and the base temperature of the sample stage is $<20$ mK.

\subsection{Electronics and wiring}
\label{sec:arrays-wiring}

The scanning SQUID sensor used in this work is a gradiometric niobium SQUID susceptometer with on-chip field coil and flux modulation coils~\cite{Kirtley2016-zz}. The bare mutual inductance between the field coil and pickup loop is 1580 $\Phi_0/\mathrm{A}$ or 3.27 pH. The field coil is used to locally apply magnetic field to the sample, while the modulation coil is used for flux-locked loop readout of the SQUID as described below.

The SQUID susceptometer is voltage biased with a 1 $\Omega$ shunt resistor located at the 3 K stage of the DR. The current through the SQUID is inductively coupled to a series SQUID array amplifier (SSA)~\cite{Huber2001-py} also at 3 K, which acts as a cryogenic transimpedance preamplifier. The SQUID is operated in a flux-locked loop (FLL) controlled by a commercial FPGA board\footnote{Red Pitaya StemLab 125-14, https://redpitaya.com/stemlab-125-14/} running open source firmware and software~\cite{Neuhaus2023-np}. The error signal for the FLL is the voltage across the SSA and the output of the FLL is a voltage proportional to the flux through the SQUID, which is fed back to the susceptometer modulation coil through a 3 $\mathrm{k}\Omega$, 30 nF (11 kHz) RC low-pass filter at room temperature to maintain a constant flux through the SQUID.

The SQUID and SSA wiring is on a loom of low resistivity twisted pairs made of copper above the 3 K stage and superconducting NbTi/CuNi from the 3 K stage to the MXC stage. The wiring for the positioners, scanner, heater, thermometer, and magnet coil is on a separate Cu + NbTi/CuNi loom. The wiring for the sample leads is on a high resistivity PhBr loom which is filtered by a QDevil QFilter-II RC and RF low-pass filter bank anchored to the MXC plate\footnote{QDevil QFilter-II, https://www.quantum-machines.co/products/qfilter/}. This filter solution is advertised to yield electron temperatures that are typically 5-15 mK above the mixing chamber temperature~\cite{van-der-Heijden2024-ii}. A more detailed discussion of sample thermalization is provided below.

The local magnetic susceptibility, transport voltage, and transport current are measured using Stanford Research SR830 lock-in amplifiers, labeled SR830${}_\mathrm{SUSC}$, SR830${}_V$, and SR830${}_I$, respectively. SR830${}_\mathrm{SUSC}$ supplies an AC current to the SQUID field coil through a 1 $\mathrm{k}\Omega$ room temperature series resistor and demodulates the SQUID flux signal at frequency $f_\mathrm{SUSC}$ ($=887.7$ Hz unless otherwise noted). SR830${}_V$ and SR830${}_I$ are phase locked together at frequency $f_\mathrm{IV}$. SR830${}_V$ supplies the source-drain current to the sample through a 50 M$\Omega$ room temperature series resistor, the $\sim$4 $\mathrm{k}\Omega$ line resistance provided by the wiring and QDevil filters, and the 1 $\mathrm{k}\Omega$ input impedance of the SR830${}_I$ $10^6$ V/A current input. The voltage across the sample is measured differentially by SR830${}_V$, while the source-drain current is measured by SR830${}_I$. In some cases, we use an SR560 voltage pre-amplifier prior to SR830${}_V$.
A fourth lock-in amplifier, SR830${}_\mathrm{CAP}$ operating at $f_\mathrm{CAP}=2189\,\mathrm{Hz}$, is used along with a General Radio 1615 capacitance bridge to measure the capacitance of the copper cantilever on which the SQUID is mounted, in order to detect mechanical contact between the SQUID and sample.

\subsection{Background subtraction and phase correction}
\label{sec:arrays-background-subtraction}

While the gradiometric, shielded design of the SQUID is intended to ensure that only the front pickup-loop/field coil pair is sensitive to magnetic flux, there is inevitably some inductive interaction between the sample and the rest of the on-chip SQUID circuit. This interaction results in an unwanted background in the phase stiffness measurements. The measured temperature-dependent background is small ($<0.1\%$ of the bare field coil/pickup loop mutual inductance [$1580\,\Phi_0/\mathrm{A}$] and $<0.4\%$ of the signal from the 5 nm thick Al film [$250\,\Phi_0/\mathrm{A}$]), but it is comparable to the magnetic response of the arrays. In addition to this temperature dependent background due to unwanted coupling between the SQUID and sample, there is a fixed offset in the SQUID field coil/pickup loop mutual inductance due to a slight asymmetry ($\sim0.2\%$) between the two field coil/pickup loop pairs. To isolate the magnetic response of the arrays at a given SQUID position, we measured the temperature dependence of the background at that position with the array gated deep into the insulating regime, where there is essentially no contrast between the array and the non-magnetic mesa etched substrate. We then subtracted this background from the $\rho_s(T)$ curves.

\begin{figure}[h]
    \centering
    \includegraphics[width=\textwidth]{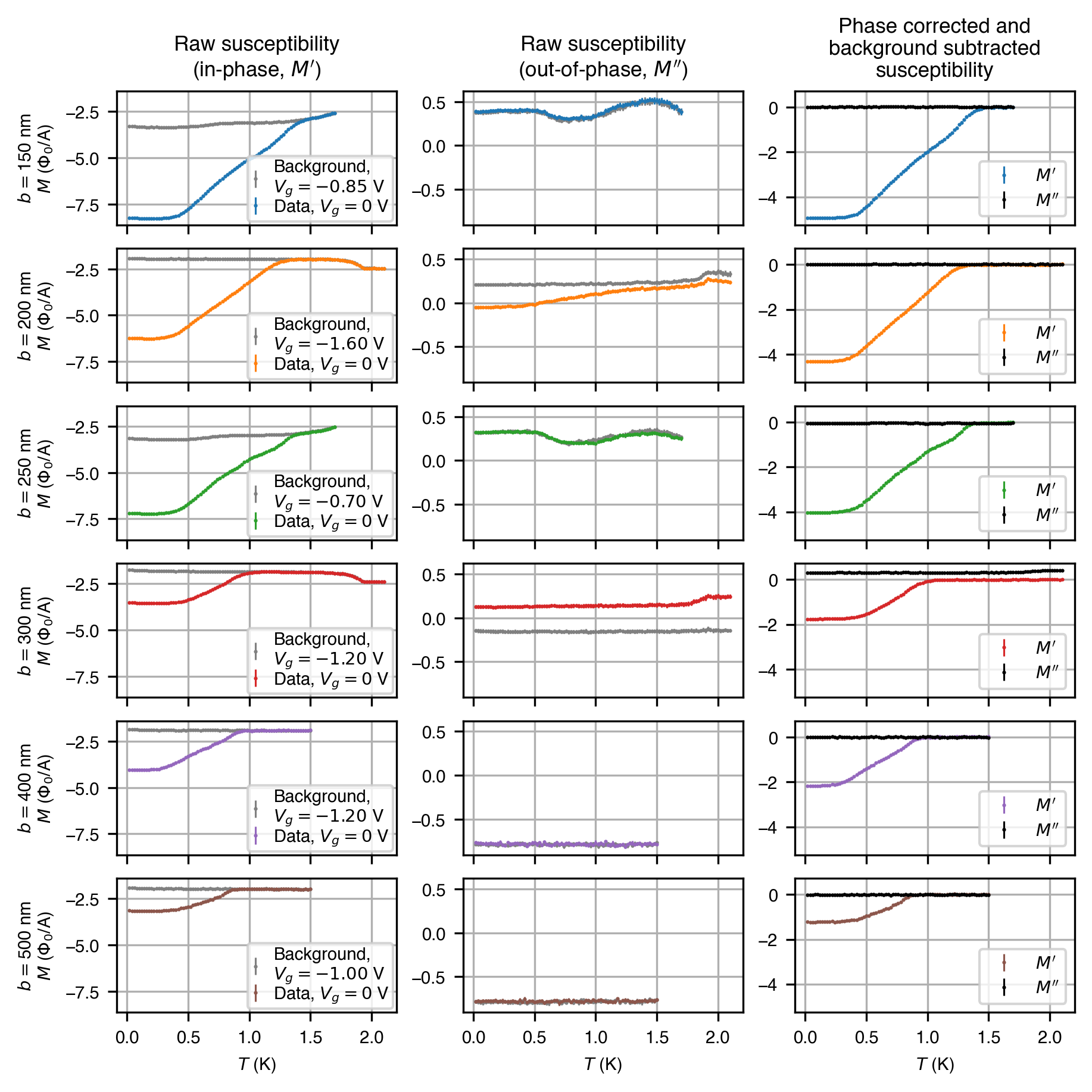}
    \caption{{\bf Background subtraction and lock-in amplifier phase correction procedure for the measured susceptibility.} Left column: raw in-phase susceptibility $M'$ measured at $V_g=0$ V (colors) and at a $V_g$ deep in the insulating regime (light gray) for all six values of the island spacing. Center column: raw out-of-phase susceptibility $M''$. Right column: The in-phase susceptibility $M'$ (colors) and out-of-phase susceptibility $M''$ (black) after background subtraction and phase offset correction. The shape of the background is correlated with the position of each array on the $\sim1\,\text{cm}\times1\,\text{cm}$ chip, as illustrated in Figure~\ref{fig:chip-layout}.} 
    \label{fig:background-subtraction}
\end{figure}

\begin{figure}[h]
    \centering
    \includegraphics[width=0.6\textwidth]{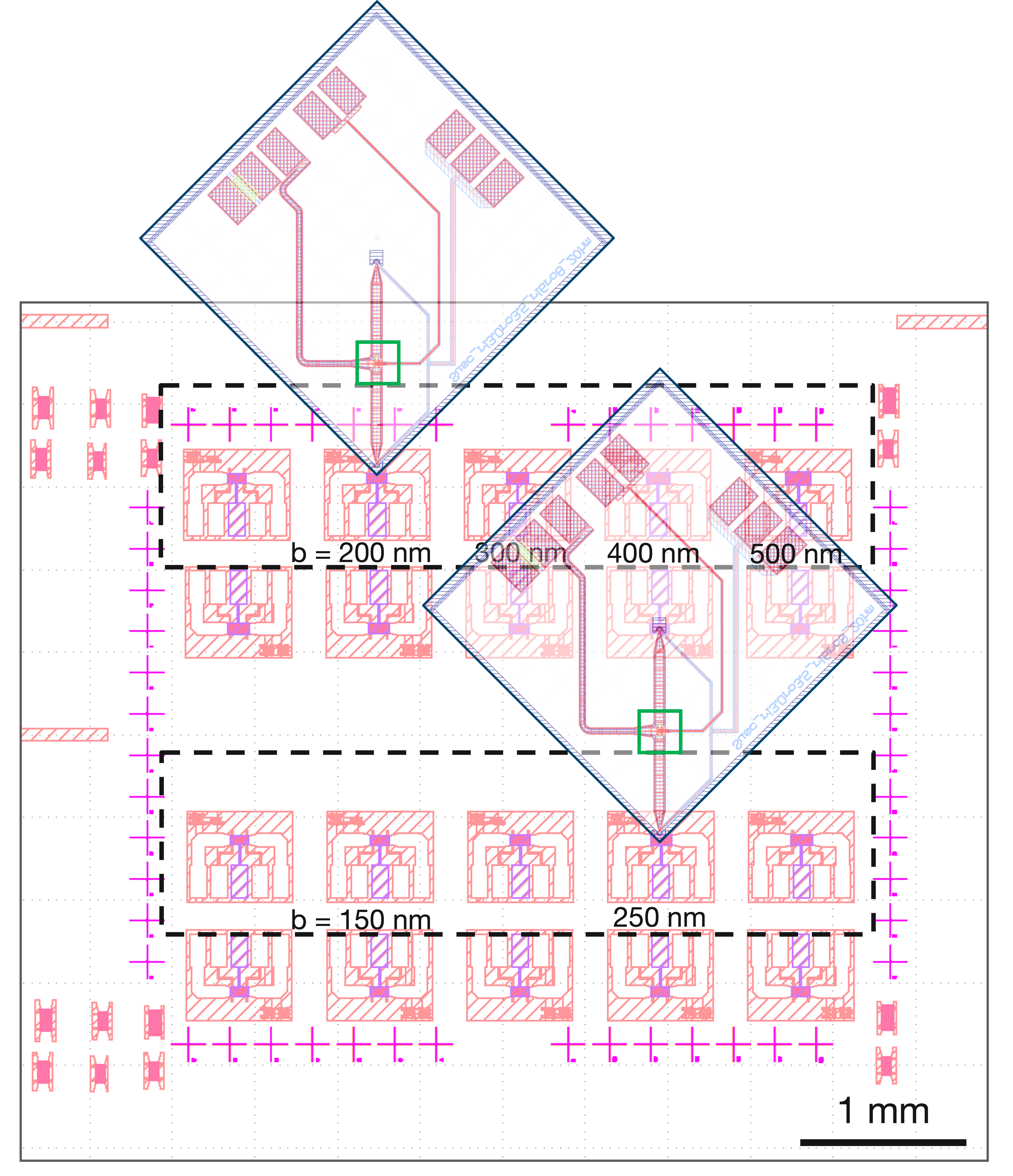}
    \caption{{\bf GDS layout of the sample chip and two copies of the SQUID chip.} The white regions on the sample chip indicate the Al film on InAs 2DEG. The hatched pink region surrounding each device is the mesa etch. The arrays with $b=200$ nm, 300 nm, 400 nm, and 500 nm are in the row of devices marked by the upper dashed box. When the SQUID is positioned over these two arrays, the modulation coil (green box) is near the edge of the chip and the back field coil/pickup loop pair is off the chip. The arrays with $b=150$ nm and 250 nm are in the row of devices marked by the lower dashed box. When the SQUID is positioned over these two arrays, the modulation coil is near the center of the chip (green boxes) and the back field coil/pickup loop pair located over the mesa etched region surrounding other devices on the chip.} 
    \label{fig:chip-layout}
\end{figure}

The background subtraction procedure is illustrated in Figure~\ref{fig:background-subtraction}. We first measure the background $M_\mathrm{bg}=M'_\mathrm{bg} + iM''_\mathrm{bg}$ with the array gated deep into the insulating regime, and bin the in-phase component $M'_\mathrm{bg}$ and out-of-phase component $M''_\mathrm{bg}$ by temperature as described above. We use the binned background and the binned data at $V_g=0$ V, $M_{\mathrm{data}0}=M'_{\mathrm{data}0} + iM''_{\mathrm{data}0}$, to determine the lock-in amplifier phase offset $\phi$. We subtract the background from the $V_g=0$ V data, yielding $M_{\mathrm{sub}0}=M_{\mathrm{data}0}-M_\mathrm{bg}$, then fit $M_{\mathrm{sub}0}$ to a linear model, $\mathrm{imag}(M_{\mathrm{sub}0})=c_0 + c_1\mathrm{real}(M_{\mathrm{sub}0})$, and define the phase offset $\phi=\tan^{-1}(c_1)$. After finding the phase offset $\phi$, we rotate the background in the complex plane by $-\phi$ to correct the phase offset: $M_\mathrm{bg}\to M_\mathrm{bg}\exp(-i\phi)$. Finally, for each gate voltage we rotate the raw time series susceptibility data in the complex plane by $-\phi$, bin the rotated data, then subtract the binned and rotated background. For a given device, the same rotation of $-\phi$ (derived from the $V_g=0$ V data) is applied for all values of $V_g$.

In addition to the unwanted background, this subtraction process will also subtract off any diamagnetic signal from the isolated Al islands. As a result, the $\rho_s(T)$ and $M(T)$ curves represent the magnetic response due only to supercurrent flowing between islands. Because the islands are much smaller than the effective penetration depth of the 5 nm Al film, the signal from isolated Al islands is close to our sensitivity, as can be seen by the lack of contrast between the array and the mesa etched substrate at very negative gate voltages, for example in the lower right panel of Figure~\ref{fig:fig5}e.

The shape of the background curves in Figure~\ref{fig:background-subtraction} (light gray) supports our interpretation that that the background is due to an unwanted SQUID-sample interaction far from the region of interest. In particular, the shape of the background is clearly correlated with the general position of the devices on the $\sim1\,\text{cm}\times 1\,\text{cm}$ sample chip, as illustrated in Figure~\ref{fig:chip-layout}. In future work, this type of background could likely be largely avoided by removing the Al in the regions between array devices. For scanning SQUID maps of the magnetic response (e.g., Figures~\ref{fig:fig1}(d,e) and \ref{fig:fig5}e), pixels acquired over the nonmagnetic substrate are averaged together and this average is subtracted to remove the offset in $M$ due to asymmetry between the front and back field coil/pickup loop pairs.

\subsection{Conversion from mutual inductance to phase stiffness}

\begin{figure}[h!]
    \centering
    \includegraphics[width=\textwidth]{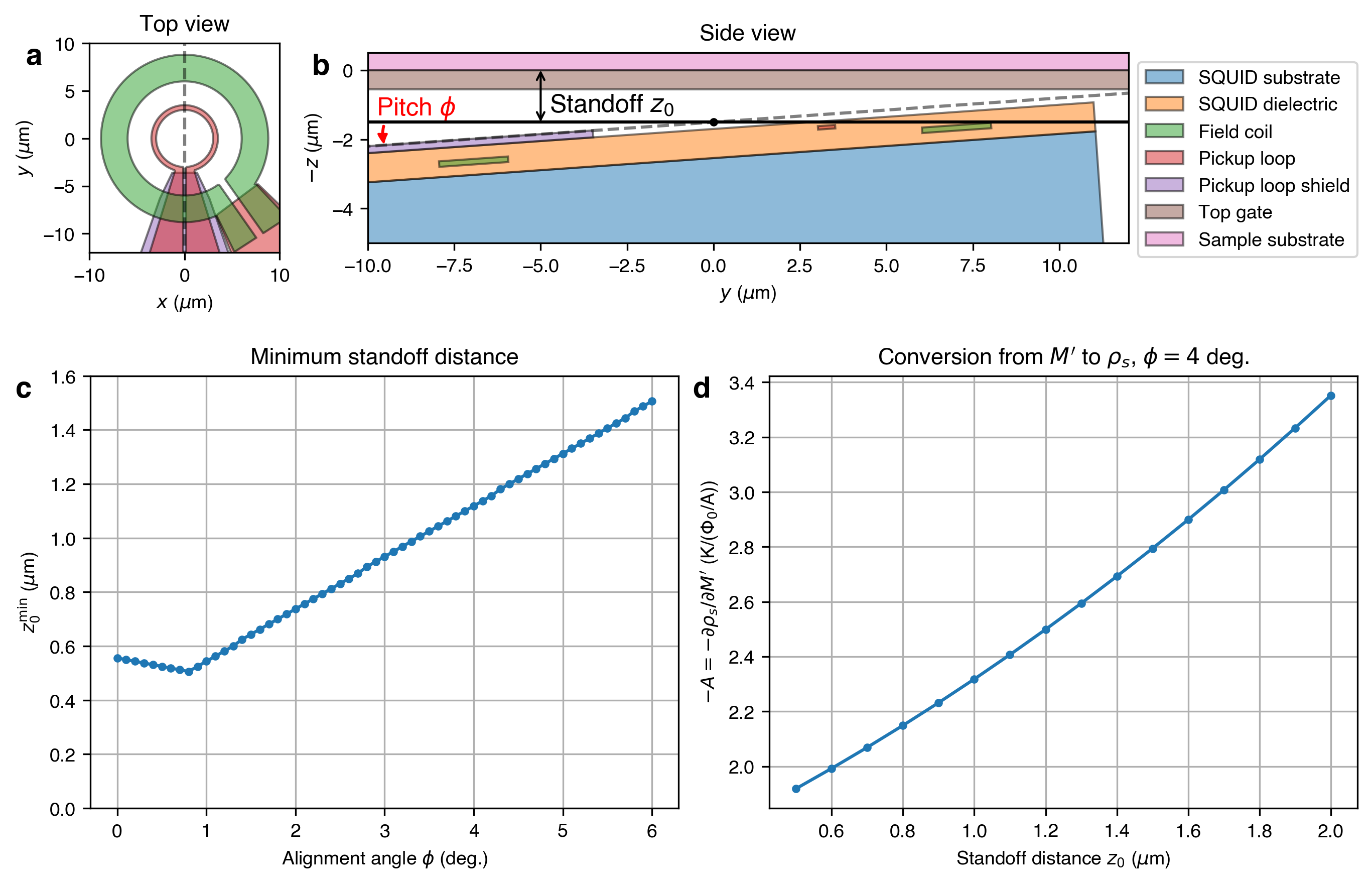}
    \caption{{\bf Simulation of the effect of SQUID standoff distance and alignment angle.} ({\bf a}) Top view of the SQUID field coil and pickup loop. ({\bf b}) Cross section of the SQUID chip near the field coil and pickup loop. ({\bf c}) Minimum possible SQUID standoff distance $z_0$ as a function of pitch angle $\phi$ (as defined in the top right panel). ({\bf d}) Simulated conversion factor from $M'$ to $\rho_s$, $-A=-\partial\rho_s/\partial M'$, as a function of standoff distance for $\phi=4^\circ$.} 
    \label{fig:standoff}
\end{figure}

The largest source of systematic uncertainty in our measurement of the superfluid stiffness is the relationship in space between the SQUID susceptometer and the sample. 
The minimum standoff distance between the sample and the field coil or pickup loop is determined by the structure of the SQUID chip near the pickup loop/field coil (Figure~\ref{fig:standoff}(a,b)) and the angle between the plane of the sample and the plane of the SQUID chip. We set this angle at room temperature using the sample stage goniometer prior to each cooldown. Based on the measured alignment angle, which can vary slightly between cooldowns, we can estimate the nominal standoff distance for a given cooldown. We estimate the uncertainty in the standoff distance to be $-0.2\,\mu\mathrm{m}$/$+0.5\,\mu\mathrm{m}$, where negative means smaller standoff distance and positive means larger. The alignment angle and estimated standoff distance for each device are shown in Table~\ref{table:alignment}.

To extract the superfluid stiffness, we model the array as a rectangular 2D superconducting film with the known lateral dimensions of the array. In the weak screening limit $\Lambda\gg r_\text{FC}$, where $r_\text{FC}$ is the radius of the SQUID field coil, the SQUID mutual inductance signal $M'$ varies linearly with the inverse effective penetration depth $\Lambda^{-1}$, $M'\propto -\Lambda^{-1}$. The constant of proportionality in general depends on the sample geometry, SQUID standoff, and (very weakly) on the SQUID alignment angle. Expressed in terms of the superfluid stiffness, we have 
\begin{equation}
    \rho_s=\frac{\hbar^2\Lambda^{-1}}{4\mu_0k_\text{B} e^2}=AM',
    \label{eq:rhos}
\end{equation}
where $A=\partial\rho_s/\partial M'$ is independent of gate voltage and temperature and has units of $\mathrm{K}/(\Phi_0 / \mathrm{A})$. We estimate $A$ based on finite element modeling~\cite{Bishop-Van_Horn2022-sy} of the SQUID-sample magnetic interaction using the known SQUID and sample geometries and the alignment angle measured at room temperature. The spatial distribution of the magnetic field generated by the SQUID field coil for a given standoff distance is shown in Figure~\ref{fig:squid-field}. The setup for the simulations used to estimate $A$ is shown in Figure~\ref{fig:standoff}, and the resulting estimated values for $A$ are shown in Table~\ref{table:alignment}.

\bgroup
\begin{table}
\caption{{\bf Estimated SQUID standoff distance and proportionality between $\rho_s$ and $M'$ based on SQUID alignment angle measured at room temperature.} The minimum standoff distance is set by the SQUID alignment angle and the thickness of the array top gate. The estimated systematic uncertainty in $z_0$ translates to a systematic uncertainty in $-A=-\partial\rho_s/\partial M'$ of $+20\%/-10\%$.\\}

\begin{tabular}{|c|c|c|c|c|}
\hline
Island spacing $b$ & Cooldown & Pitch angle $\phi$      & Standoff distance $z_0$ & $-A=-\frac{\partial\rho_s}{\partial M'}$ \\
{(}nm{)}           &          & {(}${}^\circ${)} & {(}$\mu\text{m}${)}             & {(}$\mathrm{K}/(\Phi_0/\mathrm{A})${)}  \\ \hline \hline
150                & 2        & 5                & $1.5^{+0.5}_{-0.2}$     & $2.81^{+0.56}_{-0.28}$                  \\ \hline
200                & 1        & 4                & $1.3^{+0.5}_{-0.2}$     & $2.60^{+0.52}_{-0.26}$                  \\ \hline
250                & 2        & 5                & $1.5^{+0.5}_{-0.2}$     & $2.81^{+0.56}_{-0.28}$                  \\ \hline
300                & 1        & 4                & $1.3^{+0.5}_{-0.2}$     & $2.60^{+0.52}_{-0.26}$                  \\ \hline
400                & 3        & 3.5              & $1.2^{+0.5}_{-0.2}$     & $2.47^{+0.49}_{-0.25}$                  \\ \hline
500                & 3        & 3.5              & $1.2^{+0.5}_{-0.2}$     & $2.47^{+0.49}_{-0.25}$                  \\ \hline
\end{tabular}
\label{table:alignment}
\end{table}
\egroup

As a consistency check, we consider an analytical model of the SQUID-sample interaction~\cite{Kirtley2012-od}, which predicts the following relationship in the thin film, weak screening limit:
\begin{equation}
    \frac{M'(\bar{z})}{M_0}=-\frac{r_\text{FC}}{2\Lambda}\left(1-\frac{2\bar{z}}{\sqrt{1+4\bar{z}^2}}\right).
    \label{eq:kirtley-model}
\end{equation}
Here, $\bar{z}=z_0/r_\text{FC}$ where $z_0$ is the standoff distance, $M_0$ is the mutual inductance of the field coil/pickup loop pair in the absence of the sample, and $2\Lambda$ is the Pearl length~\cite{Pearl1964-cl}. In this model~\cite{Kirtley2012-od}, the field coil and pickup loop are approximated as coaxial, coplanar 1D loops and the sample is modeled as a uniform infinite plane. Combining Eqs.~\ref{eq:rhos} and \ref{eq:kirtley-model}, we find
\begin{equation}
    \rho_s=-\frac{M'(\bar{z})}{M_0}\frac{\hbar^2}{2\mu_0k_\text{B}e^2r_\text{FC}}\left(1-\frac{2\bar{z}}{\sqrt{1+4\bar{z}^2}}\right)^{-1}.
    \label{eq:kirtley-rhos}
\end{equation}
Using $M_0=1580\,\Phi_0/\text{A}$, $r_\text{FC}=6\,\mu\mathrm{m}$, and $z_0=1.5\,\mu\mathrm{m}$, Eq.~\ref{eq:kirtley-rhos} predicts $-\partial\rho_s/\partial M'=2.38\,\text{K}/(\Phi_0/\text{A})$. This result is close to the values shown in Table~\ref{table:alignment}, which were obtained from finite element simulations of the SQUID-sample interaction.
Note that none of the experimental findings presented in the main text are particularly sensitive to the precise value of the conversion factor $A=\partial\rho_s/\partial M'$.

\subsection{Monte Carlo simulations of the classical XY model}

We evaluated the helicity modulus (Eq.~\ref{eq:helicity}) using Wolff's cluster update algorithm~\cite{Wolff1989-su}. We start by representing the state of each site by a ``spin'' or unit vector in $\mathbb{R}^2$. A site with phase $\varphi$ is represented by a vector $\vec{s}=(\cos\varphi, \sin\varphi)$. In the Wolff algorithm, a cluster of spins $C$ is constructed and flipped as follows:
\begin{enumerate}
    \item Select a random vector $\vec{r}$ from the unit circle and a random initial site $i$ for the cluster.
    \item Flip the initial spin according to $\vec{s}_i\to\vec{s}_i-2(\vec{s}_i\cdot\vec{r})\vec{r}$, mark the spin as flipped, and add it to the cluster $C$.
    \item Visit all sites $j$ connected to site $i$. With probability
    $$P_{ij}=1-\exp\left(\min\left[0, \frac{2E_\text{J}}{k_\text{B}T}(\vec{s}_i\cdot\vec{r})(\vec{s}_j\cdot\vec{r})\right]\right),$$
    flip site $j$ via $\vec{s}_j\to\vec{s}_j-2(\vec{s}_j\cdot\vec{r})\vec{r}$, mark the site as flipped, and add it to the cluster $C$.
    \item Continue by visiting each site connected to the newly added spins until no more unmarked spins can be flipped. The updated phases $\varphi_i$ are then given by the counterclockwise angle between $\vec{s}_i$ and the positive $x$-axis.
\end{enumerate}

For the simulations in Figure~\ref{fig:fig3}, we used a system with 20 rows and 40 columns. At each temperature, we initialized the system with phases distributed uniformly in the range $[-\pi,\pi)$. We then performed 10,000 ``thermalization'' cluster updates and discarded the results. After this thermalization step, we performed 100,000 ``measurement'' cluster updates, where we recorded observables such as the helicity modulus (Eq.~\ref{eq:helicity}) after each update. The Wolff algorithm has a short autocorrelation time, meaning that it efficiently generates uncorrelated samples from valid classical XY phase configurations. However, the algorithm does not support frustrated couplings, for example due to applied fields.

To investigate the effect of finite system size, open boundary conditions, and the experimental method used to probe the phase stiffness, we also performed Monte Carlo simulations using the single-flip Metropolis-Hastings (MH) algorithm~\cite{Metropolis1953-ds,Hastings1970-rz}. This approach is less efficient than cluster update methods such as Wolff's algorithm, but accommodates frustrated couplings/applied gauge fields.

In an MH update step, a single site $i$ is selected at random. A perturbation $\delta\varphi$ to the phase $\varphi$ is selected at random. In our case, we sample $\delta\varphi$ uniformly in the range $[-\pi/2,\pi/2)$. The change in the system's total energy due to this perturbation is calculated: $\delta E=E(\varphi+\delta\varphi)-E(\varphi)$. The perturbation is accepted ($\varphi_i \to \varphi_i+\delta\varphi$) with probability $\exp\left(-\max[0, \delta E/(k_\text{B}T)]\right)$. Because the MH algorithm has a long autocorrelation time, we do not calculate observables after each MH step. Instead, we perform a full MH ``pass'' consisting of $N$ MH steps, where $N$ is the number of sites in the lattice, such that each site on average experiences one trial perturbation per pass. After each MH pass, we calculate the system observables.

We simulate the SQUID mutual inductance measurement by applying a gauge field to the XY system corresponding to the magnetic vector potential generated by a current-carrying loop above the center of the array~\cite{Teitel1983-wc}. After each MH update pass, we calculate the currents flowing in the array. Then, using the Biot-Savart law, we calculate the flux through a circular pickup loop concentric with the field coil. The SQUID susceptibility signal is given by the average pickup loop flux divided by the current in the field coil.

\begin{figure}
    \centering
    \includegraphics[width=0.8\linewidth]{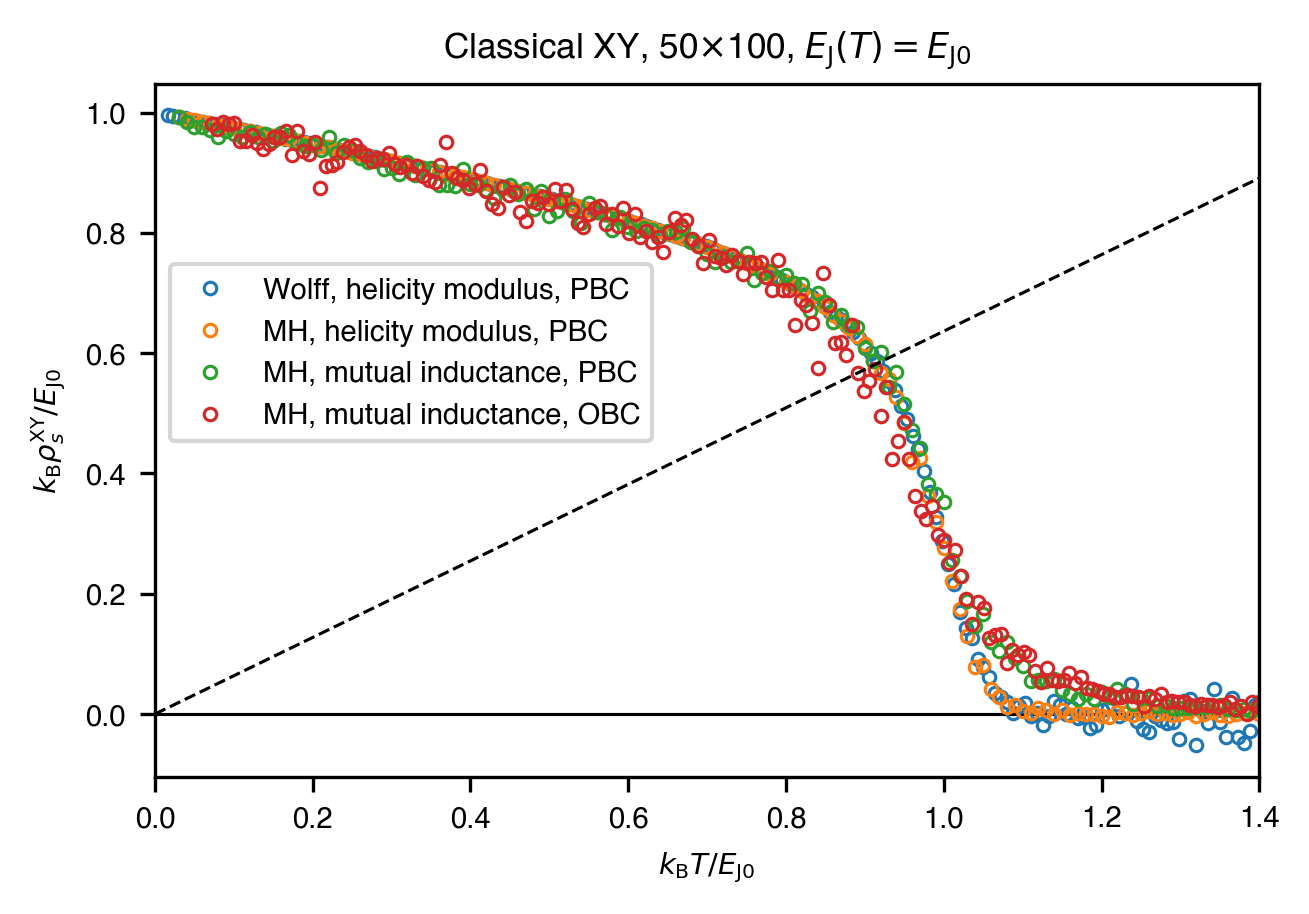}
    \caption{{\bf Monte Carlo simulations of phase stiffness in the classical XY model with temperature-independent coupling.} The blue points show the helicity modulus (Eq.~\ref{eq:helicity}) calculated using a Wolff cluster update with periodic boundary conditions (PBC). The orange points show the helicity modulus calculated using a Metropolis-Hastings (MH) update with PBC. The green points show the two-loop mutual inductance (normalized to the $T\to 0$ value) calculated using an MH update with PBC. The red points show the two-loop mutual inductance (normalized to the $T\to 0$ value) calculated using an MH update with open boundary conditions (OBC). For the Wolff simulations we performed 10,000 thermalization steps and 100,000 measurement steps. For the MH simulations we performed 100,000 thermalization passes and 1,000,000 measurement passes (each MH pass consisting of $50\times 100$ MH steps). For the two-loop mutual inductance simulation, we used an array lattice constant $(a+b)=1.2\,\mu\mathrm{m}$, field coil radius $r_\text{FC}=7.5\,\mu\mathrm{m}$, pickup loop radius $r_\text{PL}=3.5\,\mu\mathrm{m}$, standoff distance $z_0=2\,\mu\mathrm{m}$, and field coil current $I_\text{FC}=100\,\mu\text{A}$. The dashed black line indicates $\rho_s=(2/\pi)T$.}
    \label{fig:classical-xy}
\end{figure}

\begin{figure}[h]
    \centering
    \includegraphics[width=4.75in]{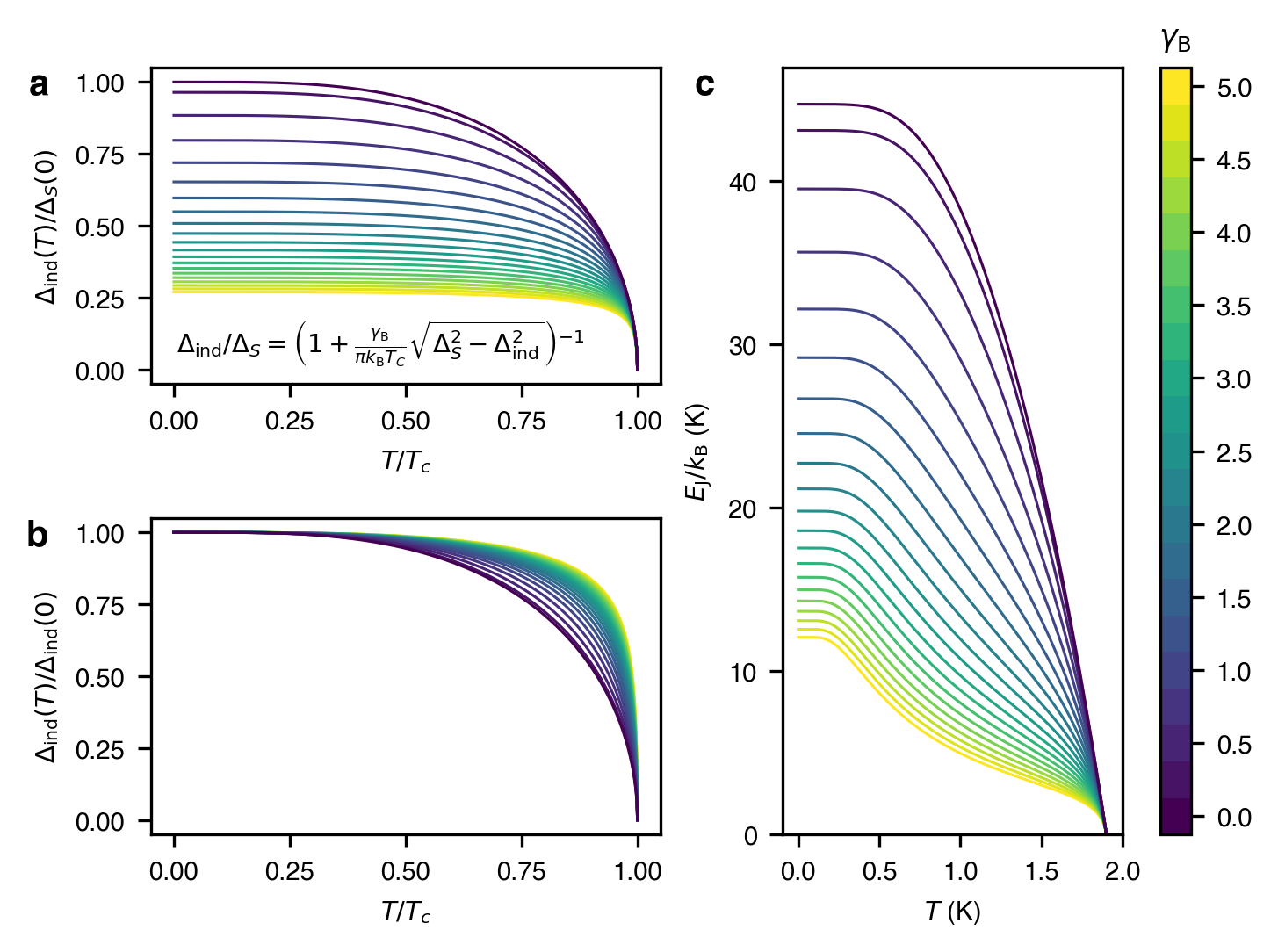}
    \caption{{\bf Theoretical estimate of the temperature-dependent proximity induced gap.} ({\bf a}) The induced gap $\Delta_\text{ind}(T)$ (Eq.~\ref{eq:induced-gap}) normalized by the zero-temperature ``parent gap'' $\Delta_S(0)$ for different values of the interface barrier strength $\gamma_\text{B}$. ({\bf b}) The induced gap $\Delta_\text{ind}(T)$ normalized by the zero-temperature induced gap $\Delta_\text{ind}(0)$ for different values of $\gamma_\text{B}$. ({\bf c}) Josephson coupling $E_\text{J}(T)/k_\text{B}$ (Eq.~\ref{eq:EJ}) for different values of $\gamma_\text{B}$ assuming $R_N=242\,\Omega/\square$ and $\Delta_S(T) = \Delta_\text{Al}(T)$ (Eq.~\ref{eq:bcs-gap}) with $T_c=T_{c,\text{Al}}=1.9$ K.}
    \label{fig:induced-gap}
\end{figure}

\begin{figure}[h]
    \centering
    \includegraphics[width=\textwidth]{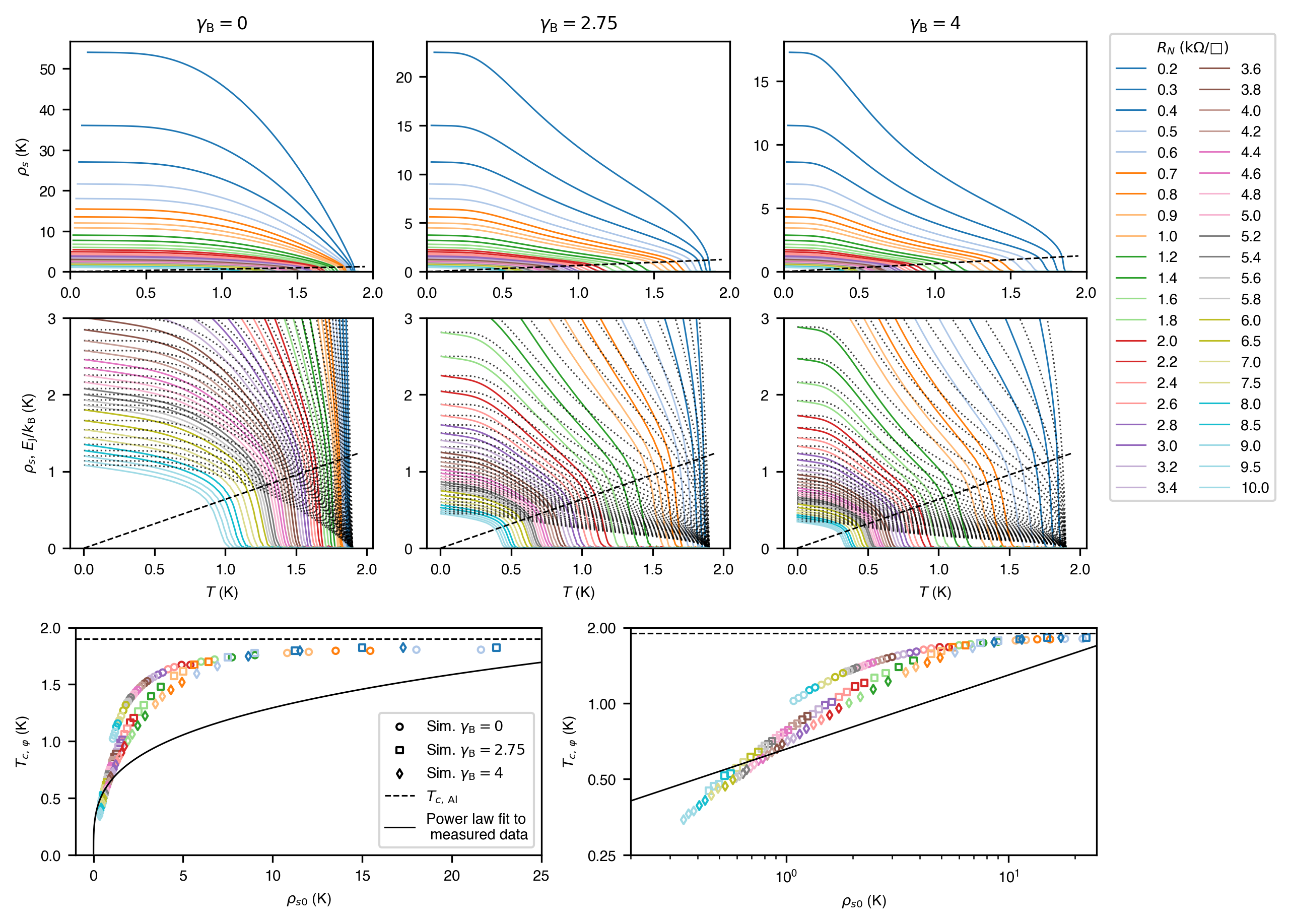}
    \caption{{\bf Classical XY simulation of the temperature-dependent phase stiffness.} ({\bf Top row}) Simulated phase stiffness $\rho_s(T)$ for different values of the normal state sheet resistance with superconductor/2DEG interface barrier strength $\gamma_\text{B}=0$ (left column), 2.75 (middle column), and 4 (right column). The dashed line is $\rho_s(T)=(2/\pi)T$. ({\bf Middle row}) The same data as the top row with the $y$ axis truncated to show the low-$\rho_{s0}$ behavior. The dotted lines show the Josephson coupling $E_\text{J}(T)/k_\text{B}$. The difference between each colored line and its corresponding dotted line represents the suppression of $\rho_s(T)$ due to thermal phase fluctuations. The dashed line is $\rho_s(T)=(2/\pi)T$. ({\bf Bottom row}) The relationship between $T_{c,\varphi}$ and $\rho_{s0}$ extracted from the classical XY simulations shown on a linear-linear scale (left) and log-log scale (right). The black dashed line shows $T_{c,\text{Al}}=1.90$ K. The solid black line shows the power law fit to the measured data: $T_{c,\varphi}=T_0^{(1-\alpha)}\rho_{s0}^\alpha$, with $T_0=0.552$ K and $\alpha=0.294$ (Fig.~\ref{fig:fig3}c).}
    \label{fig:xy-sims}
\end{figure}

Figure~\ref{fig:classical-xy} shows the phase stiffness in a $50\times100$ site classical XY system with temperature-independent coupling $E_\text{J}(T)=E_{\text{J}0}$ calculated in four different ways (see the figure caption for details). From Figure~\ref{fig:classical-xy}, we see that the two-loop mutual inductance faithfully probes the same quantity as the thermodynamic helicity modulus (Eq.~\ref{eq:helicity}). The BKT transition is slightly broadened in the mutual inductance simulation relative to the helicity modulus, most likely because the mutual inductance method effectively averages over a smaller system size (of order the size of the field coil). However, the BKT transition is still clearly visible. Similarly, the finite system size and open boundary conditions do not dramatically alter the temperature dependence of the phase stiffness.

Figure~\ref{fig:induced-gap} shows the dependence of the induced gap $\Delta_\text{ind}$ on interface barrier strength $\gamma_\text{B}$ and temperature $T/T_c$ as predicted by Eq.~\ref{eq:induced-gap}~\cite{Aminov1996-qs,Chrestin1997-vi,Schapers2001-jo,Kjaergaard2017-dh}. Figure~\ref{fig:xy-sims} show the results of Monte Carlo simulations of the classical XY model using the Josephson coupling predicted by Equations~\ref{eq:induced-gap} and \ref{eq:EJ} for $\gamma_\text{B}=0$, 2.75, and 4. These simulations show that thermal phase fluctuations alone cannot explain the measured $\rho_s(T)$ or the power law relationship between $T_{c,\varphi}$ and $\rho_{s0}$.

\subsection{Sample thermalization}
\label{sec:arrays-electron-temperature}

We do not have a direct measurement of the electron temperature of our devices, so we cannot definitively rule out the possibility that despite the filtering and thermalization strategies described above, the effective electron temperature is higher than the measured sample stage temperature. Such a failure to thermalize the array below a certain temperature could result in an apparent low temperature saturation of the resistivity and/or superfluid stiffness. Below we describe several tests we performed to assess potential sources of heating in our setup. To directly measure the electron temperature, one could fabricate gate-defined quantum dots or normal metal-insulator-superconductor (NIS) junctions on the same die as the Josephson junction arrays and perform primary electron thermometry alongside transport and magnetic imaging on the arrays. 

Note that recent work on InAs/Al Josephson junction arrays observed excess microwave noise in the anomalous metal regime, which was interpreted as a failure to thermalize the arrays below approximately 150 mK~\cite{Leonard2026-qp}. While we cannot definitively exclude such an effect here, our measurements (Figures~\ref{fig:fig5}, \ref{fig:RS-linear-200}, \ref{fig:RS-linear-300}, \ref{fig:RS-linear-400}, \ref{fig:sample150-scans}, \ref{fig:sample250-scans}, \ref{fig:sample300-scans}, \ref{fig:sample400-scans}, \ref{fig:sample200-5-80nA}, and \ref{fig:sample300-1V}) strongly suggest that in our devices anomalous metal transport is related to the onset of large-scale spatial inhomogeneity as opposed to nonequilibrium effects. In particular, in contrast to \cite{Leonard2026-qp}, the observed anomalous metal transport in our system is largely insensitive to the magnitude of the source-drain current in the range studied here.

To test for heating of the arrays by the SQUID (due to the dissipative voltage biased SQUID readout, the current in the modulation coil provided by the FLL, or the current in the field coil provided by the lock-in amplifier), we grounded all connections to the SQUID and SSA at room temperature and moved the SQUID approximately $100\,\mu\mathrm{m}$ away from the array. We observed no significant difference in the measured transport (RMS voltage across the array as a function of RMS current through the array) at base temperature in the anomalous metal and insulating regimes with the SQUID biased normally and positioned close to the sample vs. grounded and positioned far from the array.

The electron system could also in principle be heated by the electronics for driving the piezoelectric positioners and scanners, as these instruments are connected to unfiltered low-resistivity copper wiring looms. The Attocube coarse positioners are grounded at room temperature during all measurements. The piezoelectric scanner is driven by a $\pm$ 10 V 16-bit digital to analog converter (DAC, National Instruments USB-6363) through the ANC250 $20\times$ high voltage amplifier, which has a specified output noise of $20\,\mu\mathrm{V}$ RMS. To test for heating by the DAC and amplifier, we used the coarse positioners to position the SQUID $\sim 1\,\mu\mathrm{m}$ above the center of the sample, then grounded the connections to both the Attocubes and the piezo scanner at room temperature. We observed no change in the temperature below which the superfluid stiffness saturates with the piezo scanner grounded vs. operating normally.

\subsection{Additional data}

\subsubsection*{Scanning electron microscope images}

\begin{figure}[h]
    \centering
    \includegraphics[width=\textwidth]{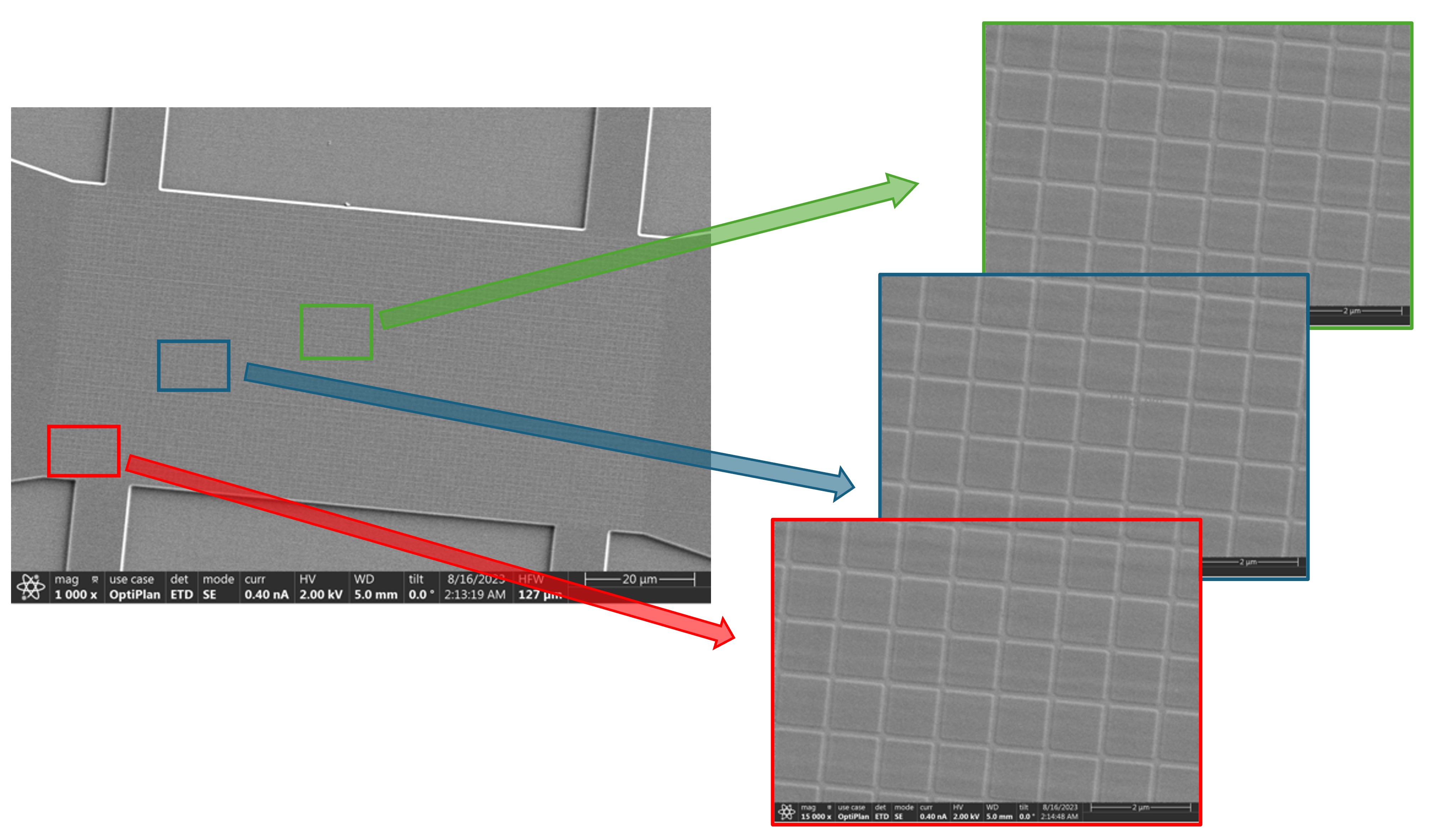}
    \caption{{\bf Scanning electron microscope (SEM) images of a test device showing uniformity of the island size and spacing across the array.} The images were taken after the mesa etch used to define the Hall bar, the Al etch used to define the islands, and the ALD growth of Al${}_2$O${}_3$ dielectric, but before deposition of the Ti/Au top gate. This test device has island size $a=1\,\mu\mathrm{m}$ and island spacing $b=100$ nm.}
    \label{fig:arrays-sem}
\end{figure}

Figure~\ref{fig:arrays-sem} shows scanning electron microscope (SEM) images of a test device, demonstrating uniformity in the Al island spacing across the array.

\subsubsection*{Temperature sweeps}

\begin{figure}[h]
    \centering
    \includegraphics[width=0.6\linewidth]{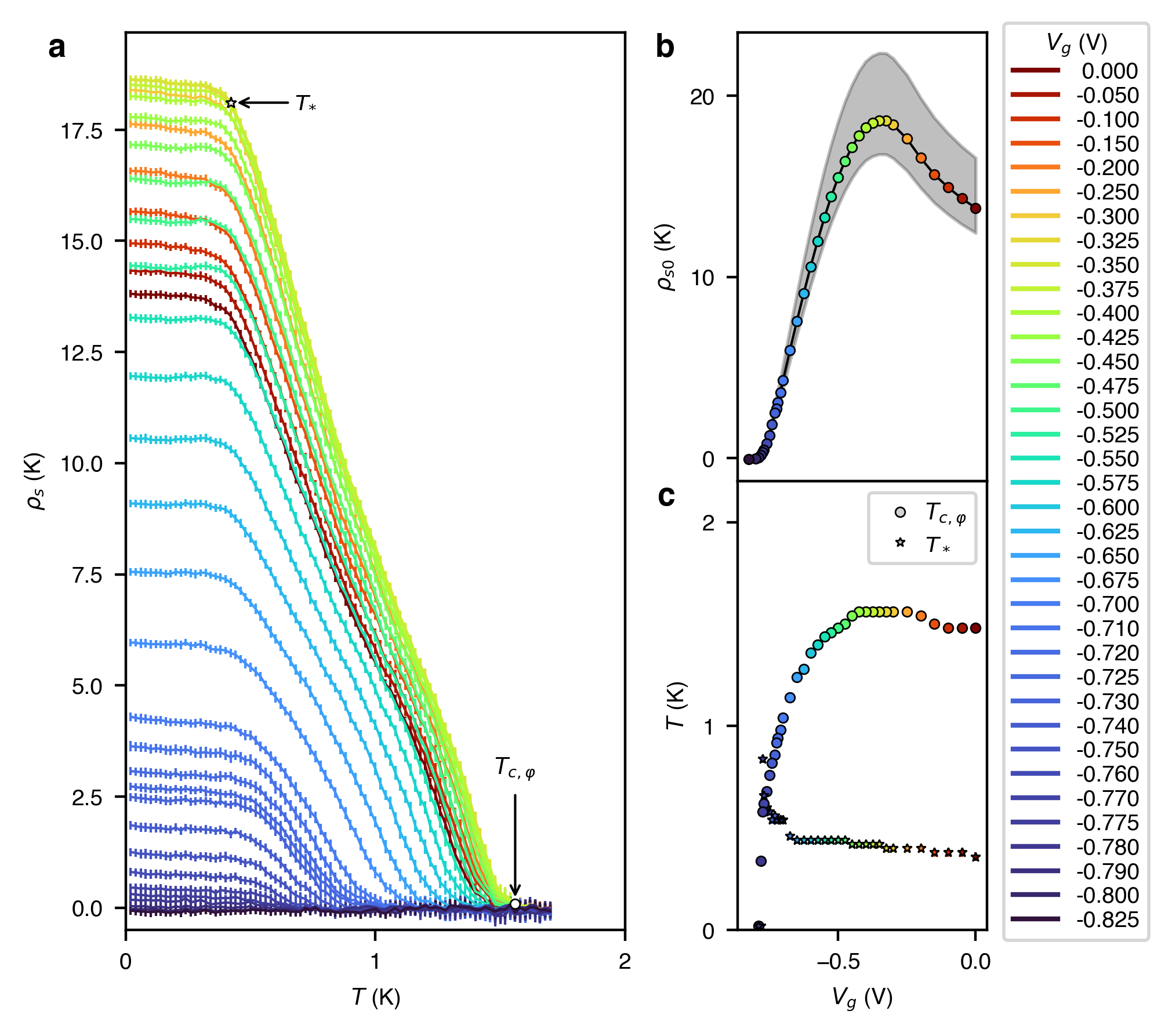}
    \caption{{\bf Phase stiffness data for $b=150$ nm.}}
    \label{fig:sweep-150}
\end{figure}

\begin{figure}[h]
    \centering
    \includegraphics[width=0.6\linewidth]{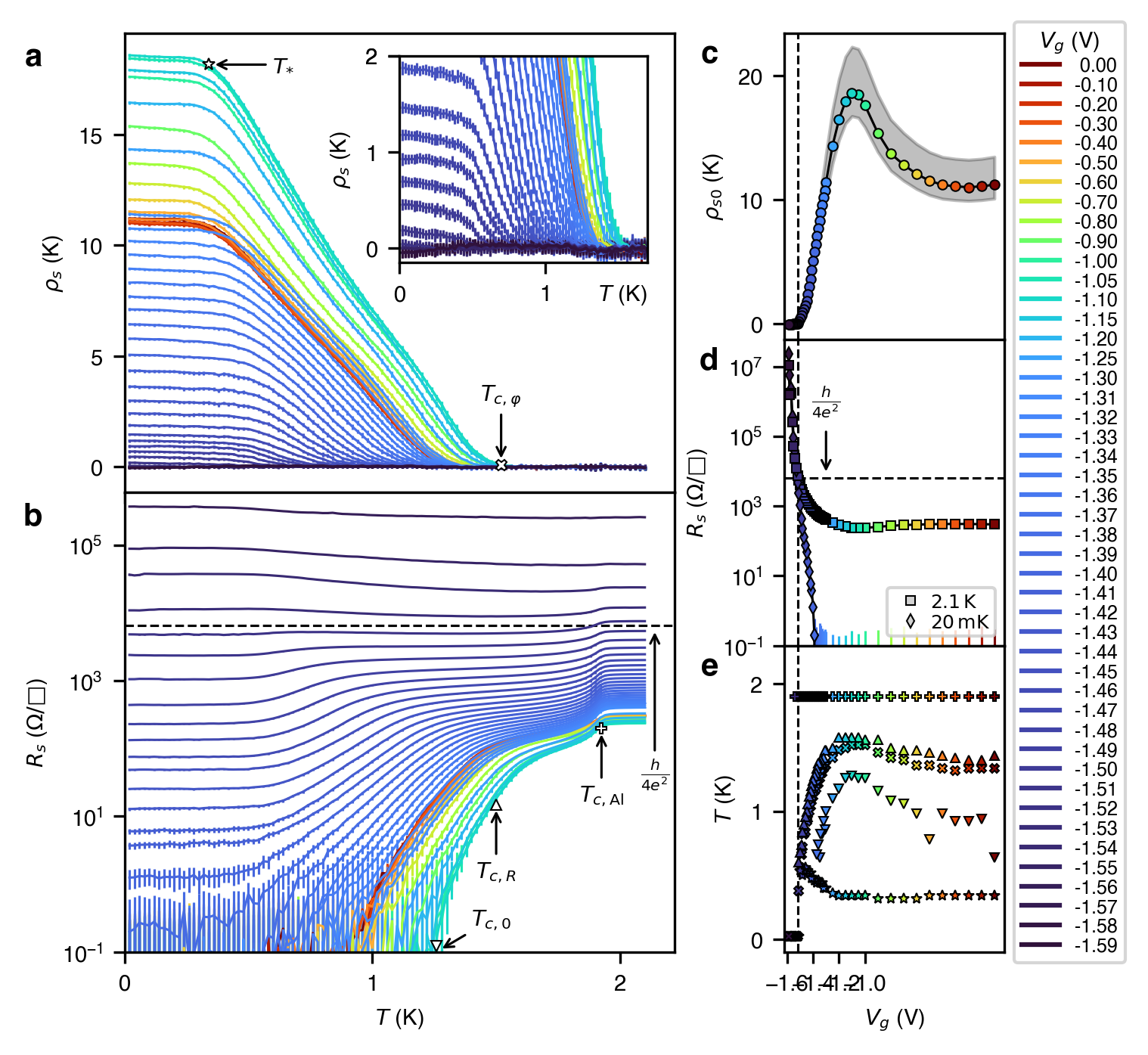}
    \caption{{\bf Phase stiffness and sheet resistance data for $b=200$ nm.} The data for $V_g\leq -1.05$ V is shown in Figure~\ref{fig:fig4}. In the superconducting and anomalous metal regimes, $V_g\geq -1.55$ V, $R_s$ was measured using a current biased four-terminal setup with RMS source-drain current $<5$ nA at frequency 137.77 Hz. In the insulating regime, $V_g < -1.55$ V, $R_s$ was measured using a two-terminal setup with RMS voltage bias $<0.5$ mV and source-drain current $<1.4$ nA at 17.777 Hz. This is the same dataset that is shown in Fig.~\ref{fig:fig4}.
    The width of the array is $60\,\mu\mathrm{m}$ and the minimum distance between the voltage probes is $86\,\mu\mathrm{m}$, so $R_s=R_{xx}\times 60/86$, where $R_{xx}$ is the measured longitudinal resistance.
    See the caption of Figure~\ref{fig:fig4} in the main text for more details.}
    \label{fig:sweep-200}
\end{figure}

\begin{figure}[h]
    \centering
    \includegraphics[width=0.6\linewidth]{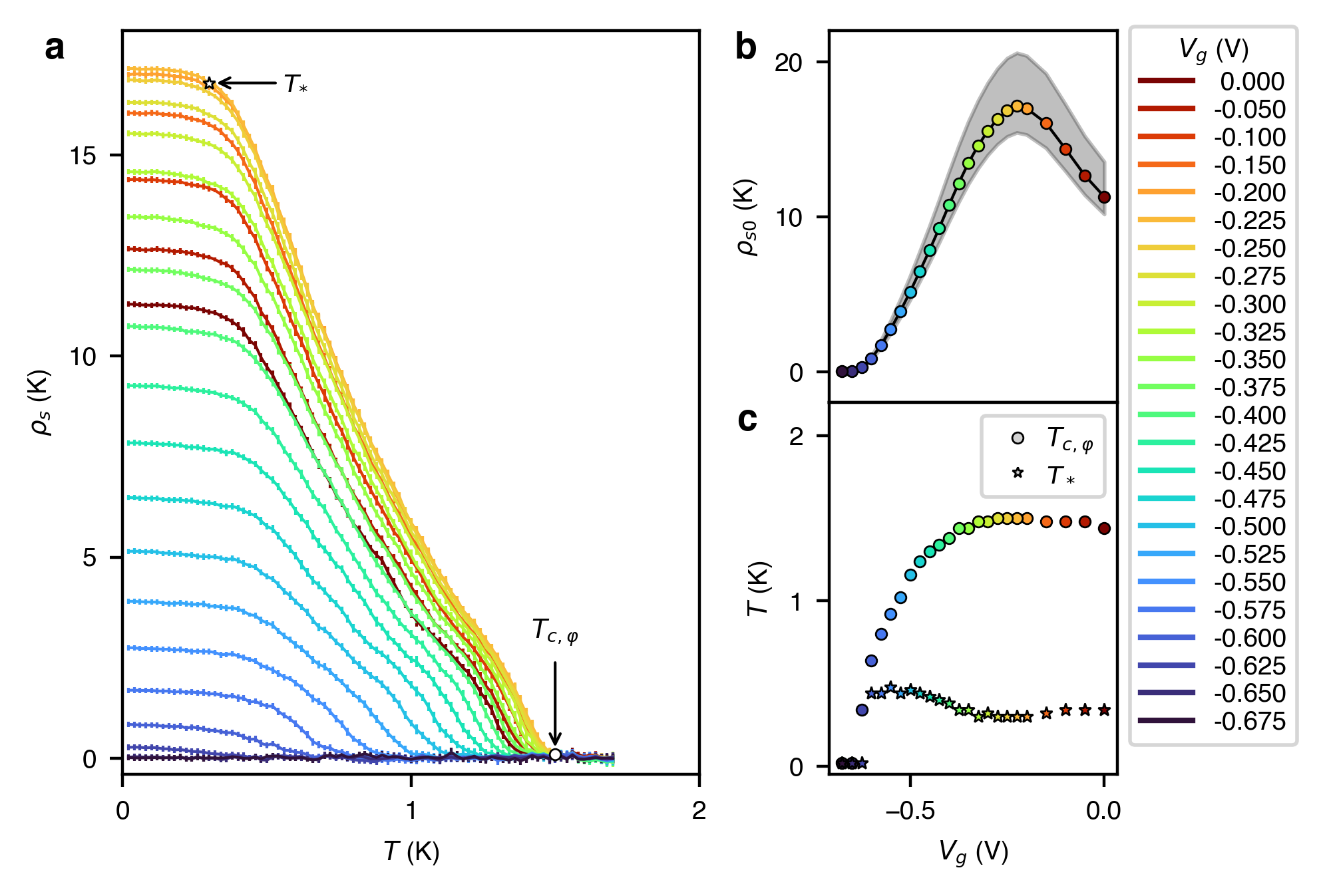}
    \caption{{\bf Phase stiffness data for $b=250$ nm.}}
    \label{fig:sweep-250}
\end{figure}

\begin{figure}[h]
    \centering
    \includegraphics[width=0.6\textwidth]{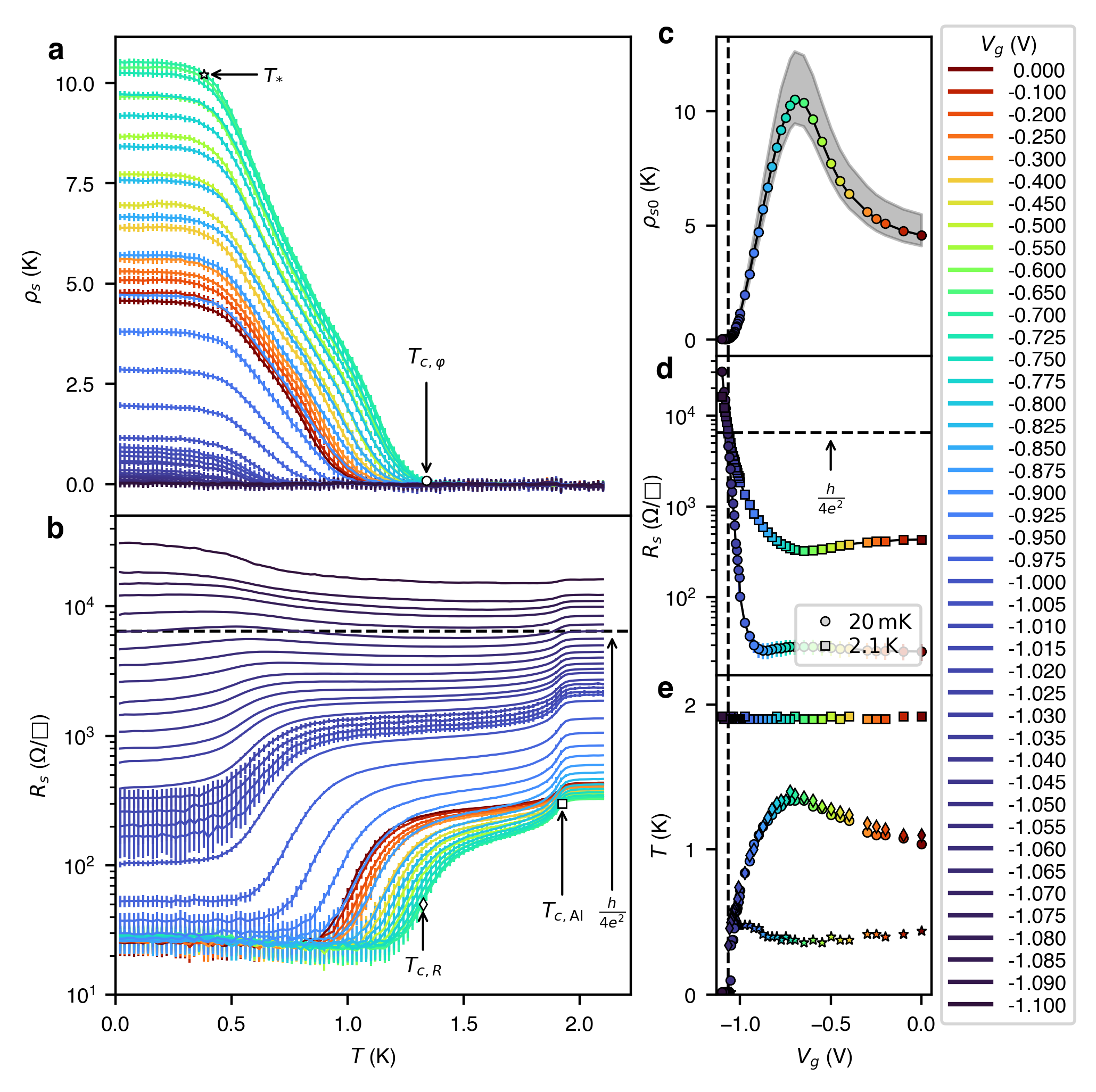}
    \caption{
    {\bf Phase stiffness and sheet resistance data for $b=300$ nm.}
    ({\bf a}) phase stiffness $\rho_s$ and ({\bf b}) sheet resistance $R_s$ as a function of temperature $T$ and gate voltage $V_g$. ({\bf c}) $\rho_s$ at the sample base temperature $T=20$ mK as a function of gate voltage $V_g$. ({\bf d}) Normal state (squares, $T=2.1$ K) and low temperature (circles, $T=20$ mK) sheet resistance. ({\bf e}) Gate voltage dependence of measured temperature scales. $T_{c,\text{Al}}$ (squares): critical temperature of the Al film, identified as the location of the maximum in $\partial R_s/\partial T$ for $T>1.7$ K. $T_{c,R}$ (diamonds): resistive transition of the array, identified as the location of the maximum in $\partial R_s/\partial T$ for $T<1.7$ K. $T_{c,\varphi}$ (circles): onset temperature for measurable diamagnetic response.
    The horizontal dashed line in (b,d) indicates the quantum of resistance for Cooper pairs, $R_Q=h/(4e^2)$, and the vertical dashed line in (c--e) indicates the value of $V_g$ at which the normal state sheet resistance $R_N=R_s(T=2.1\,\mathrm{K})$ is closest to $R_Q$.
    $R_s$ was measured using a current biased four-terminal setup with RMS source-drain current $<10$ nA at frequency 589.7 Hz. The width of the array is $65\,\mu\mathrm{m}$ and the minimum distance between the voltage probes is $94\,\mu\mathrm{m}$, so $R_s=R_{xx}\times 65/94$, where $R_{xx}$ is the measured longitudinal resistance. The $R_s(T)$ curves with large error bars ($R_{s0}\approx(150-400)\,\Omega/\square$) were measured with RMS source-drain current $<1$ nA, hence the worse signal to noise ratio. For this device, the minimum measurable resistance was $\sim$ 25 $\Omega$, likely due to an issue with one of the voltage contacts.
    }
    \label{fig:sweep-300}
\end{figure}

\begin{figure}[h]
    \centering
    \includegraphics[width=0.6\linewidth]{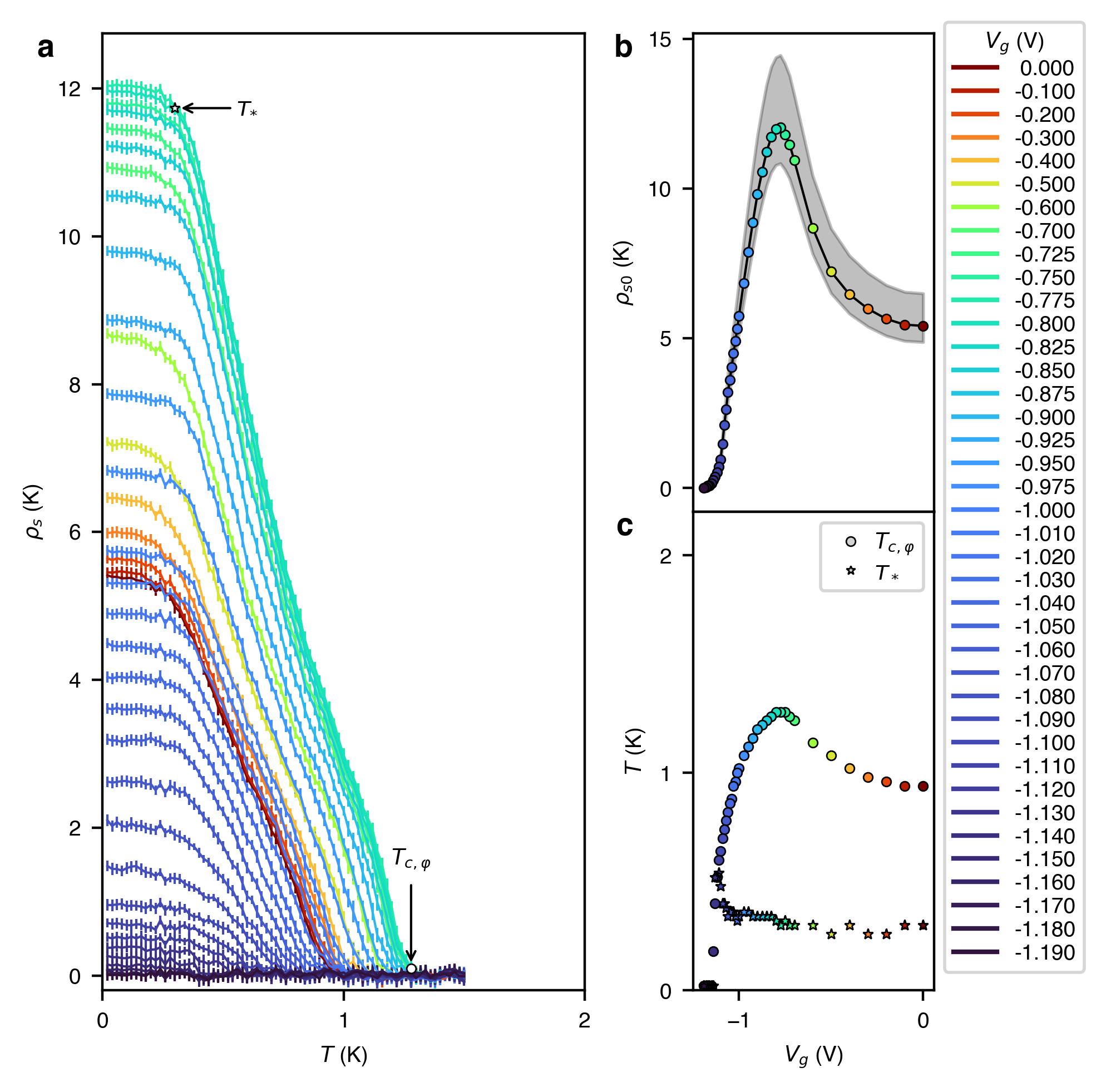}
    \caption{{\bf Phase stiffness data for $b=400$ nm.}}
    \label{fig:sweep-400}
\end{figure}

\begin{figure}[h]
    \centering
    \includegraphics[width=0.6\linewidth]{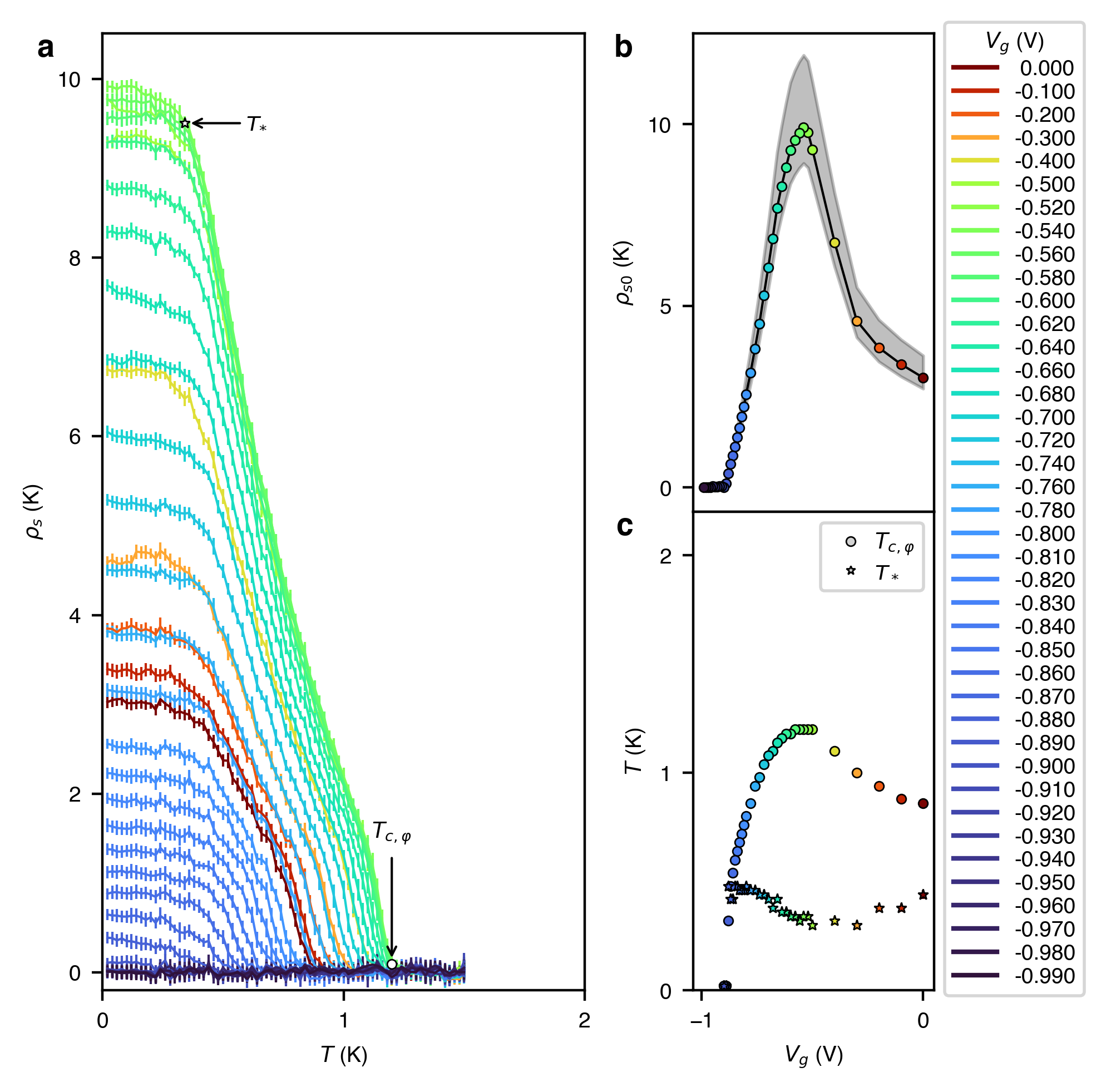}
    \caption{{\bf Phase stiffness data for $b=500$ nm.}}
    \label{fig:sweep-500}
\end{figure}

Figures~\ref{fig:fig2}, \ref{fig:sweep-150}, \ref{fig:sweep-200}, \ref{fig:sweep-250}, \ref{fig:sweep-300}, \ref{fig:sweep-400}, and \ref{fig:sweep-500} show the full temperature-dependent phase stiffness datasets for sample $b=150,$ 200, 250, 300, 400, and 500 nm. The vertical errorbars in these plots indicate the statistical uncertainty due to noise in the measurement. The systematic uncertainty in the superfluid stiffness due to the SQUID-sample standoff is denoted by the gray shaded region in the plots of $\rho_s(T=20\,\text{mK})$ vs. $V_g$.

\begin{figure}[h]
    \centering
    \includegraphics[width=0.6\linewidth]{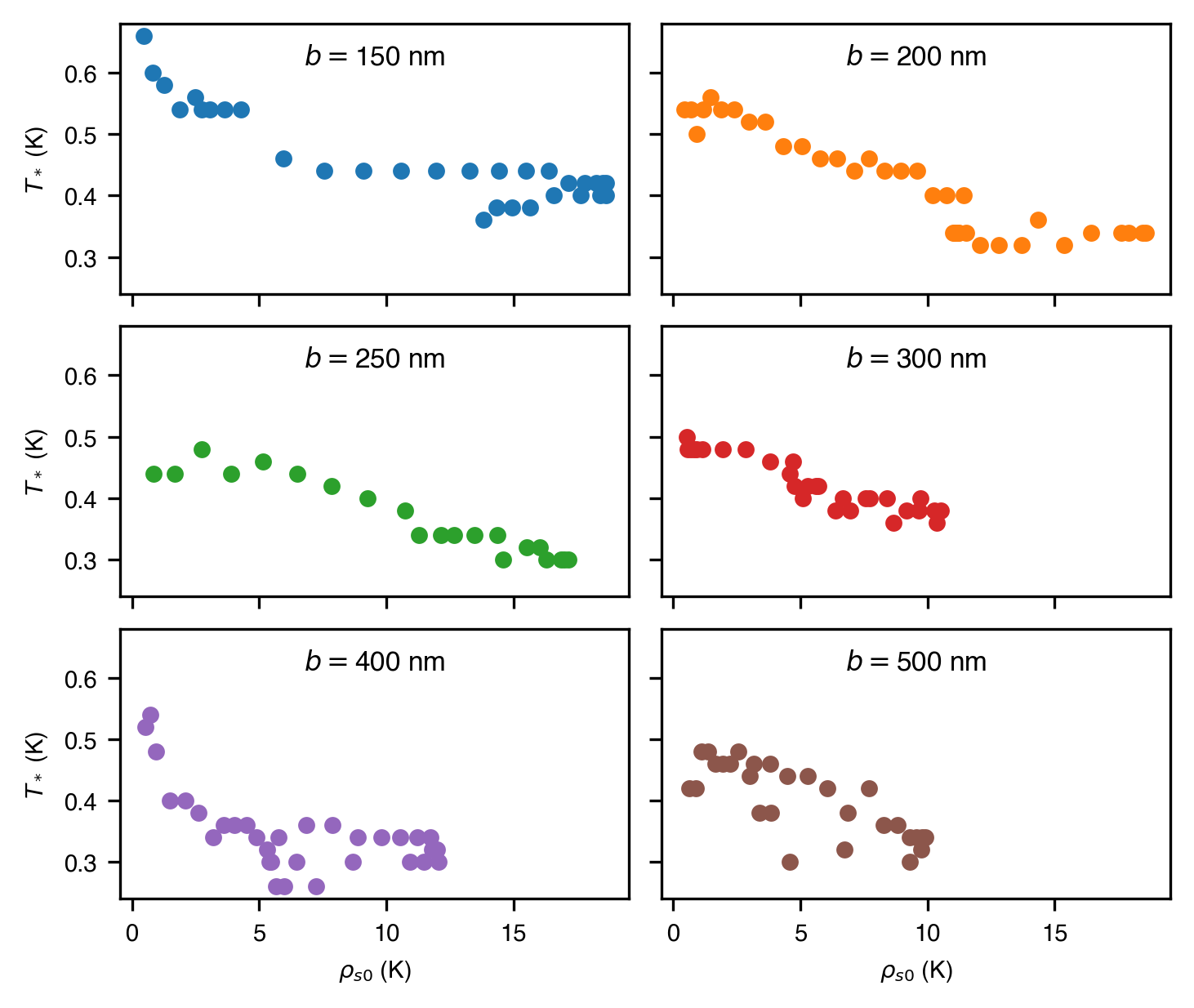}
    \caption{{\bf Crossover temperature vs. zero-temperature phase stiffness.} $T_*$ vs. $\rho_{s0}$ for all six arrays. The crossover temperature is negatively correlated with the zero-temperature phase stiffness and normal state conductivity (not shown).}
    \label{fig:T_star_vs_rhos0}
\end{figure}

The crossover temperature $T_*$ is identified empirically by a deviation from the low-temperature linear trend in $\rho_s(T)$. For each $V_g$, we perform a linear fit to $\rho_s(T)$ for $T<0.2$ K and define $T_*$ to be the lowest temperature for which $\rho_s(T)$ deviates from this low temperature fit by more than 0.2 K. For each array, the crossover temperature $T_*$ is negatively correlated with the $T\to 0$ phase stiffness $\rho_{s0}$, as shown in Figure~\ref{fig:T_star_vs_rhos0}.

\subsubsection*{$T_{c,\varphi}$ vs. $\rho_{s0}$ on a log scale}
\begin{figure}[h]
    \centering
    \includegraphics[width=0.5\linewidth]{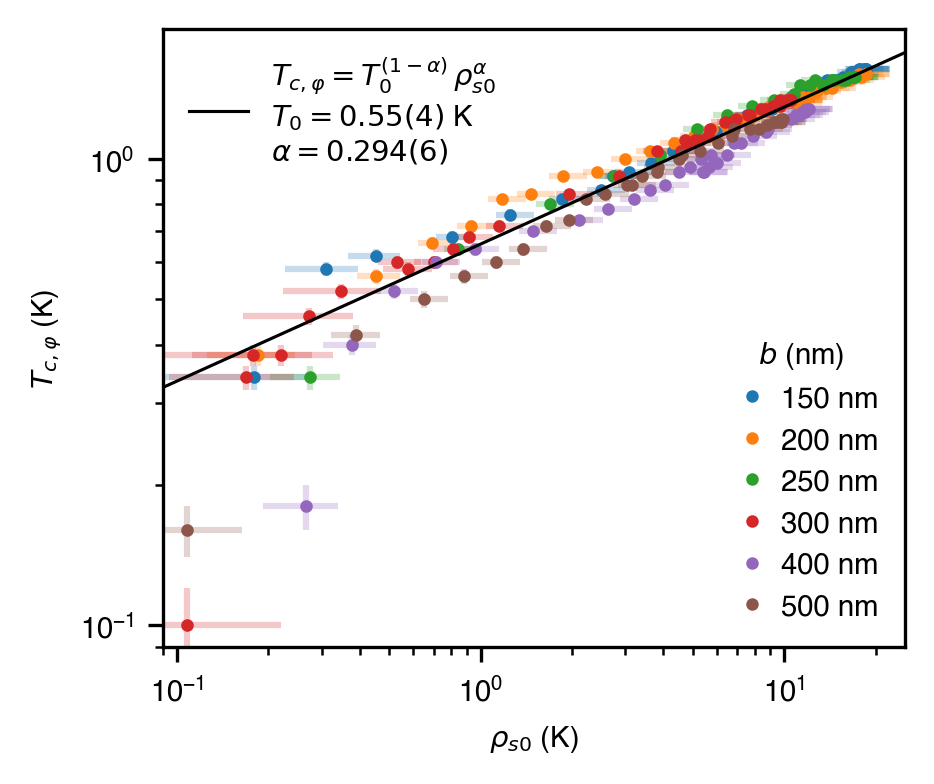}
    \caption{$T_{c,\varphi}(V_g)$ vs. $\rho_{s0}(V_g)$ for all six arrays on a log-log scale. A fit of $T_{c,\varphi}=T_0^{(1-\alpha)}\rho_{s0}^\alpha$ to the aggregated dataset gives $T_0=0.55(4)$ K and $\alpha=0.294(6)$. The spread between devices exceeds this uncertainty, so we regard the power law as an approximate empirical scaling.}
    \label{fig:fig3-loglog}
\end{figure}

Figure~\ref{fig:fig3-loglog} shows $T_{c,\varphi}$ vs. $\rho_{s0}$ on a log scale.

\subsubsection*{Transport on a linear scale}

\begin{figure}[h]
    \centering
    \includegraphics[width=0.6\linewidth]{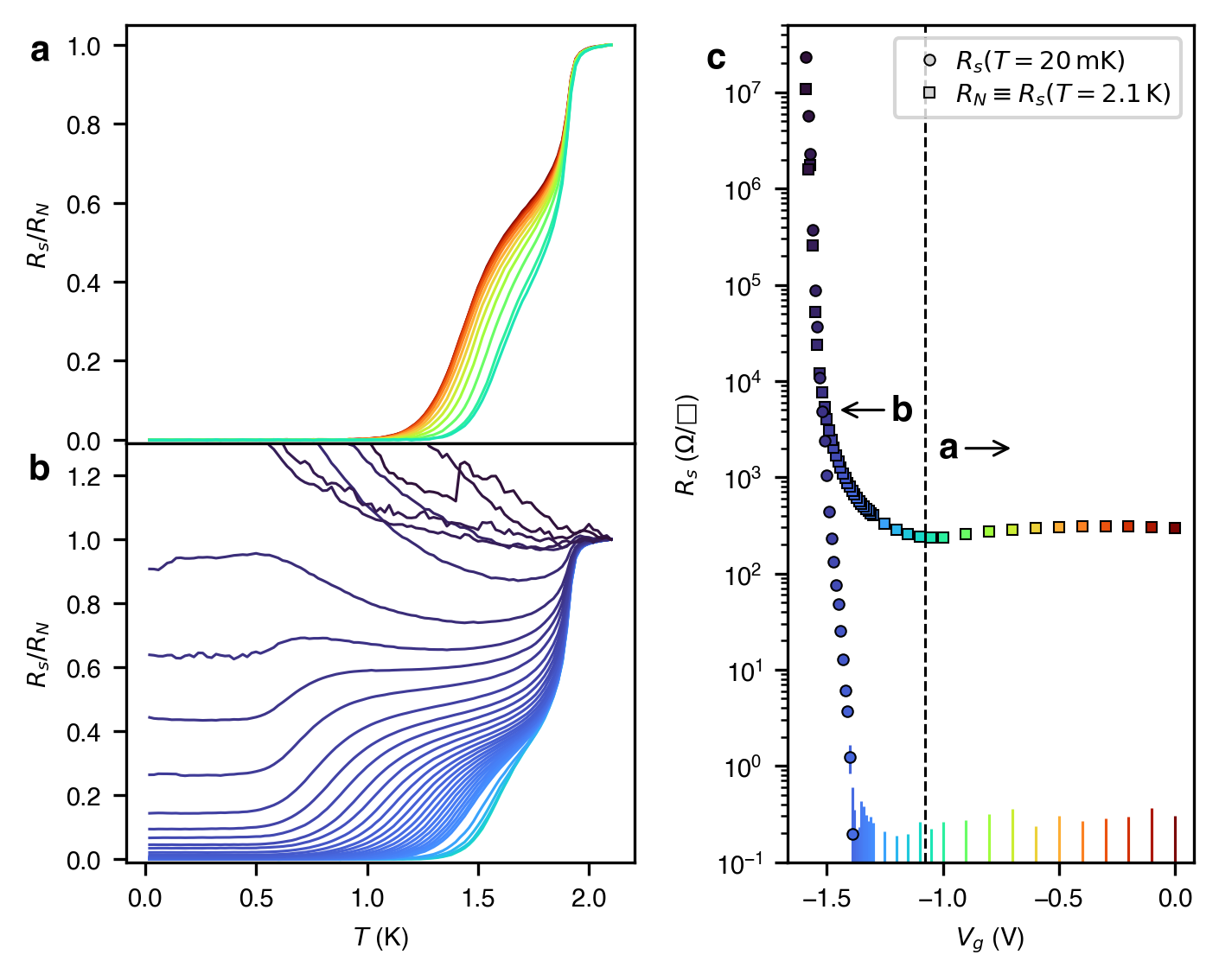}
    \caption{
    {\bf Temperature-dependent transport for $b=200$ nm on a linear scale.}
    Normalized sheet resistance $R_s/R_N=R_s(T)/R_s(2.1\,\text{K})$ for $V_g\geq V_{g,\text{peak}}$ ({\bf a}) and $V_g<V_{g,\text{peak}}$ ({\bf b}). ({\bf c}) Low-temperature (circles) and normal state (squares) sheet resistance as a function of gate voltage.
    In the superconducting and anomalous metal regimes, $V_g\geq -1.55$ V, $R_s$ was measured using a current biased four-terminal setup with RMS source-drain current $<5$ nA at frequency 137.77 Hz. In the insulating regime, $V_g < -1.55$ V, $R_s$ was measured using a two-terminal setup with RMS voltage bias $<0.5$ mV and source-drain current $<1.4$ nA at 17.777 Hz. This is the same dataset that is shown in Fig.~\ref{fig:fig4}b.
    }
    \label{fig:RS-linear-200}
\end{figure}

\begin{figure}[h]
    \centering
    \includegraphics[width=0.6\linewidth]{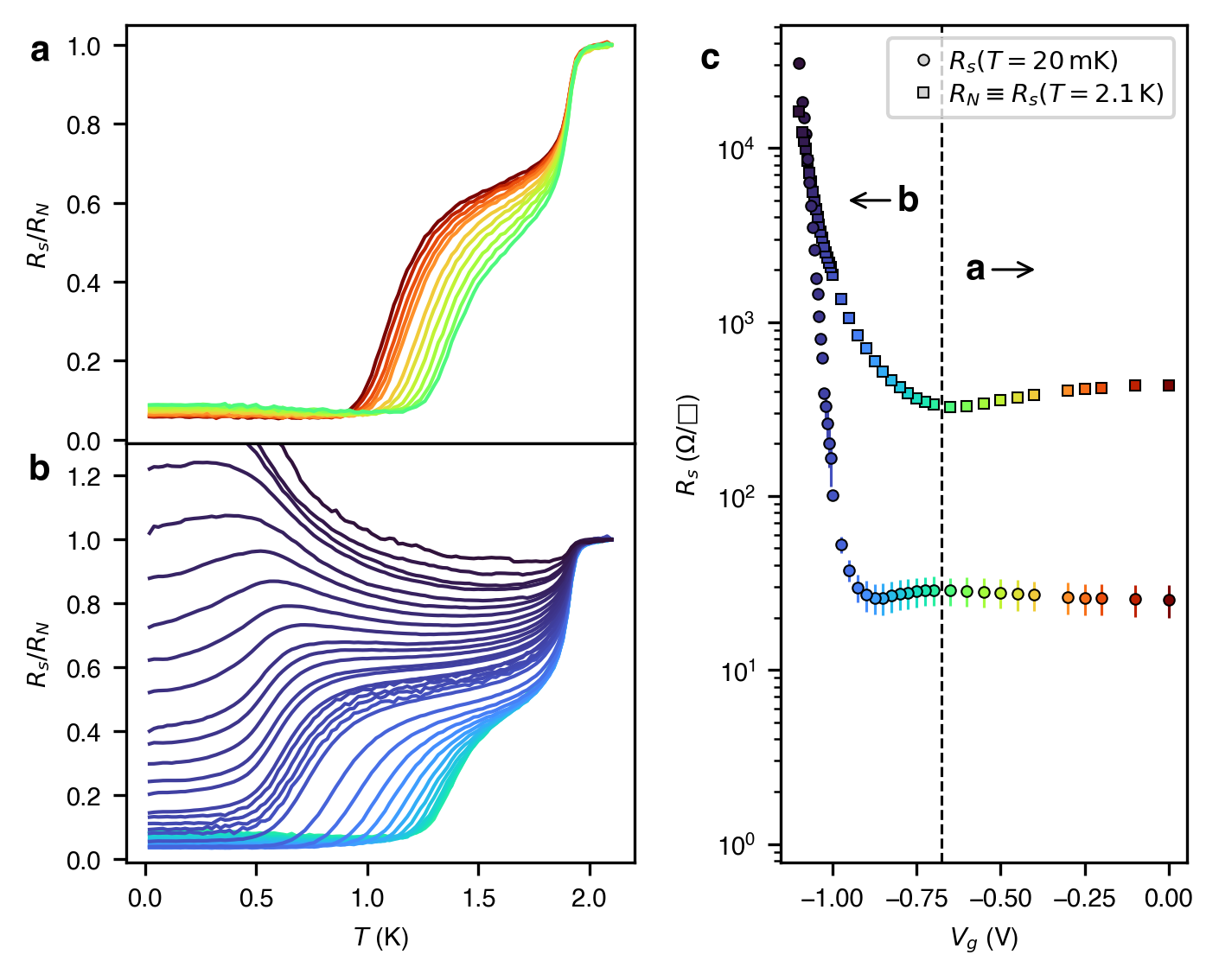}
    \caption{
    {\bf Temperature-dependent transport for $b=300$ nm on a linear scale.}
    Normalized sheet resistance $R_s/R_N=R_s(T)/R_s(2.1\,\text{K})$ for $V_g\geq V_{g,\text{peak}}$ ({\bf a}) and $V_g<V_{g,\text{peak}}$ ({\bf b}). ({\bf c}) Low-temperature (circles) and normal state (squares) sheet resistance as a function of gate voltage. For this device, the minimum measurable resistance was $\sim$ 25 $\Omega$, likely to an issue with one of the voltage contacts. $R_s$ was measured using a current biased four-terminal setup with RMS source-drain current $<10$ nA at frequency 589.7 Hz. This is the same dataset that is shown in Fig.~\ref{fig:sweep-300}.
    }
    \label{fig:RS-linear-300}
\end{figure}

\begin{figure}[h]
    \centering
    \includegraphics[width=0.6\linewidth]{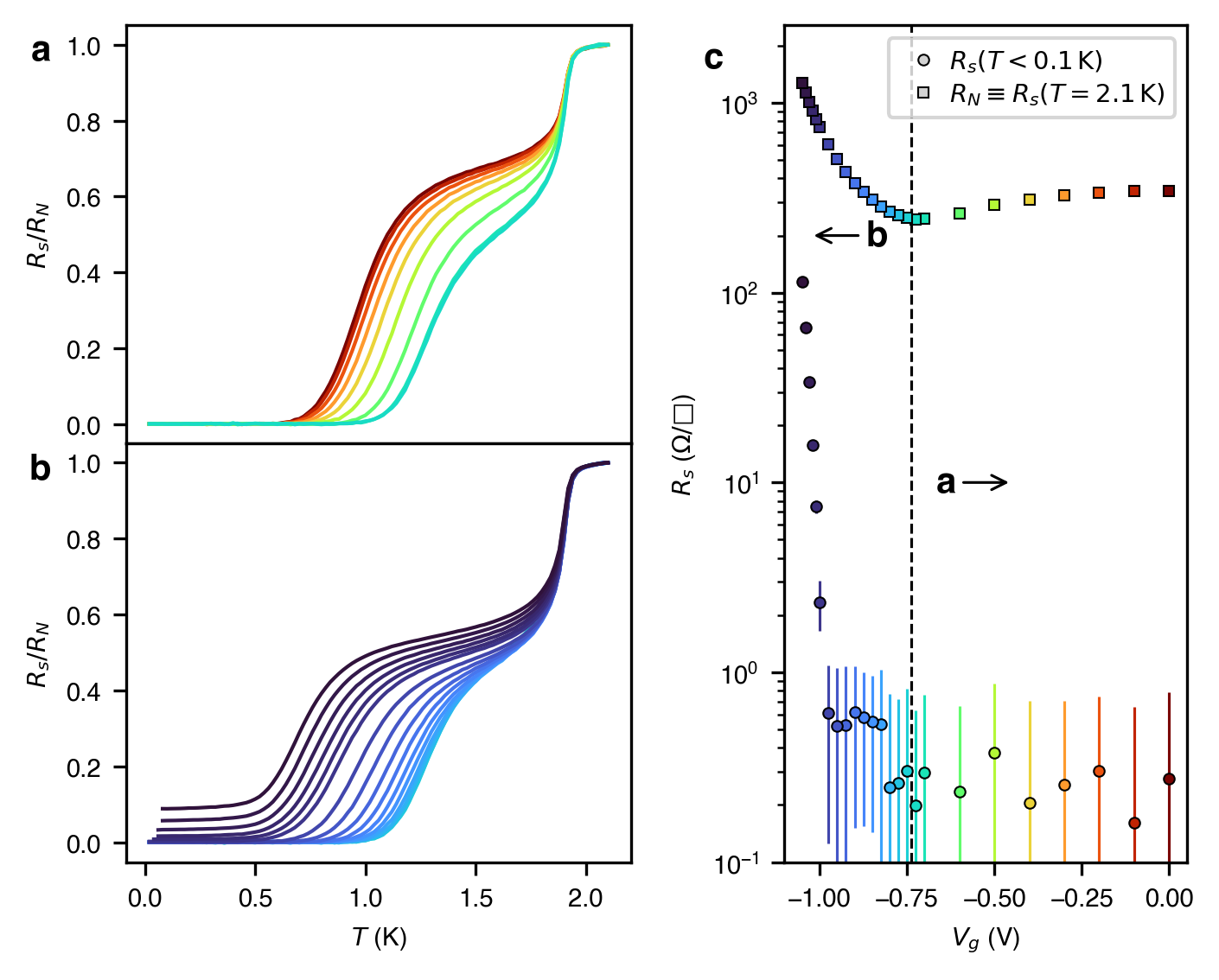}
    \caption{{\bf Temperature-dependent transport for $b=400$ nm on a linear scale.} Normalized sheet resistance $R_s/R_N=R_s(T)/R_s(2.1\,\text{K})$ for $V_g\geq V_{g,\text{peak}}$ ({\bf a}) and $V_g<V_{g,\text{peak}}$ ({\bf b}). ({\bf c}) Low-temperature (circles) and normal state (squares) sheet resistance as a function of gate voltage. Measurement of this device was interrupted by a clog in the dilution refrigerator, which required warming the fridge to room temperature. One of the wire bonds broke during this thermal cycle, hence the truncated dataset. The phase stiffness data presented for this device is from the second cooldown of the device. $R_s$ was measured with an RMS source-drain current of 5 nA at a frequency of 17.777 Hz.}
    \label{fig:RS-linear-400}
\end{figure}

Figures~\ref{fig:RS-linear-200}, \ref{fig:RS-linear-300}, and \ref{fig:RS-linear-400} show the measured temperature-dependent transport for $b=200$ nm, 300 nm, and 400 nm on a linear scale.

\subsubsection*{Spatial inhomogeneity at low carrier density}

\begin{figure}
    \centering
    \includegraphics[width=\linewidth]{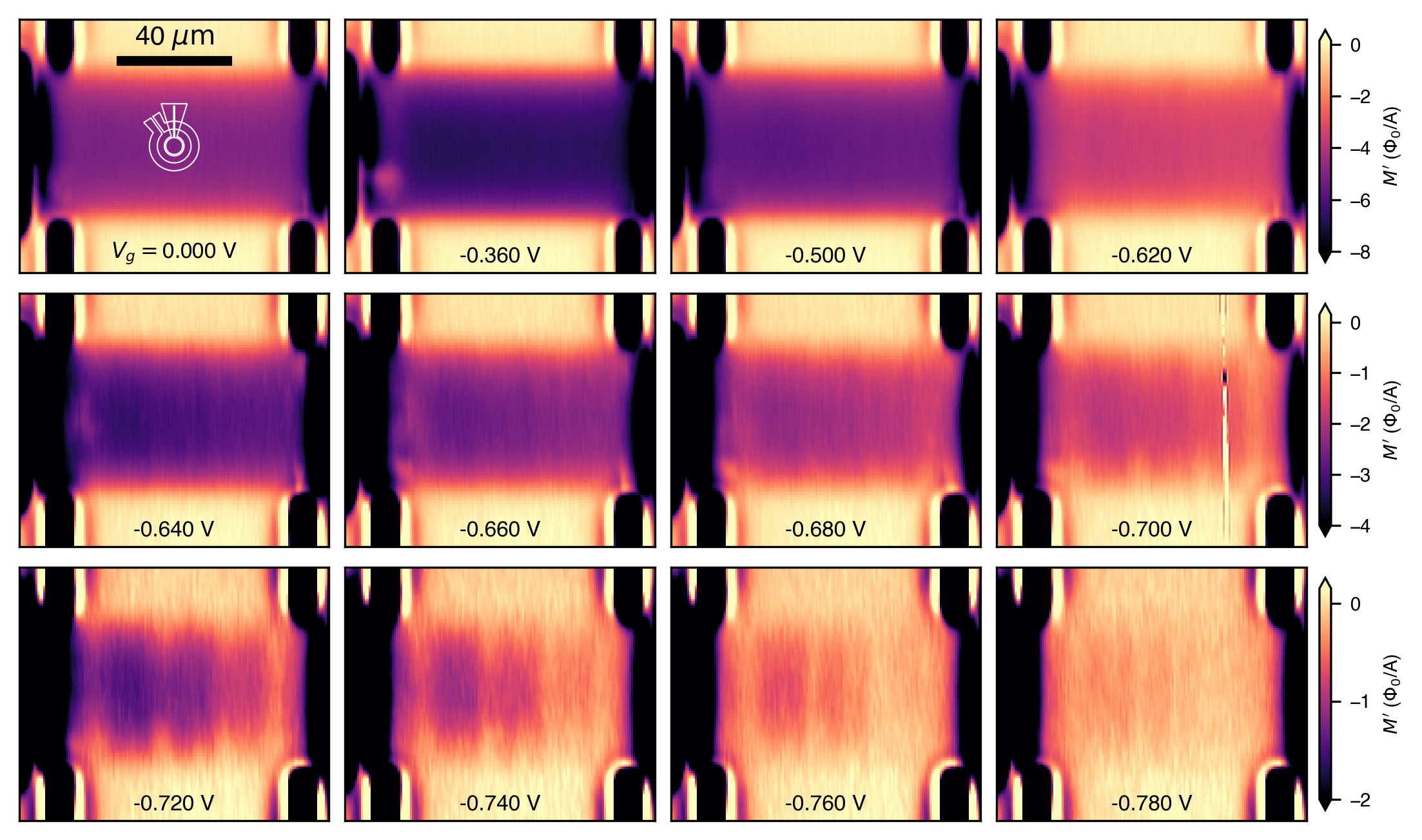}
    \caption{
    {\bf Spatial inhomogeneity at low carrier density in the $b=150$ nm array.}
    Scanning SQUID maps of the in-phase magnetic response $M'$ measured at $T=30$ mK at gate voltages from 0 V to -0.78 V. At large carrier density (from $V_g=0$ V to $V_g=V_{g,\text{peak}}\approx -0.35$ V), the magnetic response is spatially uniform. There is a vortex trapped in the lower left corner of the array, which is visible as a light spot in the $V_g=-0.36$ V panel. For $V_g$ well below $V_{g,\text{peak}}$, $M'$ exhibits stripe-like spatial inhomogeneity. All four transport leads were grounded at room temperature for this measurement, i.e., there was no applied source-drain current. The scale bar in the upper left applies to all panels. The SQUID pickup loop and field coil geometry is drawn in the upper left. The color scale for each row is indicated by the colorbar to the right of the row.
    }
    \label{fig:sample150-scans}
\end{figure}

Figures~\ref{fig:sample150-scans}, \ref{fig:sample250-scans}, \ref{fig:sample300-scans}, and \ref{fig:sample400-scans} show scanning SQUID maps of the low-temperature in-phase magnetic response $M'$ for the $b=150$ nm, 250 nm, 300 nm, and 400 nm arrays, respectively. All devices are spatially uniform at high carrier density and become inhomogeneous at low carrier density.

\begin{figure}
    \centering
    \includegraphics[width=\linewidth]{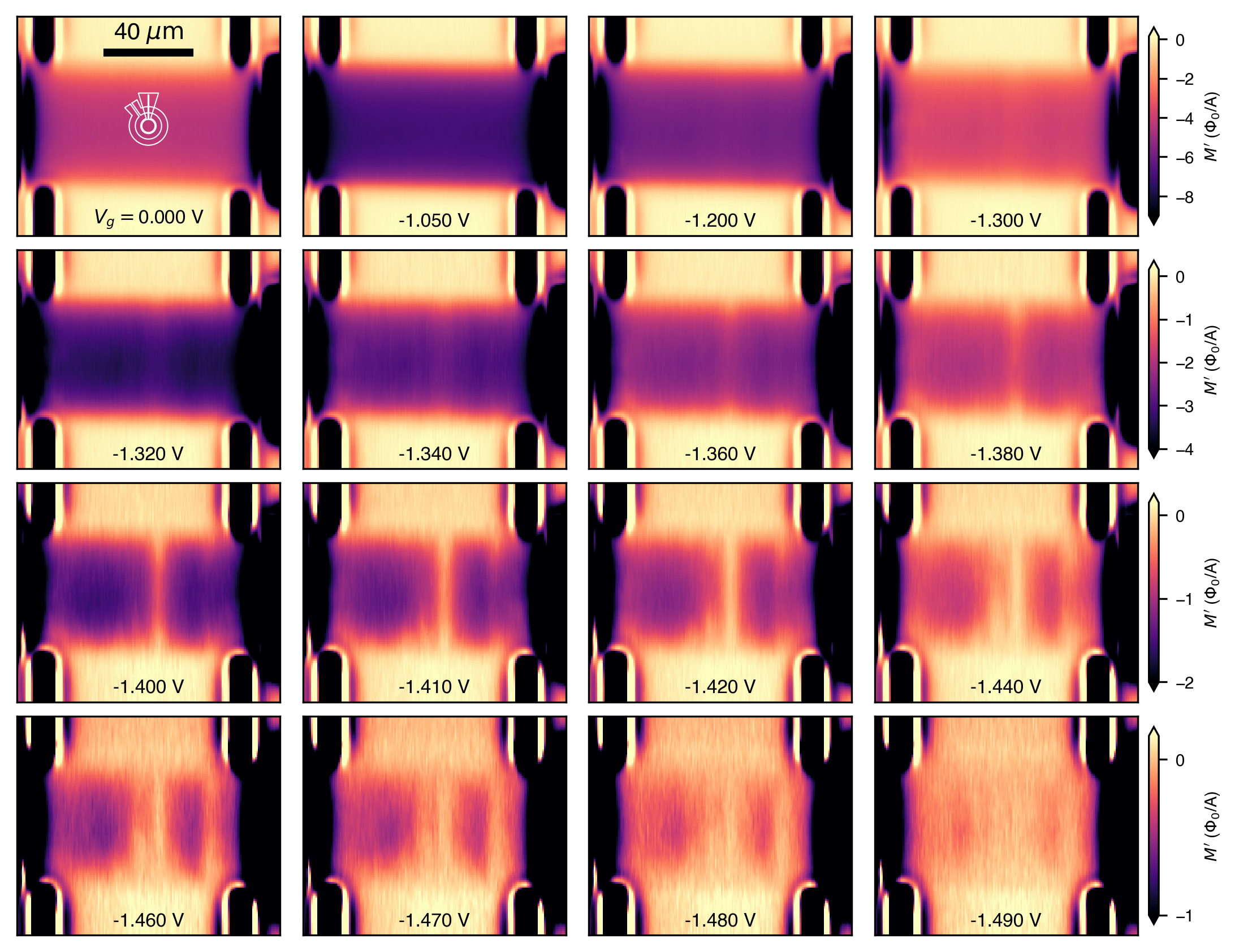}
    \caption{
    {\bf Spatial inhomogeneity at low carrier density in the $b=200$ nm array.}
    Scanning SQUID maps of the in-phase magnetic response $M'$ measured at $T=20$ mK at gate voltages from 0 V to $-1.49$ V. At large carrier density (from $V_g=0$ V to $V_g=-1.05\,\mathrm{V}\approx V_{g,\text{peak}}$), the magnetic response is spatially uniform. For $V_g$ well below $V_{g,\text{peak}}$, $M'$ exhibits stripe-like spatial inhomogeneity. This measurement was performed with an applied transport current of 80 nA RMS, but the inhomogeneity is not sensitive to or caused by the applied current (see Figure~\ref{fig:sample200-5-80nA}). The scale bar in the upper left applies to all panels. The SQUID pickup loop and field coil geometry is drawn in the upper left. The color scale for each row is indicated by the colorbar to the right of the row.
    }
    \label{fig:sample200-scans}
\end{figure}

\begin{figure}
    \centering
    \includegraphics[width=\linewidth]{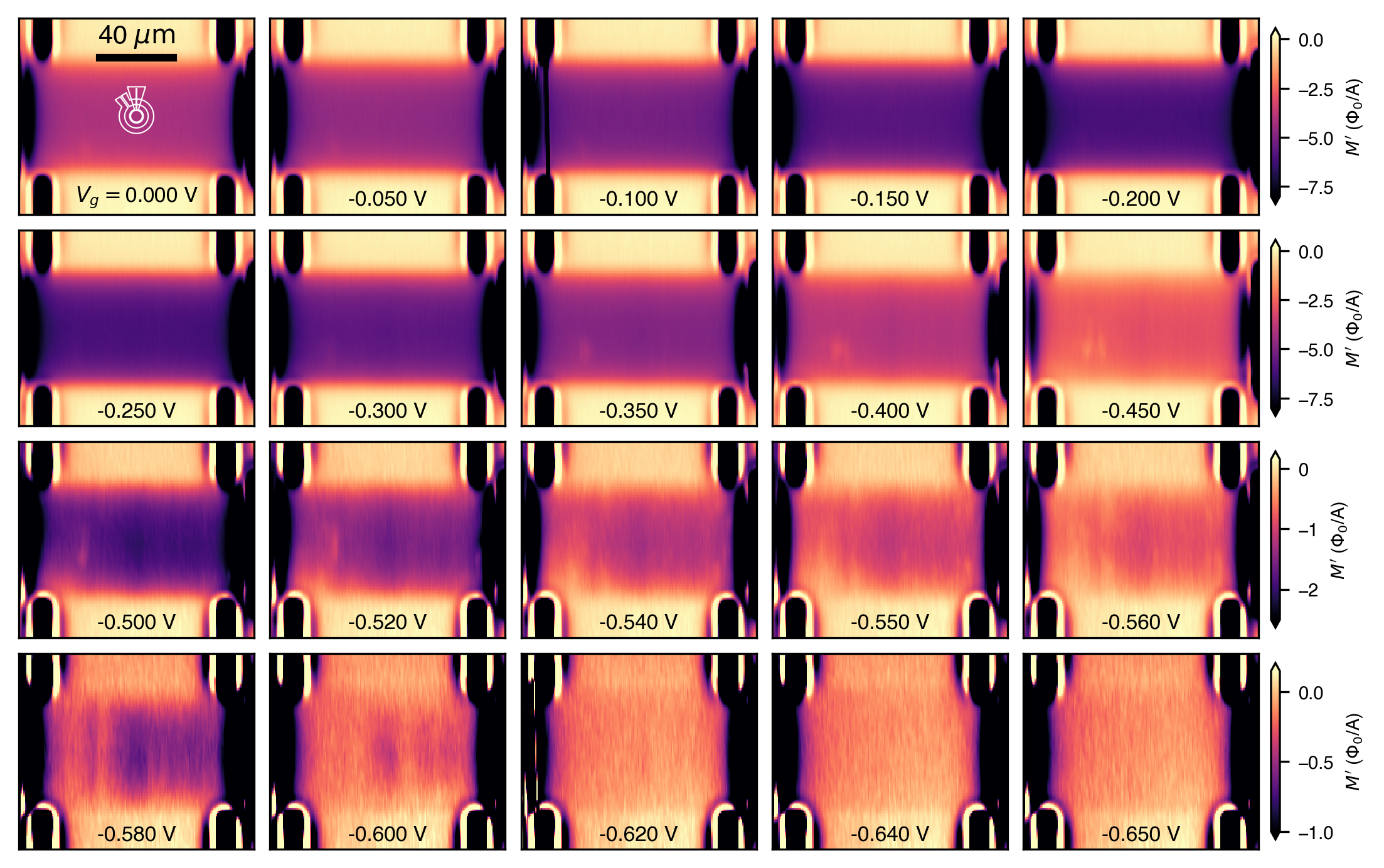}
    \caption{
    {\bf Spatial inhomogeneity at low carrier density in the $b=250$ nm array.}
    Scanning SQUID maps of the in-phase magnetic response $M'$ measured at $T=30$ mK at gate voltages from 0 V to $-0.65$ V. At large carrier density (from $V_g=0$ V to $V_g=V_{g,\text{peak}}\approx -0.225$ V), the magnetic response is spatially uniform. For $V_g$ well below $V_{g,\text{peak}}$, $M'$ exhibits stripe-like spatial inhomogeneity. All four transport leads were grounded at room temperature for this measurement, i.e., there was no applied source-drain current. The scale bar in the upper left applies to all panels. The SQUID pickup loop and field coil geometry is drawn in the upper left. The color scale for each row is indicated by the colorbar to the right of the row.
    }
    \label{fig:sample250-scans}
\end{figure}

\begin{figure}
    \centering
    \includegraphics[width=\linewidth]{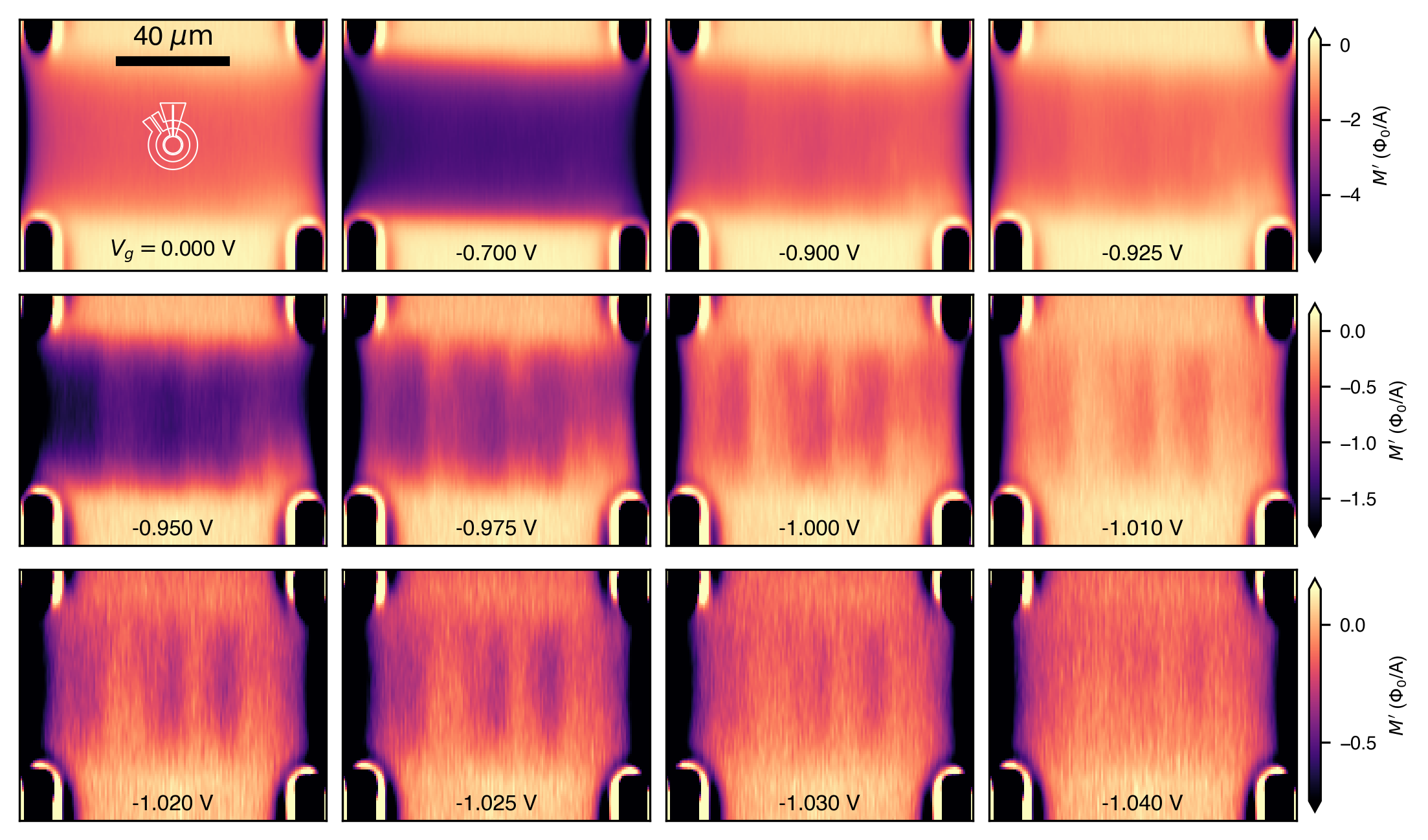}
    \caption{
    {\bf Spatial inhomogeneity at low carrier density in the $b=300$ nm array.}
    Scanning SQUID maps of the in-phase magnetic response $M'$ measured at $T= 20$ mK at gate voltages from 0 V to $-1.04$ V. At large carrier density ($V_g=0$ V and $V_g=V_{g,\text{peak}}=-0.775$ V), the magnetic response is spatially uniform. For $V_g$ well below $V_{g,\text{peak}}$, $M'$ exhibits stripe-like spatial inhomogeneity. For $V_g\leq$ -1.05 V, the magnetic response of the array is below our sensitivity for the chosen measurement parameters. All four transport leads were grounded at room temperature for this measurement, i.e., there was no applied source-drain current. The scale bar in the upper left applies to all panels. The SQUID pickup loop and field coil geometry is drawn in the upper left. The color scale for each row is indicated by the colorbar to the right of the row.
    }
    \label{fig:sample300-scans}
\end{figure}

\begin{figure}
    \centering
    \includegraphics[width=\linewidth]{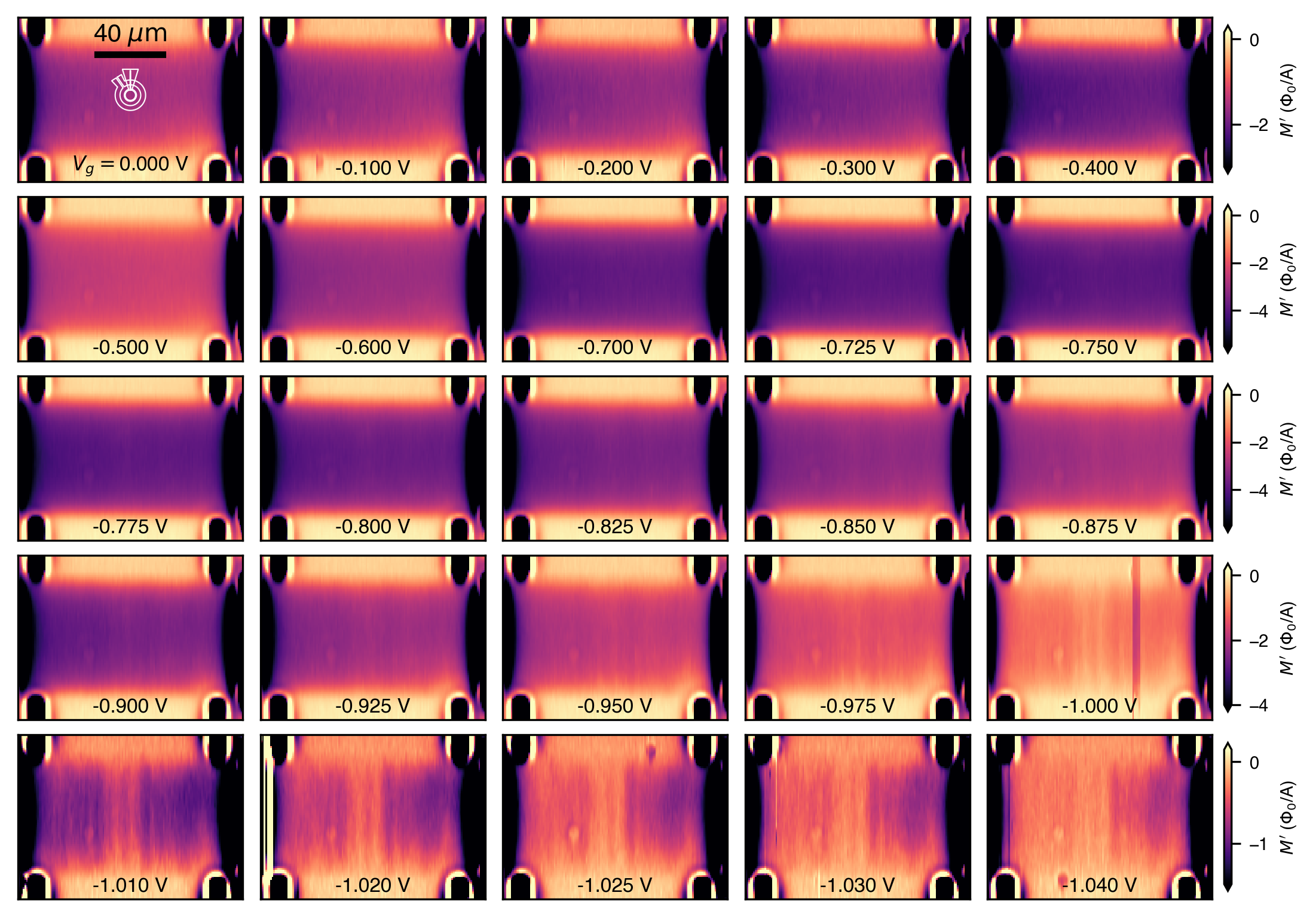}
    \caption{
    {\bf Spatial inhomogeneity at low carrier density in the $b=400$ nm array.}
    Scanning SQUID maps of the in-phase magnetic response $M'$ measured at $T=20$ mK at gate voltages from 0 V to $-1.040$ V. At large carrier density (from $V_g=0$ V to $V_g=V_{g,\text{peak}}\approx -0.750$ V), the magnetic response is spatially uniform. For $V_g$ well below $V_{g,\text{peak}}$, $M'$ exhibits stripe-like spatial inhomogeneity. This measurement was performed with an RMS source-drain current of 5 nA. The scale bar in the upper left applies to all panels. The SQUID pickup loop and field coil geometry is drawn in the upper left. The color scale for each row is indicated by the colorbar to the right of the row.
    }
    \label{fig:sample400-scans}
\end{figure}

\begin{figure}
    \centering
    \includegraphics[width=\linewidth]{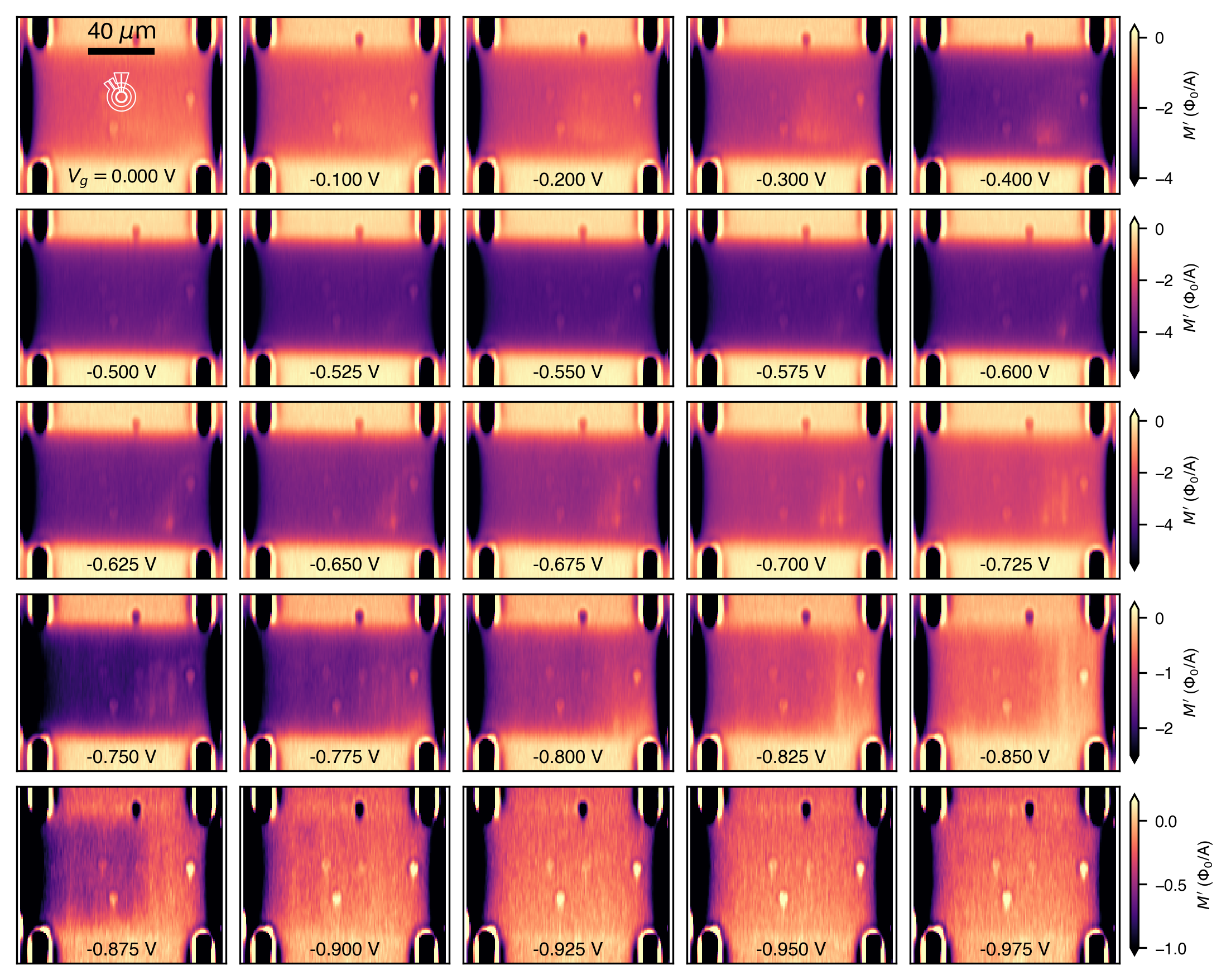}
    \caption{
    {\bf Spatial inhomogeneity at low carrier density in the $b=500$ nm array.}
    Scanning SQUID maps of the in-phase magnetic response $M'$ measured at $T=20$ mK at gate voltages from 0 V to $-0.975$ V. The scale bar in the upper left applies to all panels. The light ``tadpole'' features (reduced diamagnetic signal) are due to the SQUID coming into contact with debris on the top gate surface. Corresponding features are visible in the capacitance measurement channel that detects mechanical contact between the SQUID and sample (not shown). The dark ``tadpole'' feature near the upper edge of the array is a piece of diamagnetic debris, most likely Al that was deposited on the sample surface during wire bonding or handling. Large-scale spatial inhomogeneity is visible from $V_g=-0.8$ V to $V_g=-0.9$ V. The SQUID pickup loop and field coil geometry is drawn in the upper left. The color scale for each row is indicated by the colorbar to the right of the row.
    }
    \label{fig:sample500-scans}
\end{figure}

\end{appendices}


\end{document}